\documentclass{aa}
\pdfoutput = 1
\usepackage{graphicx}
\usepackage{txfonts}
\usepackage{natbib}

\newdimen\captwidth
\newdimen\figwidth
\newcommand{\fomb}{Fom~b}
\newcommand{\fomc}{Fom~c}
\newcommand{\fom}{Fomalhaut}
\newcommand{\mjup}{M_\mathrm{Jup}}
\newcommand{\msun}{M_\odot}
\let\dy=\displaystyle
\newcommand{\rd}{\mathrm{d}}
\newcommand{\excs}{\extracolsep{\fill}}
\begin{document}
\title{An independent determination of Fomalhaut b's orbit and the dynamical effects on the outer dust belt}
\author{H. Beust \inst{1} \and
  J.-C. Augereau \inst{1} \and  A. Bonsor \inst{1}
  \and J.R. Graham \inst{3} \and P. Kalas \inst{3}
   J. Lebreton \inst{1} \and A.-M. Lagrange \inst{1}
   \and S. Ertel \inst{1} \and V. Faramaz \inst{1} \and
   P. Th\'ebault \inst{2}}
\institute{UJF-Grenoble 1 / CNRS-INSU, Institut de Plan\'etologie et
  d'Astrophysique de Grenoble (IPAG) UMR 5274, Grenoble, F-38041,
  France
\and Observatoire de Paris, Section de Meudon, F-92195 Meudon Principal
  Cedex, France
\and Department of Astronomy, University of California at Berkeley,
Berkeley, CA 94720, USA}
\date{Received 8 July 2013; Accepted 13 November 2013}
\offprints{H. Beust}
\mail{Herve.Beust@obs.ujf-grenoble.fr}
\titlerunning{Orbital determination of Fomalhaut b}
\authorrunning{H. Beust et al.}
\abstract{The nearby star Fomalhaut harbours a cold, moderately
eccentric ($e\sim0.1$) dust belt with a sharp inner edge near 133 au.
A low-mass, common proper motion companion, \fom~b (Fom b), was discovered
near the inner edge and was identified as a planet candidate that
could account for the belt morphology. However, the most recent orbit
determination based on four epochs of astrometry over eight years reveals
a highly eccentric orbit ($e=0.8\pm0.1$) that appears to cross the belt
in the sky plane projection.
}
{We perform here a full orbital determination based on the available
astrometric data to independently validate the orbit estimates previously
presented. Adopting our values for the orbital elements and
their associated uncertainties, we then study the dynamical interaction
between the planet and the dust ring, to check whether the proposed
disk sculpting scenario by \fomb\ is plausible.}
{We used a dedicated MCMC code to derive the statistical distributions
of the orbital elements of \fomb. Then we used symplectic N-body
integration to investigate the dynamics of the dust belt, as
perturbed by a single planet. Different attempts were made assuming
different masses for \fomb. We also performed a semi-analytical study
to explain our results.}
{Our results are in good agreement with others
regarding the orbit of \fomb.  We find that the orbit is highly eccentric,
is close to apsidally aligned with the belt, and has a mutual inclination
relative to the belt plane of $<29\degr$ (67\%\ confidence). If coplanar,
this orbit crosses the disk. Our dynamical study then reveals that the
observed planet could sculpt a
transient belt configuration with a similar eccentricity to what is
observed, but it would not be simultaneously apsidally aligned with
the planet. This transient configuration only occurs a short time after
the planet is placed on such an orbit (assuming an initially circular disk),
a time that is inversely proportional to the planet's mass, and that is in
any case much less than the 440\,Myr age of the star.}
{We constrain how long the observed dust belt could have
survived with \fomb\ on its
current orbit, as a function of its possible mass. This
analysis leads us to conclude that \fomb\ is likely to have low
mass, that it is unlikely to be responsible for the sculpting of the
belt, and that it supports the hypothesis of a more massive, less eccentric
planet companion \fom~c.}
\maketitle
\section{Introduction}
The presence of circumstellar dust orbiting the nearby
\citep[$d=7.7\,$pc;][]{mam12,van07} A3V star
\fom\ ($\alpha\;$Psa, HD 216956, HIP 113368) has been known for a long
time through its thermal emission \citep{aum85}. The spatial structure
of its debris disk was furthermore specified by direct imaging
\citep{hol03,kal05}. HST coronographic images by \citet{kal05}
have revealed a large dust belt in optical scattered light,
extending between 133\,au
and 158\,au and modeled as a moderately eccentric ring ($e=0.11\pm0.1$) with a
$13.4\pm1.0\,$au offset between its centre and the star.
The investigators suggest that an undetected planet could account for
these features, as supported by numerical \citep{del05} and
semi-analytic studies \citep{qui06}.

\citet{kal08} then reported the detection of a planet candidate
(subsequently termed Fomalhaut~b, hereafter \fomb)
orbiting the star at 119\,au, only 18\,au inside the dust belt,
thus strongly supporting its
putative shepherding role for the inner edge of the belt.
The optical detections of \fomb\ with HST/ACS were confirmed by two
independent analyses of the data \citep{cur12,gal13}.
Since \fomb\ was $not$ detected at infrared wavelengths
\citep{kal08, mar09,jan12},
it has been suggested that \fomb\ could represent starlight reflected
from dust grains, possibly bound to a planet
in the form of a large planetary ring \citep{kal08} or a cloud due to the
collisional erosion of irregular planetary satellites \citep{ken11}.

The mass and orbit of \fomb\ continues to require better constraints.
An accurate knowledge of these parameters would clearly help
define its interaction with the dust ring orbiting
\fom. It is not possible to constrain \fomb's mass (hereafter $m$)
from photometry because the emission detected is likely dominated by
the circumplanetary dust scattering. Dynamical modeling of its
interaction with its environment is
therefore a valuable way to derive constraints. \citet{kal08} give a
conservative upper limit $m<3\,$Jupiter masses (hereafter $\mjup$),
while \citet{chi09} reduces it to possibly $0.5\,\mjup$, under the
assumption that the planet is responsible for the sculpting of the dust
ring. Based on photometric estimates,
\citet{cur12} claim $m<2\,\mjup$, but other recent studies (dynamical
or photometric) suggest a possibly much lower mass in the super-Earth regime
\citep{jan12,ken11,gal13}. According to \citet{jan12}, the recent
non-detection of \fomb\ at $\lambda=4.5\,\mu$m in thermal infrared
excludes any Jovian-sized planet, and is rather compatible with
a $\sim 10\,M_\oplus$ object.

Based on the first two epochs of HST detections in 2004 and 2006,
separated by only 1.7 years, \fomb's orbit was initially thought to be
nearly circular or moderately eccentric
\citep[$e=0.11$--0.13][]{chi09} and coplanar with the outer dust belt,
as its orbital motion was detected nearly parallel to its inner
edge. This constraint was deduced assuming that \fomb\ is responsible
for the belt's inner edge sculpting. This assumption was nevertheless
recently questioned by \citet{bol12} who suggest the presence of other
shepherding planets, in particular outside the outer edge of the ring.
Fom b was recovered at a third (2010) and fourth (2012) epoch using
HST/STIS coronagraphy \citep{kal13},
allowing accurate measurements of its sky-plane motion when all four
epochs of astrometry are combined. These investigators independently
developed a
Markov chain Monte Carlo code to estimate the orbital elements \citep{gra13},
producing a surprising result that the orbit of Fomalhaut b is highly eccentric,
and will appear to cross the dust belt in the sky plane projection.

The purpose of this paper is first to perform an independent analysis
of the available
astrometric data of \fomb\ to derive refined orbital constraints using
a Markov-Chain Monte Carlo (MCMC) method that was developed by one
of us (H. Beust) and already used to fit $\beta\:$Pic b's
orbit \citep{chau12}.  This independent analysis confirms the
eccentric nature of
the orbit, and that it is very probably coplanar with the disk and
apsidally aligned (Sect.~2). In Sect.~3 we numerically
investigate the dynamics of the \fom\ system including \fomb\ and the dust
belt. We
present in Sect.~4 a semi-analytical study to explain the numerical result we
derive. Our conclusions are presented in Sect.~5.
\section{Orbital fitting}
\subsection{Astrometric data}
\begin{table}[t]
\caption{Summary of compiled astrometric data of \fomb\ relative to \fom}
\label{astrom}
\begin{tabular*}{\columnwidth}{@{\excs}lllll}     % 3 columns
\hline\hline\noalign{\smallskip}
UT Date & \multicolumn{2}{c}{Declination ($\delta$, mas)} &
\multicolumn{2}{c}{Right Ascension ($\alpha$, mas)}\\
& \multicolumn{2}{c}{$\overbrace{\mbox{\rule{3truecm}{0truecm}}}$} &
\multicolumn{2}{c}{$\overbrace{\mbox{\rule{3truecm}{0truecm}}}$}\\
& \multicolumn{1}{c}{K13} &  \multicolumn{1}{c}{G13} &
\multicolumn{1}{c}{K13} &  \multicolumn{1}{c}{G13}\\
\noalign{\smallskip}\hline\noalign{\smallskip}
Oct. 25/26, 2004 & $9175\pm17$ & $9190\pm20$ & $-8587\pm24$ & $-8590\pm20$\\
Jul. 17/20, 2006 & $9365\pm19$ & $9360\pm20$ & $-8597\pm22$ & $-8640\pm20$\\
Sep. 13, 2010  & $9822\pm44$ & $9790\pm30$ & $-8828\pm42$ & $-8810\pm30$\\
May 29/31, 2012  & $10016\pm37$ && $-8915\pm35$ &\\
\hline\noalign{\smallskip}
\multicolumn{5}{l}{K13 = \citet{kal13}; G13 = \citet{gal13}}
\end{tabular*}
\end{table}
\fomb\ was observed with HST/ACS/HRC and HST/STIS at four epochs in
2004, 2006, 2010 and 2012. A detailed analysis and the corresponding
astrometric data are given in \citet{kal13}. \citet{gal13} also give
independently derived astrometric measurement for all epochs
before 2012. All these data
are summarised in Table~\ref{astrom}. While both sets of data are
mutually compatible within their respective error bars, we note a
slight difference between data from \citet{kal13} and those from
\citet{gal13}. To check the sensitivity of our orbital determination,
we chose then to perform our orbital analysis with two independent
sets of data: a first one with all data from \citet{kal13}, and a
second one with the \citet{gal13} data for the 2004, 2006 and 2010
data points, and the 2012 measurement from \citet{kal13}.
\subsection{Orbital fit}
The detected orbital motion with four epochs is in principle sufficient
to try a first orbital determination. This is nevertheless not a
straightforward task. Given the long expected orbital period of
\fomb\ (hundreds of years), our four astrometric epochs cover only a
tiny part of the orbit. We thus expect any orbital determination to
come with large error bars. In this context, a standard least-square
fitting procedure like Levenberg-Marquardt \citep{press92} may produce
meaningless results with huge error bars, as the $\chi^2$ surface is
probably very chaotic with many local minima. This was confirmed by our
first attempts. Therefore we moved to a more robust statistical
approach using the Markov-Chain Monte Carlo (MCMC) Bayesian analysis
technique \citep{ford05,ford06}. This technique applied to astrometric
orbits was already successfully used to constrain the orbit of the
giant planet $\beta\:$Pic b \citep{chau12}.
We use here the same code for \fomb. We assume
$d=7.7\,$pc and $M=1.92\,\msun$ for the distance and the mass
of \fom\ \citep{van07,mam12}
\begin{figure*}
\makebox[\textwidth]{
\includegraphics[width=0.33\textwidth]{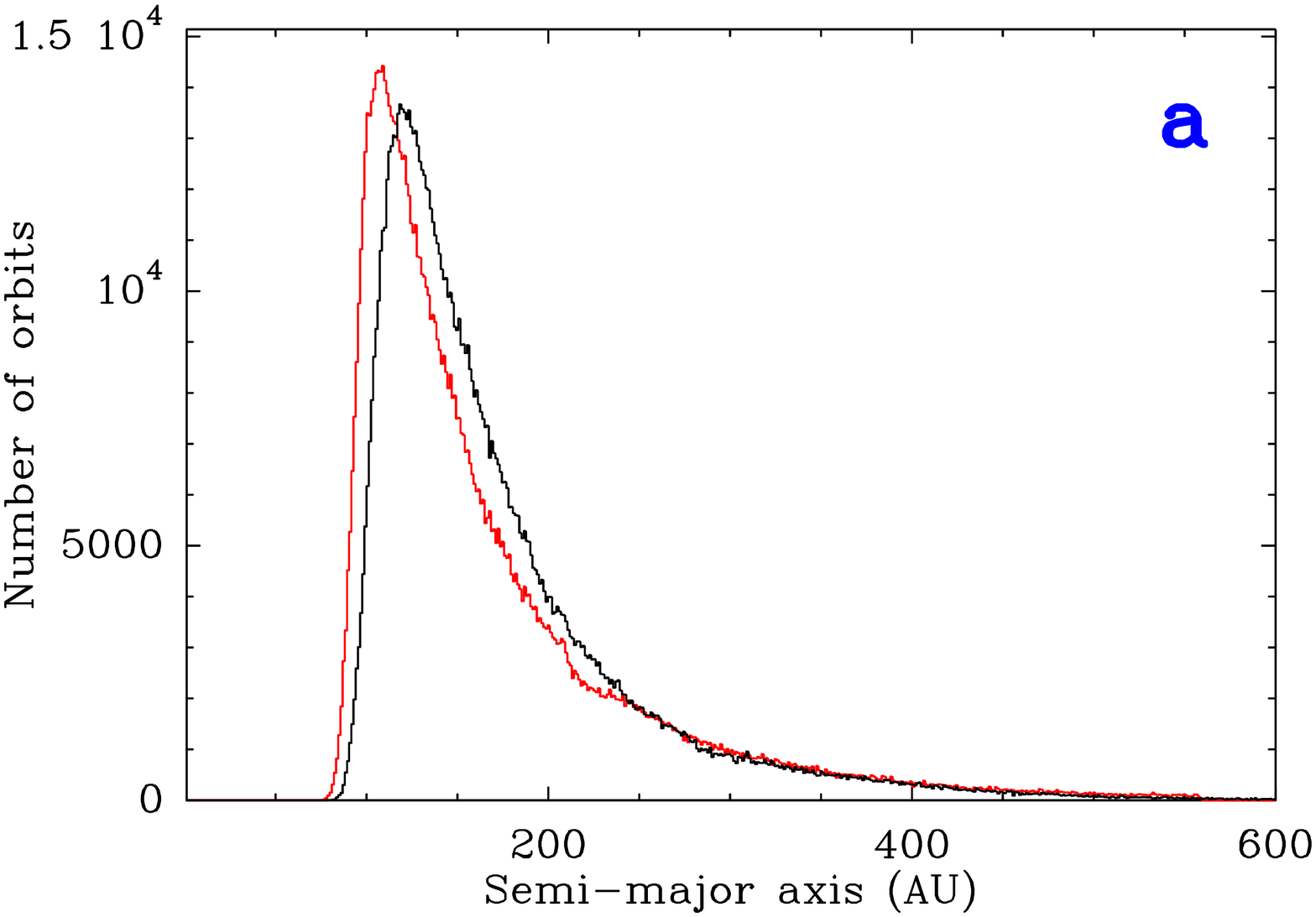} \hfil
\includegraphics[width=0.33\textwidth]{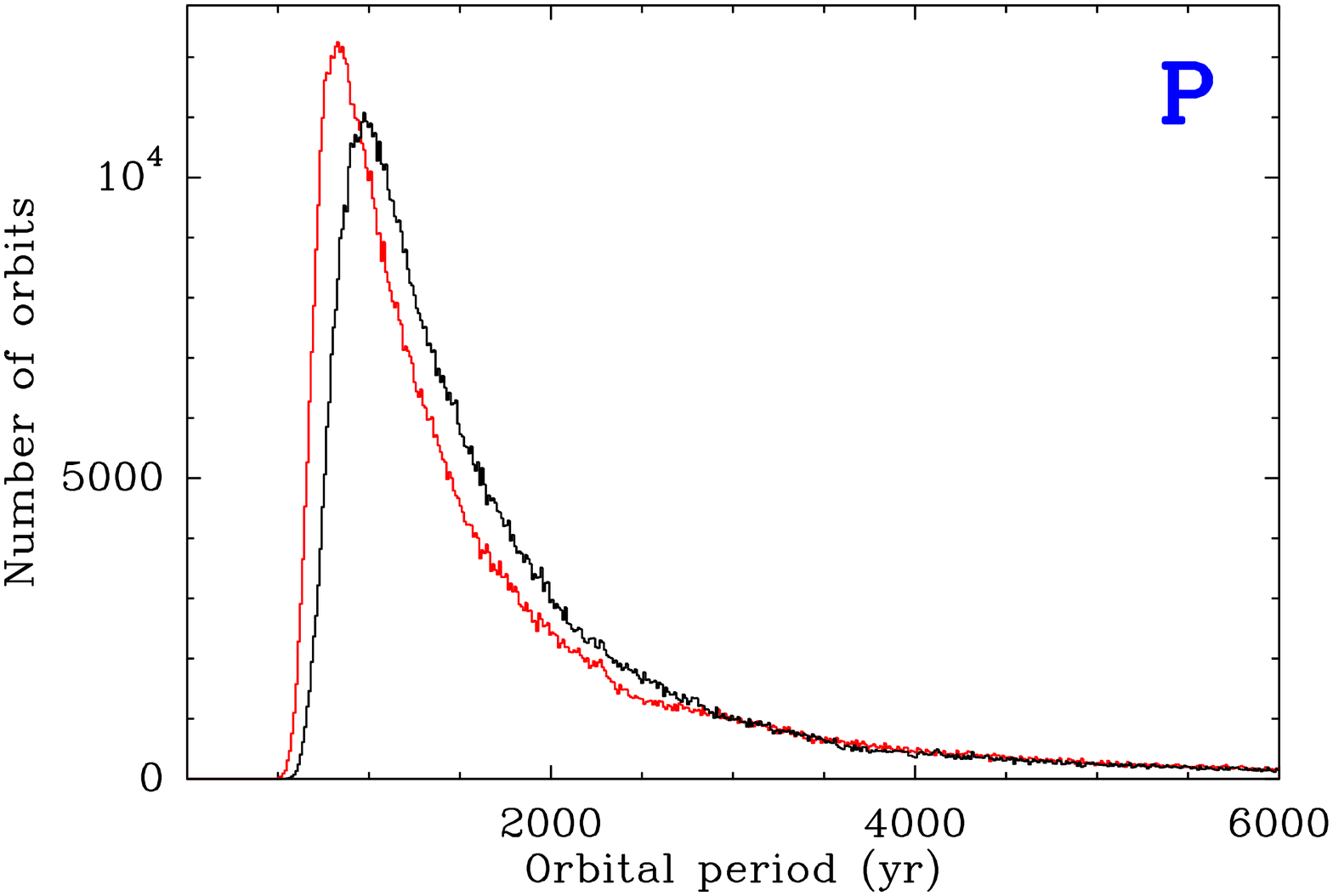} \hfil
\includegraphics[width=0.33\textwidth]{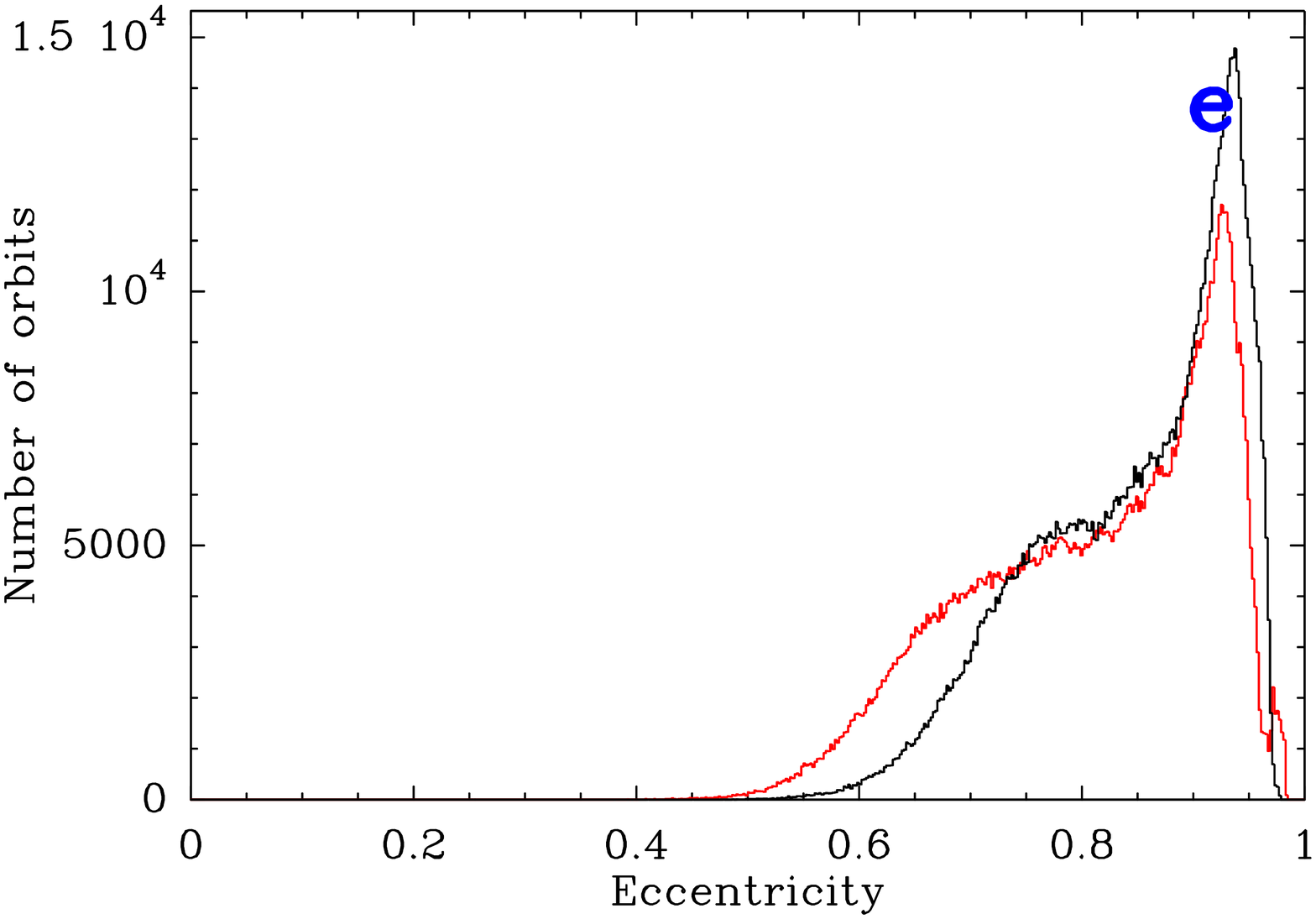}}
\makebox[\textwidth]{
\includegraphics[width=0.33\textwidth]{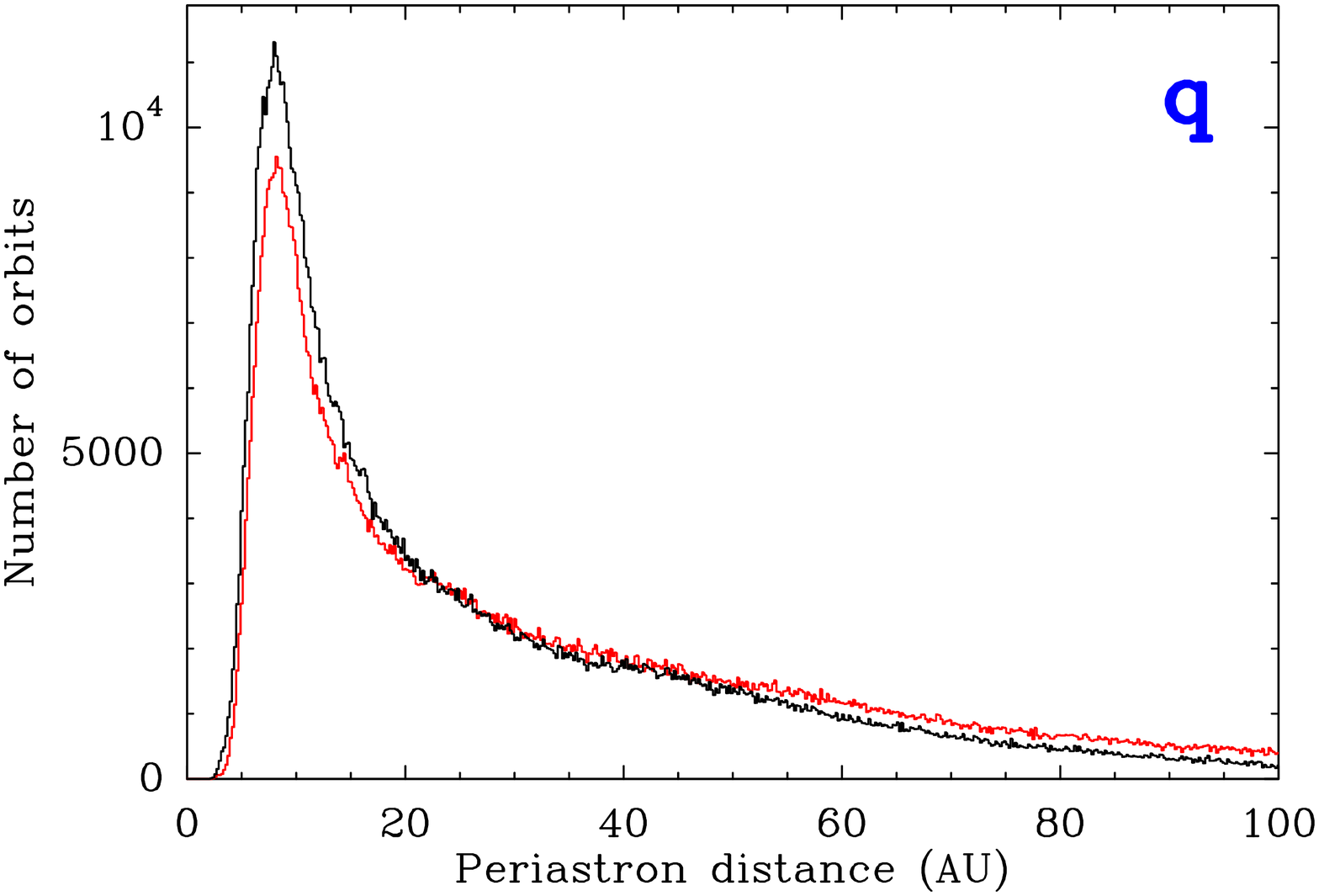} \hfil
\includegraphics[width=0.33\textwidth]{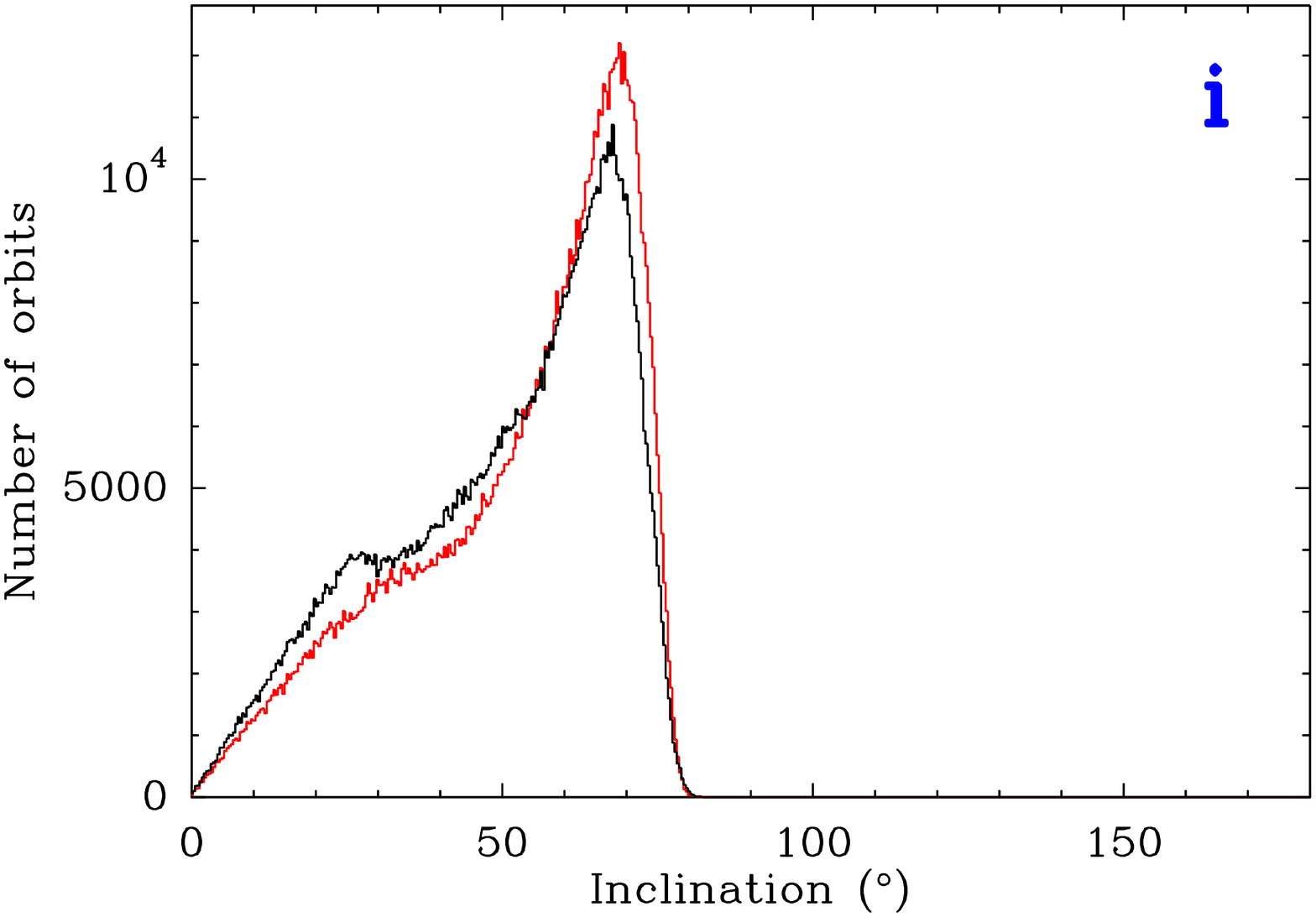} \hfil
\includegraphics[width=0.33\textwidth]{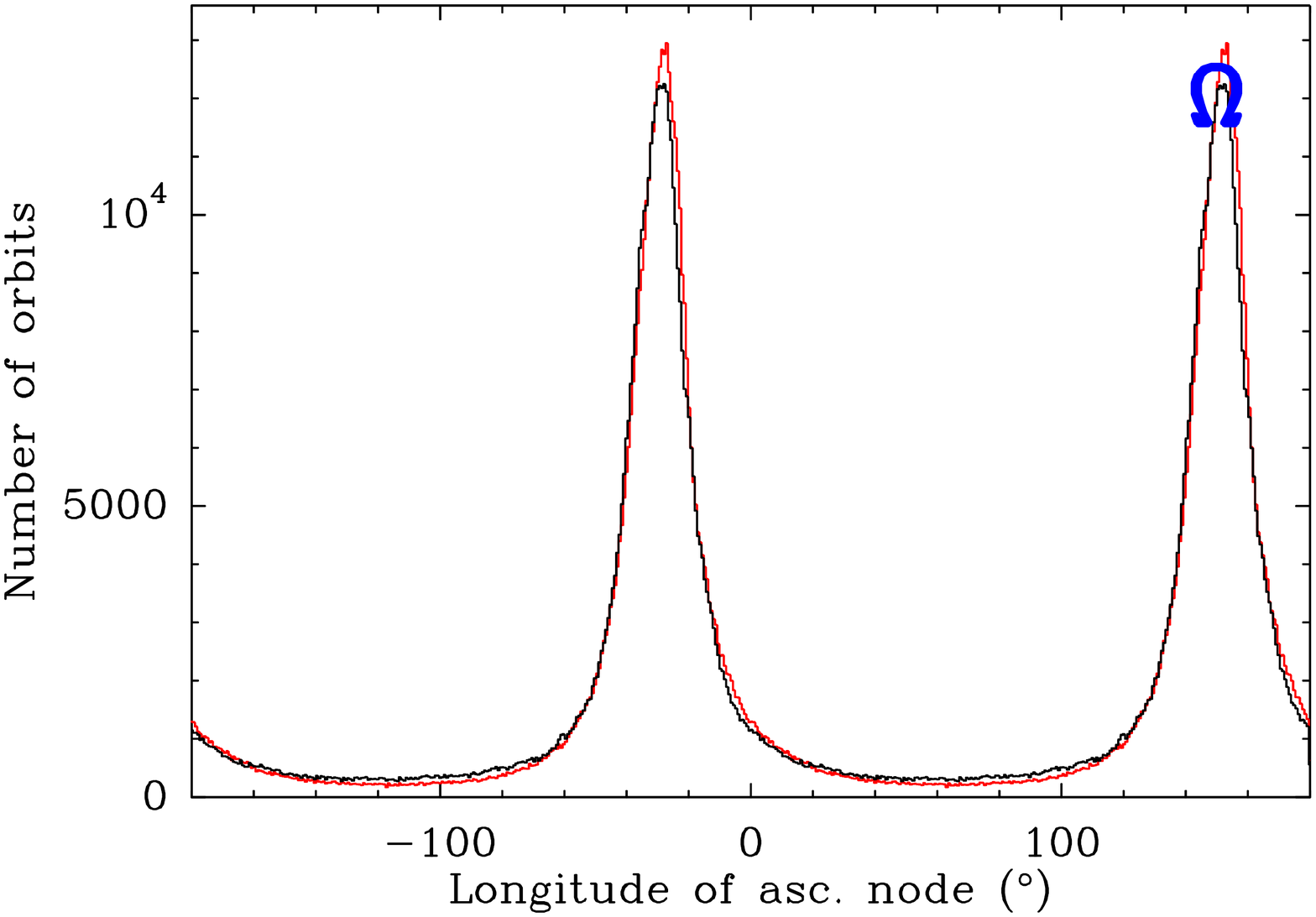}}
\makebox[\textwidth]{
\includegraphics[width=0.33\textwidth]{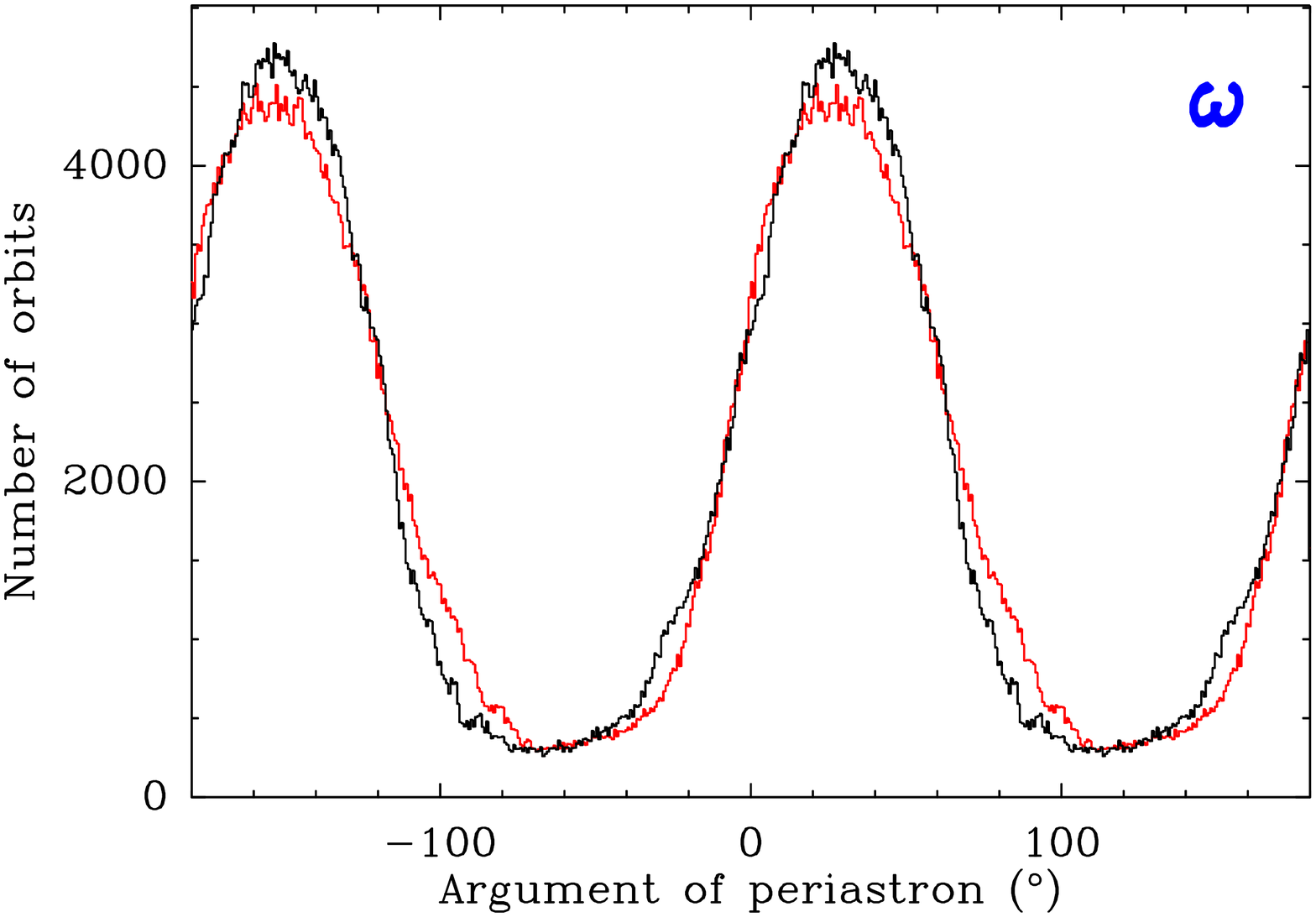} \hfil
\includegraphics[width=0.33\textwidth]{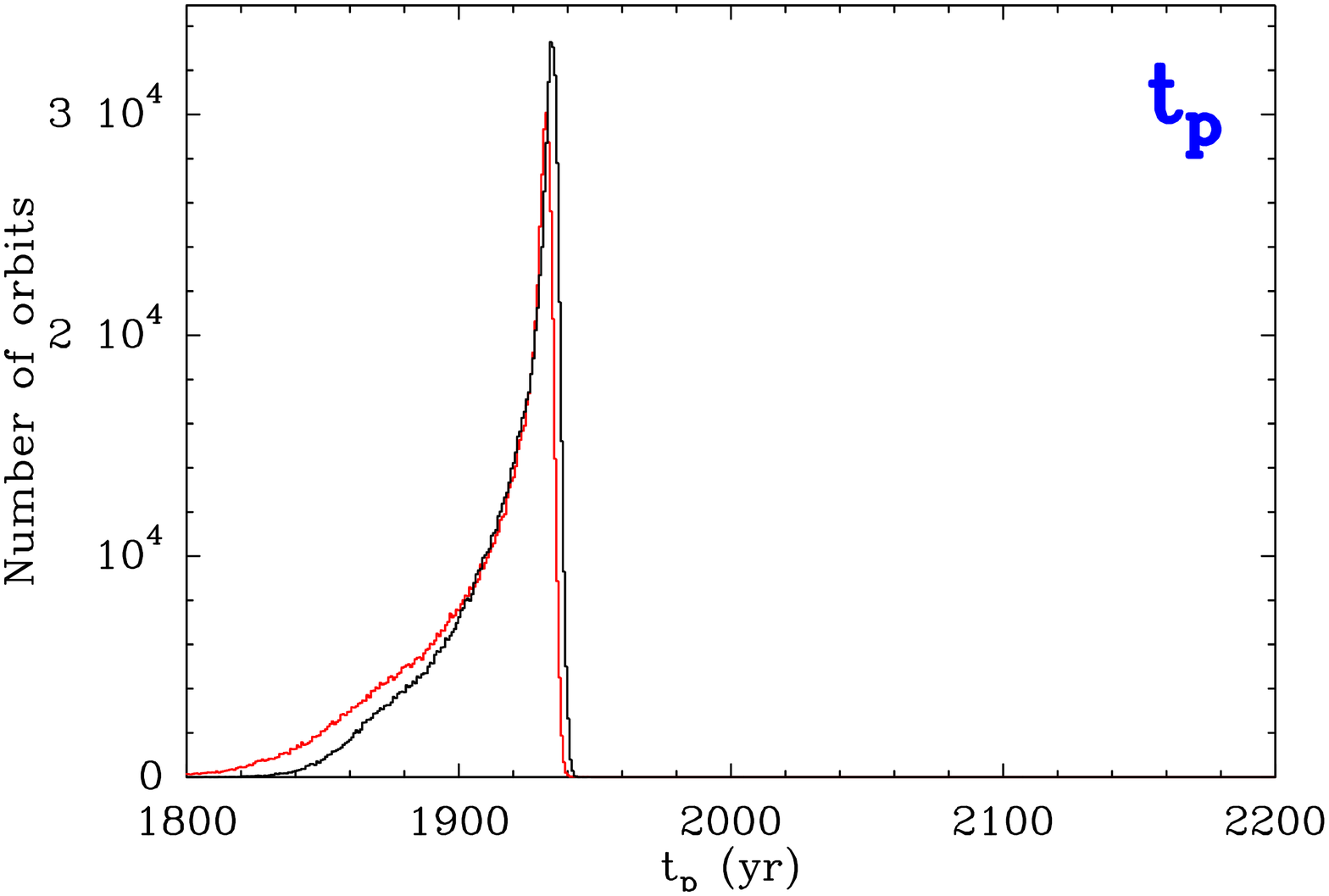} \hfil
\includegraphics[width=0.33\textwidth]{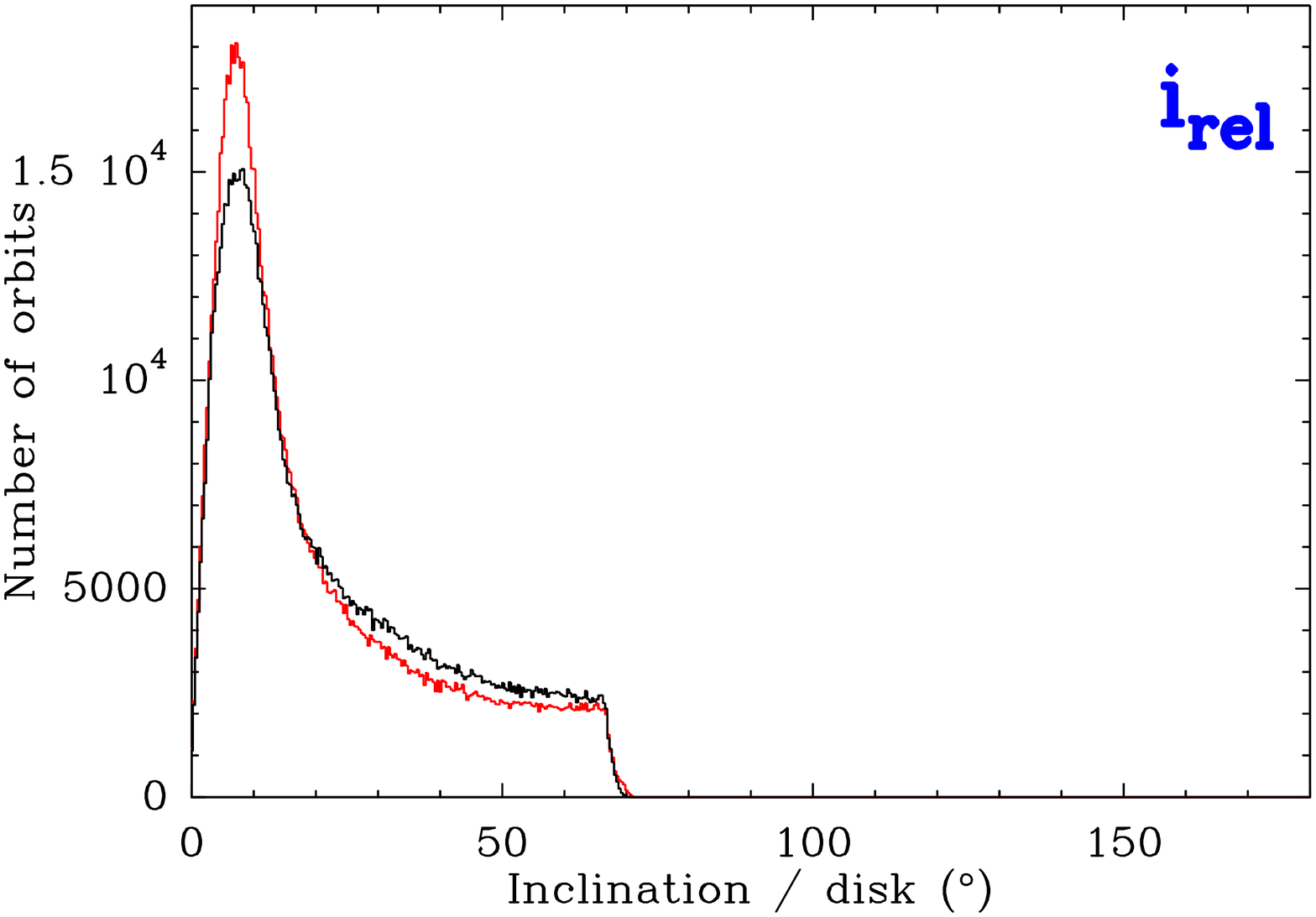}}
\caption[]{Resulting MCMC distribution of the orbital elements
of \fomb's orbit.
In all plots, the black curves correspond to the first run using the full
\citet{kal13} data, and the red one to the second run using \citet{gal13} data
before 2012.
Upper row, from left to right: semi-major axis ($a$), orbital period ($P$),
eccentricity ($e$); second row, id: periastron ($q$), inclination ($i$),
longitude of ascending node ($\Omega$); third row, id: argument of periastron
($\omega$), time for periastron passage ($t_p$), and inclination relative
to the disk plane ($i_\mathrm{rel}$)}
\label{1dmcmc}
\end{figure*}
\begin{figure*}
\makebox[\textwidth]{
\includegraphics[width=0.49\textwidth]{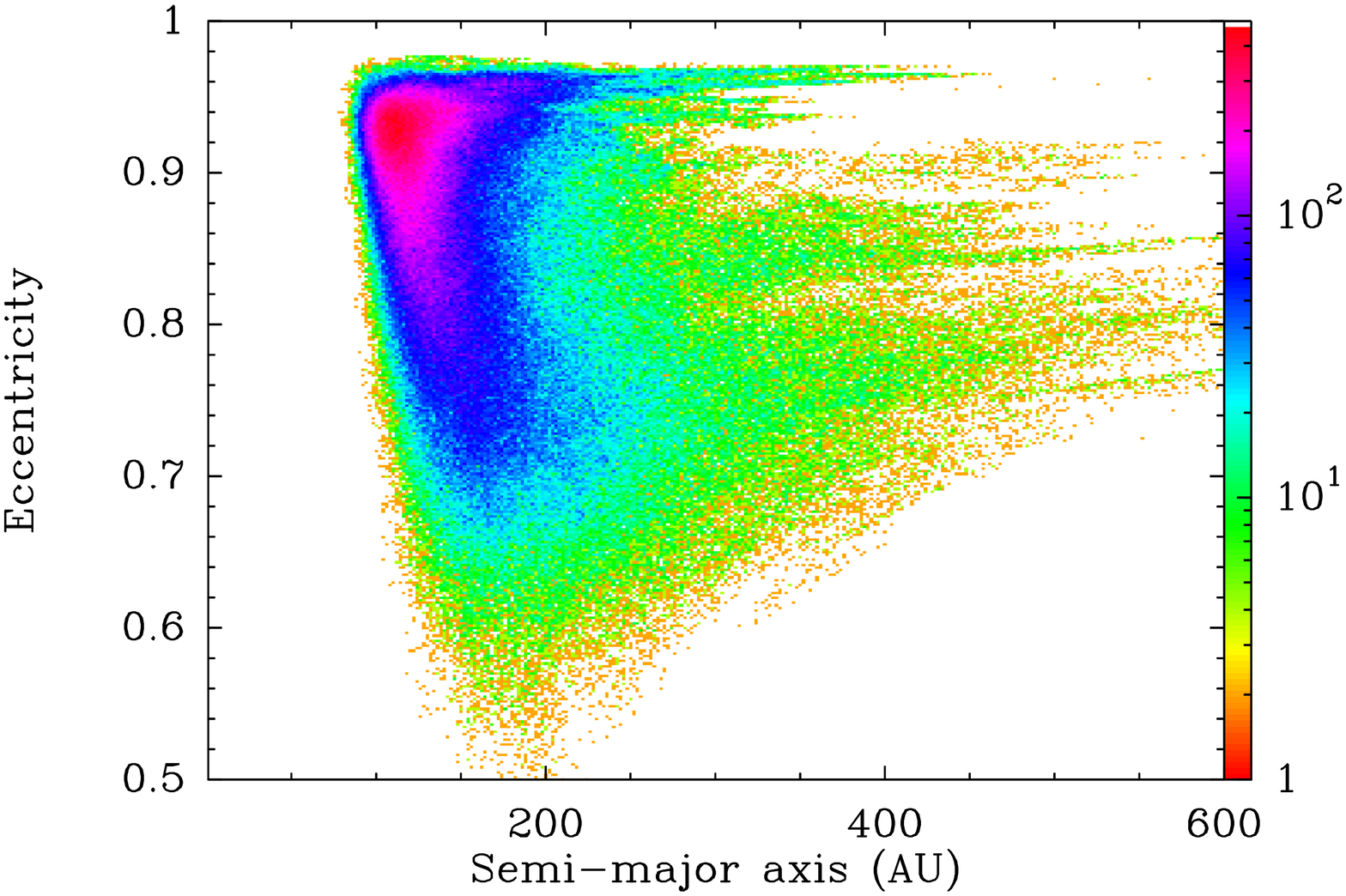} \hfil
\includegraphics[width=0.49\textwidth]{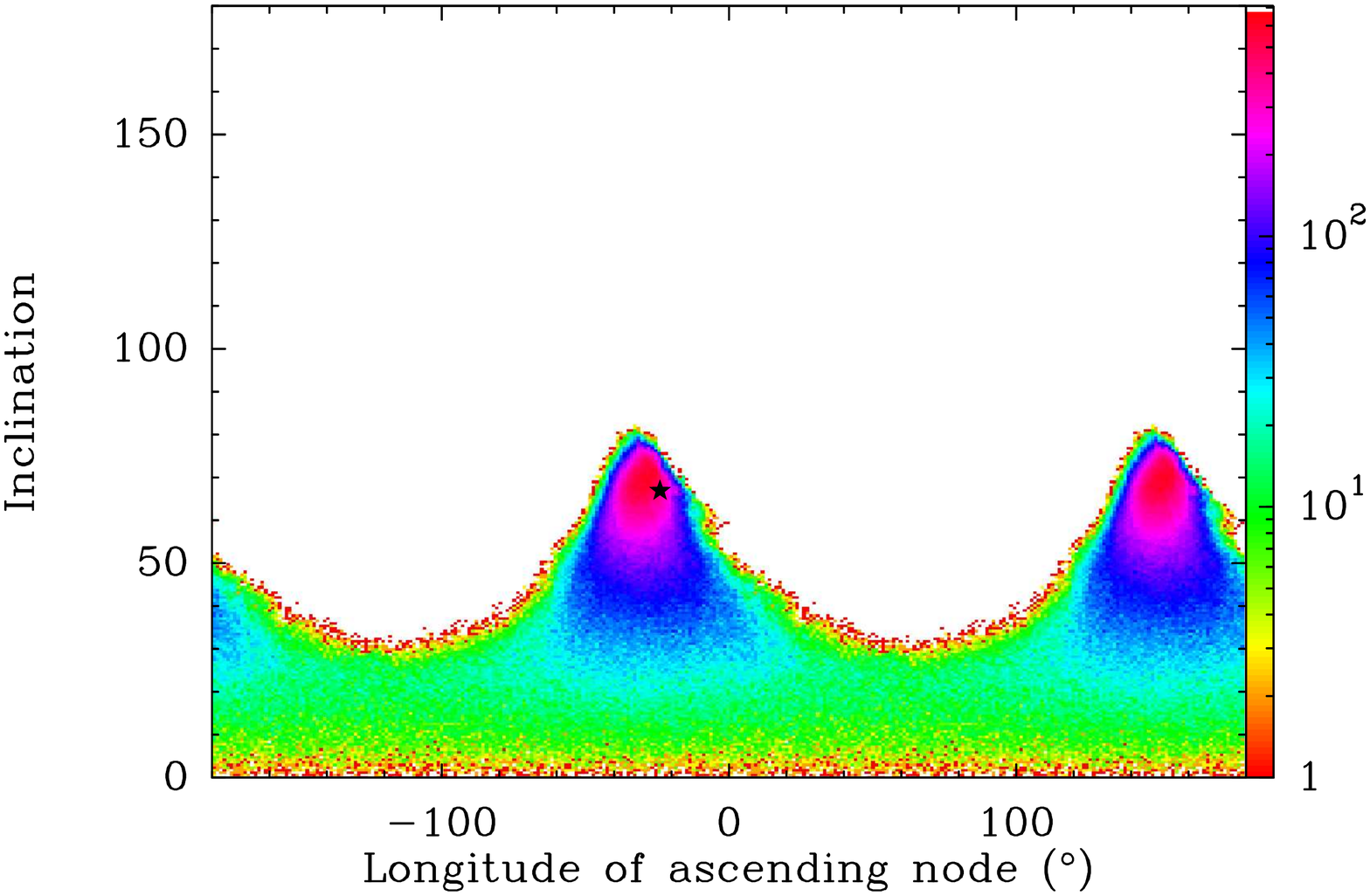}}
\caption[]{Resulting 2D MCMC distribution of \fomb's orbital elements
for two couples of parameters, for the run with the full \citet{kal13} data:
semi-major axis ($a$) -- eccentricity ($e$) (left); longitude
of ascending node ($\Omega$) --
inclination ($i$) (right). The color scale represents the joint 2D density of
solutions for the considered set of parameters. In the right plot,
the star indicates the corresponding location of the mid-plane of the dust disk
\citep{kal05}. The same plot using \citet{gal13} data is almost identical}
\label{2dmcmc}
\end{figure*}
\begin{table*}[t]
\caption{Summary of various statistical parameters resulting from the
MCMC distribution of \fomb's orbital solutions (run with \citet{kal13} data)}
\label{intervals}
\begin{tabular*}{\textwidth}{@{\excs}llllll}     % 5 columns
\hline\hline\noalign{\smallskip}
Parameter & Peak value & Median & 67\%\ confidence interval & 95\%\
confidence interval & \citet{kal13} 95\%\ interval\\
\noalign{\smallskip}\hline\noalign{\smallskip}
Semi-major axis ($a$, au) & 120 & 160 & 81 -- 193
& 81 -- 415 & 126.6 -- 242.9\\
Orbital Period ($P$, yr) & 999 & 1522 & 554 -- 2028 &
554 -- 5116 & 1028 -- 2732\\
Eccentricity ($e$) & 0.94 & 0.87 & 0.82 -- 0.98 & 0.69 -- 0.98 &
0.694 -- 0.952\\
Periastron ($q$, au) & 7.8 & 20. & 2.7 -- 33 & 2.7 -- 77.8 &
6 -- 74\\
Time for periastron ($t_p$, yr AD) & 1935 & 1922  & 1910 -- 1944 &
1869 -- 1944 & 1800 -- 2000 \\
Inclination ($i$, $\degr$) & 67 & 55 & 43 -- 81 & 15. -- 81. &
31.9 -- 71.5\\
Inclination relative to the disk ($i_\mathrm{rel}$, $\degr$) & 6.1 &
17. & 0 -- 29 & 0 -- 61 & 5 -- 29\\
Argument of periastron ($\omega$, $\degr$) & \multicolumn{2}{l}
{$-148$ or $33$}&&& $-19.2$ -- 52.9\\
Longitude of ascending node ($\Omega$, $\degr$) & \multicolumn{2}{l}
{$-28$ or $152$}&&& 141.1 -- 172.8\\
\hline\noalign{\smallskip}
\end{tabular*}
\end{table*}
After convergence of the Markov chains (10 simultaneously), a sample
of 500,000 orbits (out of $\sim 10^7$)
is picked up randomly in the chains of orbital
solutions. This sample is assumed to represent the probability
(posterior) distribution of \fomb's orbit. This distribution is
presented in Figs.~\ref{1dmcmc} and \ref{2dmcmc}.

Figure~\ref{1dmcmc} shows histograms of the distribution of individual
orbital elements. In each plot we show two histograms. The black one
corresponds to the first MCMC run (using \citet{kal13} data), and the
red one corresponds to the second run (using \citet{gal13} data for
epochs before 2012).
The reference frame $OXYZ$ with respect to which the
orbit is referred to is chosen as usual in such a way that the $OZ$
axis points towards the Earth (hence the $OXY$ plane corresponds to
the plane of the sky); the $OX$ axis points towards North.
In the framework of this formalism,
the astrometric position of the planet relative to the central star reads:
\begin{eqnarray}
x & = & \Delta\delta\;=\;r\left(\cos(\omega+f)\cos\Omega-\sin(\omega+f)
\cos i\sin\Omega\right)\qquad,\label{xmodel}\\
y & = & \Delta\alpha\;=\;r\left(\cos(\omega+f)\sin\Omega+\sin(\omega+f)
\cos i\cos\Omega\right)\qquad,\label{ymodel}
\end{eqnarray}
where $\Omega$ is the longitude of the ascending node (measured
counterclockwise from North), $\omega$ is the argument of periastron,
$i$ is the inclination, $f$ is the true anomaly, and
$r=a(1-e^2)/(1+e\cos f)$, where $a$ stands for the semi-major axis and
$e$ for the eccentricity.  With this convention, an $i=0$ inclination
would correspond to a prograde pole-on orbit, $i=90\degr$ to a edge-on
viewed orbit (like $\beta\:$Pictoris b), and $i=180\degr$ to a pole-on
retrograde orbit.

In Figure~\ref{1dmcmc}, the distributions of $\Omega$ and $\omega$
appear twofold, with two distinct peaks separated by $180\degr$. This
is due to a degeneracy in the Keplerian formalism. It can be seen from
Eqs.~\ref{xmodel} and \ref{ymodel} that changing simultaneously
$\Omega$ and $\omega$ to $\Omega+\pi$ and $\omega+\pi$ leads to the
same orbital model. Consequently these orbital parameters are only determined
with a $\pm 180\degr$ degeneracy. However, their sum $\Omega+\omega$
and difference $\Omega-\omega$ are unambiguously determined. It is
easy to rewrite Eqs.~\ref{xmodel} and \ref{ymodel} as a function of
$\Omega+\omega$ and $\Omega-\omega$ instead of $\omega$ and
$\Omega$:
\begin{eqnarray}
x & = & r\left(\cos^2\frac{i}{2}\cos(\Omega+\omega+f)
+\sin^2\frac{i}{2}\cos(f+\omega-\Omega)\right)\qquad,\label{xmod2}\\
y & = & r\left(\cos^2\frac{i}{2}\sin(\Omega+\omega+f)-
\sin^2\frac{i}{2}\sin(f+\omega-\Omega)\right)\qquad,\label{ymod2}
\end{eqnarray}
We used those formulas in our MCMC code, which in fact
fits $\Omega+\omega$ and $\Omega-\omega$. This avoids erratic changes
in the solution between degenerate solutions, and subsequently ensures
convergence of the chains. So, each time an orbital solution is taken
in the chains with fitted values for $\Omega+\omega$ and
$\Omega-\omega$, it results in two solutions with similar orbital
parameters but different $(\Omega,\omega)$ sets. This is why we have
dual peaks distributions for $\Omega$ and $\omega$.

The only way to eliminate the degeneracy is to obtain information regarding
the $OZ$ axis. This can be radial velocity data points, or information
about which side of the orbit in foreground in the images. In the case
of $\beta\:$Pictoris, information about the Keplerian gas disk help to fix
the ambiguity. But here no such information is available for \fom, so
we keep all possible solutions.

Figure~\ref{1dmcmc} shows orbital distributions with well
identified peaks, although this could appear surprising given the
paucity of the data points.
Detailed statistical parameters such as
peak values and confidence intervals for various parameters are given
in Table~\ref{intervals}.  The semi-major axis appears to peak at
$\sim110$--$120\,$au, a value comparable to the present day location of
\fomb\ with respect to the star, but surprisingly, the eccentricity is
very high. The peak of the eccentricity distribution is
$\sim0.92$--$0.94$ (depending on the data set taken),
virtually all solutions have
$e\ga0.5$--$0.6$, and even $e\ga0.8$ with a 70\%\ confidence
level. It must be noted that the eccentricity distribution never extends up to
$e=1$. No solution with $e\ge 0.98$ is derived in the
distribution. Thus we are confident in the fact that \fomb\ is
actually bound to \fom, although it may be on a very eccentric orbit. As a
consequence of this high eccentricity, the periastron value of the orbit is
small with a peak value of $7$--$8\,$au, and subsequently the apoastron is
$\ga 200\,$au with a high confidence level. Figure~\ref{2dmcmc} (left) shows
a 2D joint probability map for $a$
and $e$, for the first run only.
We clearly see a peak of solutions around ($a=120\,$au, $e=0.94$).
A similar plot built with the data from the second run would appear
nearly identical with a peak around ($a=110\,$au, $e=0.92$).

There are indeed very few differences between the histograms derived from the
two independent runs. The
semi-major axis distribution appears slightly shifted towards shorter values
in the second run (red curves, use of
\citet{gal13} data), with a peak appearing at $a=110\,$au instead of
$a=120\,$au. Similarly, the eccentricity peaks at $e=0.92$ in the
second run instead of $e=0.94$. These are the only noticeable
differences between the two resulting distributions, all remaining
differences barely reaching the level of the noise in the
histograms. The differences are in all cases far below the bulk
uncertainty on the corresponding parameters and therefore not very significant.
We may therefore consider our
orbital determination as robust. In the following, the dynamical study
are performed with a \fomb\ orbit with $a=120\,$au and $e=0.94$,
i.e. corresponding to the peak values in the first run. The use of
$a=110\,$au and $e=0.92$ (the peak values for the second run) appears
not to change anything noticeable to the dynamical behaviour we
describe below.

The inclination distribution in Fig.~\ref{1dmcmc} shows that all
solutions are with $i<90\degr$, confirming a prograde orbit. The
inclination peaks at $66.7\degr$, a value very close to the disk
inclination quoted by \citet{kal05}. The longitude of ascending node
exhibits (due to the quoted degeneracy) two peaks separated by
$180\degr$ at $-27.8\degr$ and $152.2\degr$. This is again very close
to the $PA=156\degr$ of the belt ellipse quoted by \citet{kal05}. As
our longitudes of nodes are counted counterclockwise from North like
PAs, these similarities of values are a strong indication in favour of
a coplanarity, or near coplanarity between the dust belt and \fomb's
orbit. We therefore plot in Figure~\ref{1dmcmc} the statistical
distribution of the mutual inclination $i_\mathrm{irel}$ between
\fomb's orbit and the dust disk, assuming the inclination and PA
values given by \citet{kal05}. We see a sharp peak at $6.3\degr$,
which clearly suggests quasi-coplanarity. The fact that the
peak is not at $i_\mathrm{rel}=0$ does not necessarily indicate a
non-coplanarity. Due to the error bars on the disk orbit parameters,
strict coplanarity ($i_\mathrm{rel}=0$) is just less probable than a
few degrees offset. If the direction vector perpendicular to \fomb's
orbit was drawn randomly on a sphere, the natural statistical
distribution for $i_\mathrm{rel}=0$ would be $\propto \sin
i_\mathrm{rel}$. This is equivalent to saying that the coplanar
configuration would be the least probable one if the orientation of
\fomb's orbital plane was distributed randomly. Now, if we consider
that error bars on the determination on the dust ring orbital plane
and on our determination of \fomb's orbital plane lead to an
uncertainty of $\sim 10\degr$ on the determination of
$i_\mathrm{rel}$, this means that we add a stochastic component to our
measurement of $i_\mathrm{rel}$, which should follow the $\propto \sin
i_\mathrm{rel}$ distribution, at least up to $\sim 10\degr$. This is
enough to create a peak in the MCMC distribution of $i_\mathrm{rel}$
that appears offset from the pure coplanar configuration by a few
degrees.

Note also that
when computing $i_\mathrm{rel}$, due to the Keplerian degeneracy, two
mutual inclinations could be deduced for each solution.  We
systematically chose the lower one. This shows also up in
Fig.~\ref{2dmcmc} (right), which shows a 2D joint probability map for
$\Omega$ and $i$, for the first run (full \citet{kal13} data).
We clearly identify the two peaks. The star
indicates now the corresponding values for the dust disk taken from
\citet{kal05}, which fall very close to the peaks of the
distribution. This unambiguously suggests coplanarity or
quasi-coplanarity.

The argument of periastron $\omega$ peaks at $-148.3$ or
$31.7$. \citet{kal05} report that the periastron of the elliptic dust
belt is at PA=170$\degr$. Taking into account the PA of the disk and
its inclination, we derive an argument of periastron
$\omega_\mathrm{disk}=-148.9\degr$, which is extremely close to our peak
value of $\omega$. While this could be considered a strong indication
for \fomb's orbit to be apsidally aligned with the elliptic dust belt,
the real alignment may not be this perfect given the uncertainties of
$\omega$ and $\omega_\mathrm{disk}$. The
uncertainty on $\omega_{disk}$ is roughly $\pm25\degr$, and that on
our $\omega$ determination is comparable. A a result the agreement within
less than $1\degr$ between both values could be a pure
coincidence. All we can stress looking at the whole $\omega$
distribution is that we have apsidal alignment within less than
$\pm30$--$40\degr$ with a good level of confidence ($\sim70\%$).

The conclusions is that we confirm the
orbital determination of
\fomb\ independently inferred by \citet{kal13}. The inclination distributions
are compatible (within a sign convention in \citet{kal13}), as well
as the $\Omega$ and $\omega$ distribution, although only single
peak distributions are given in \citet{kal13}. The shapes of the
semi-major axis and eccentricity distributions are noticeably similar.
The eccentricity and semi-major axis intervals are very similar, except the
that our semi-major axis distribution extends a bit lower and our
eccentricity distribution a bit higher (Table~\ref{intervals}).
We also confirm that \fomb's orbit is very probably
nearly coplanar and apsidally aligned within a few tens of degrees
with the dust belt.
\section{Numerical study assuming coplanarity}
\subsection{Pericenter glow for low eccentricity orbits}
We thus conclude like \citet{kal13} that a dust belt crossing orbit
for \fomb\ is consistent with the data. This
automatically raises the question of the long-term stability of this
configuration. Thus we move now to a dynamical study to address this issue.
fomb's orbit turns out to be nearly coplanar and apsidally aligned with
the elliptic dust ring. This is a strong indication for a pericenter
glow phenomenon.

Pericenter glow occurs when a disk of planetesimals orbiting a star is
secularly perturbed by a planet moving on an eccentric orbit.
We briefly recall here the theory, which is described in
detail in \citet{wya99} and \citet{wya05}. We consider the motion of a
planetesimal perturbed by the planet. We use Laplace-Lagrange theory,
based on an expansion of the disturbing function in ascending powers
of eccentricities and inclinations and a truncation to second order,
assuming that eccentricities and inclinations remain
low \citep{mur99}. This causes the secular system to become
linear. The analytical solution for the planetesimal eccentricity can
be described for the eccentricity variables as
\begin{equation}
z(t)\equiv e\times\exp(I\varpi)=
Be'+e_\mathrm{p}\exp\left(I(At+\beta_0)\right)\qquad.
\label{eqt}
\end{equation}
Here $z(t)$ is the complex eccentricity and $I^2=-1$;
$e$ is the planetesimal's eccentricity while $e'$ is
that of the planet; $\varpi=\omega+\Omega$ is the longitude
of periastron with respect to the direction of the planet's periastron.
$B$ and $A$
are coefficients that depend on the orbital configuration of the two
bodies via Laplace coefficients \citep[see][for details]{wya05}. The
first term in Eq.~(\ref{eqt}) is a fixed forced eccentricity due to
the eccentricity of the perturbing planet. The second term is a proper
oscillating term with additional parameters $e_\mathrm{p}$ and
$\beta_0$ that depend on the initial conditions.

Consider now that we start with an initially cold disk, i.e.,
planetesimals on circular orbits ($z(0)=0$). This could be the case at
the end of the protoplanetary phase, because before the disappearance
of the gas, the eccentricity of all solid particles tend to be
damped by gas drag. Then obviously $\beta_0=\pi$
and $e_\mathrm{p}=Be'$, so that the full solution now reads
\begin{equation}
z(t)=Be'\left(1-\exp(IAt)\right)\qquad.
\label{eqt2}
\end{equation}
The complex eccentricity $z(t)$ describes a circle path in complex
plane with radius $Be'$, centered on the point $(Be',0)$. It results
from Eq.~(\ref{eqt2}) that the maximum eccentricity $e_\mathrm{max}=2Be'$ is
reached for $At\equiv\pi [2\pi]$, when $\varpi=0$. This means that the
maximum eccentricity in the secular evolution is reached when the
planetesimal is  apsidally aligned with the planet.

When $e'\neq0$, \citet{wya05} showed
that a steady-state regime is reached after a
transient phase
characterised by spiral structures. In the steady-state regime all
planetesimals are at various phases on their secular eccentricity
cycle, but those which are close to their peak eccentricity are
approximately apsidally aligned with the planet. The global result is
an elliptic dust ring apsidally aligned with the planet.

From an observational point of view, the pericenter side of the ring
appears more luminous, thanks to a more efficient scattering of
stellar light by the dust particles produced by the planetesimals. The
same applies also to thermal emission, as grains are hotter near
pericenter. This phenomenon termed pericenter glow was invoked to
explain many observed asymmetric global structures in debris disks,
such as HR\,4796 \citep{wya99,moe11}, HD\,141569 \citep{wya05} or more
recently $\zeta^2\;$Reticuli \citep{far13}. Concerning \fom, the
dynamical study by \citet{chi09}, based on a moderately eccentric
orbit of \fomb\ shepherding the dust ring were made in this framework.

\fom's dust ring and \fomb's orbit
share many characteristics that are typical of pericenter glow: an
eccentric ring with an offset centre, coplanar and apsidally aligned
with \fomb. It is therefore tempting to invoke it here. But the
linear theory outlined above holds for moderately eccentric orbits that do
not cross each other. Here with $e=0.94$, we are far from any linear
regime. It is then important to characterise what happens
in the high eccentricity regime. This must be done numerically.
\subsection{Pericenter glow phenomenon with highly eccentric perturbers}
We present now a numerical study of the \fom\ system, to properly address
the case of high eccentricity orbits. We take an initial
ring of $10^5$ massless particles (i.e., planetesimals) between 110\,au and
170\,au, i.e., extending wider than the observed ring, and we add a
planet orbiting on an orbit corresponding to our best fit:
$a=120\,$au and $e=0.94$. The initial eccentricities of the particles
are randomly sorted between 0 and 0.05, while their inclinations with
respect to the planet's orbital plane are chosen between 0 and
$3\degr$. The dynamics of this system is integrated using the
symplectic N-body code \textsc{Swift\_rmvs} \citep{ld94} which takes
into account close encounters between the planet and the disk
particles. The integration is extended up to 500\,Myr, i.e, a bit longer than
the estimated age of \fom\ \citep[440\,Myr;][]{mam12}.

Taking into account close encounters is indeed important here. As the
planet's orbit crosses the disk we expect to
have many encounters. The perturbing action of the planet onto the
disk particles is twofold: all particles crossing the planet's path
within a few Hill radii undergo a close encounter that most of the time
scatters them out of the disk; but as long as the particles do not
encounter the planet, they are subject to a secular evolution more or
less comparable to the one described in the previous section. We
expect any global shaping of the disk to be due to secular
perturbations rather than close encounters, as close encounters rather
have a destructive effect on the disk.

The balance between the two effects (secular and close encounters)
depends actually on the mass of \fomb, which we will consider
to be a free parameter.
\subsubsection{Massive planet}
\begin{figure*}
\makebox[\textwidth]{
\includegraphics[width=0.33\textwidth]{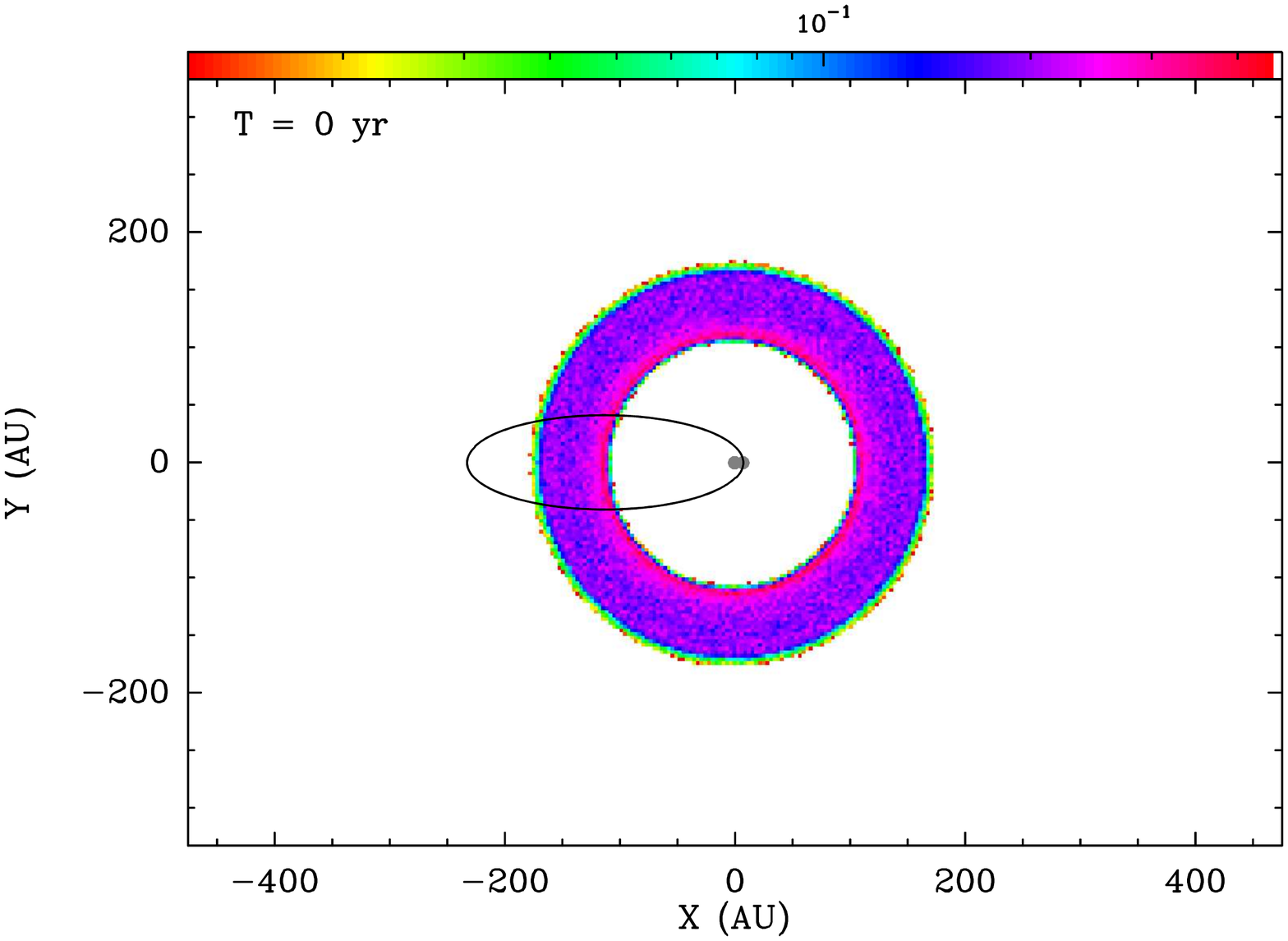} \hfil
\includegraphics[width=0.33\textwidth]{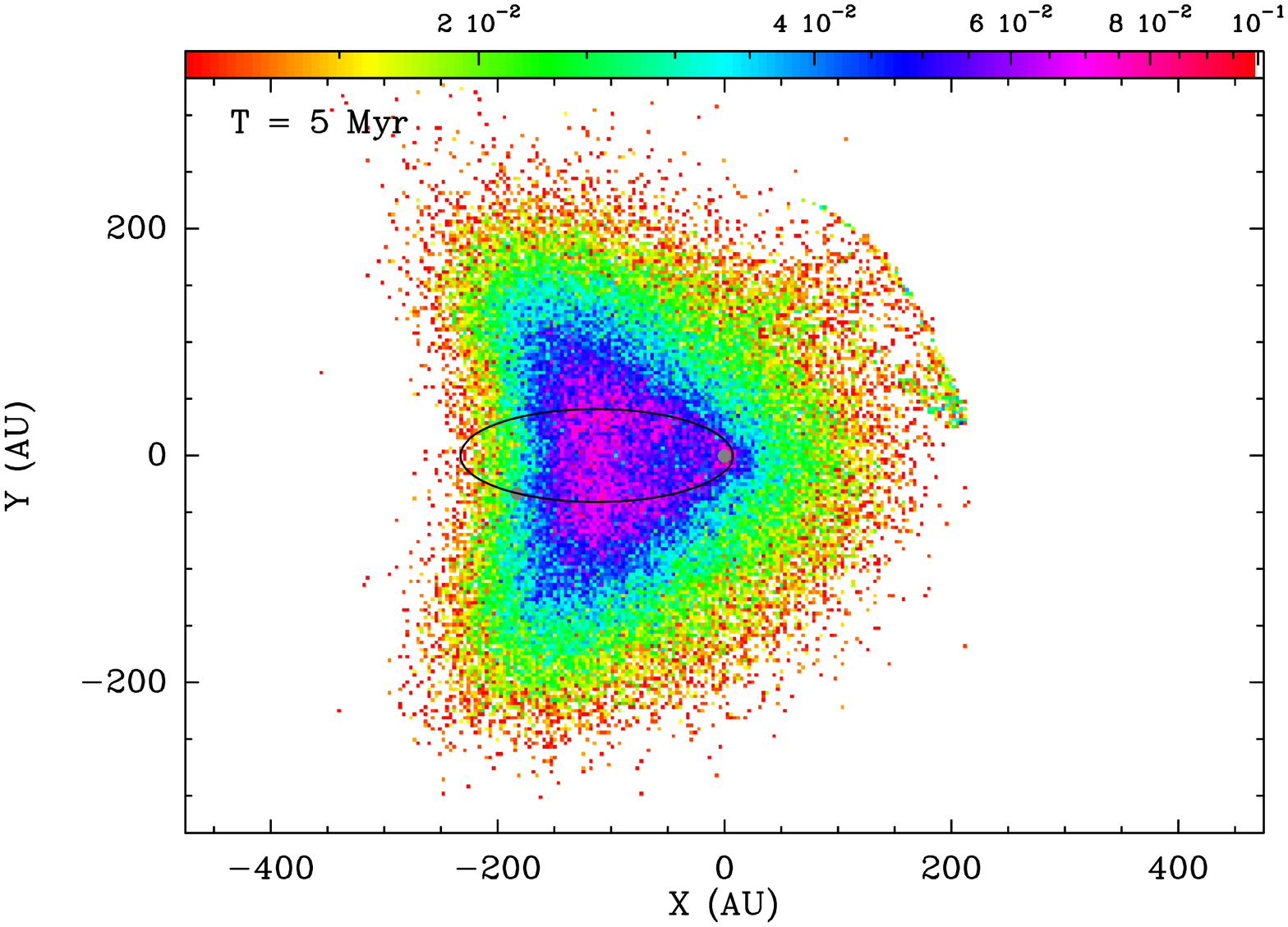} \hfil
\includegraphics[width=0.33\textwidth]{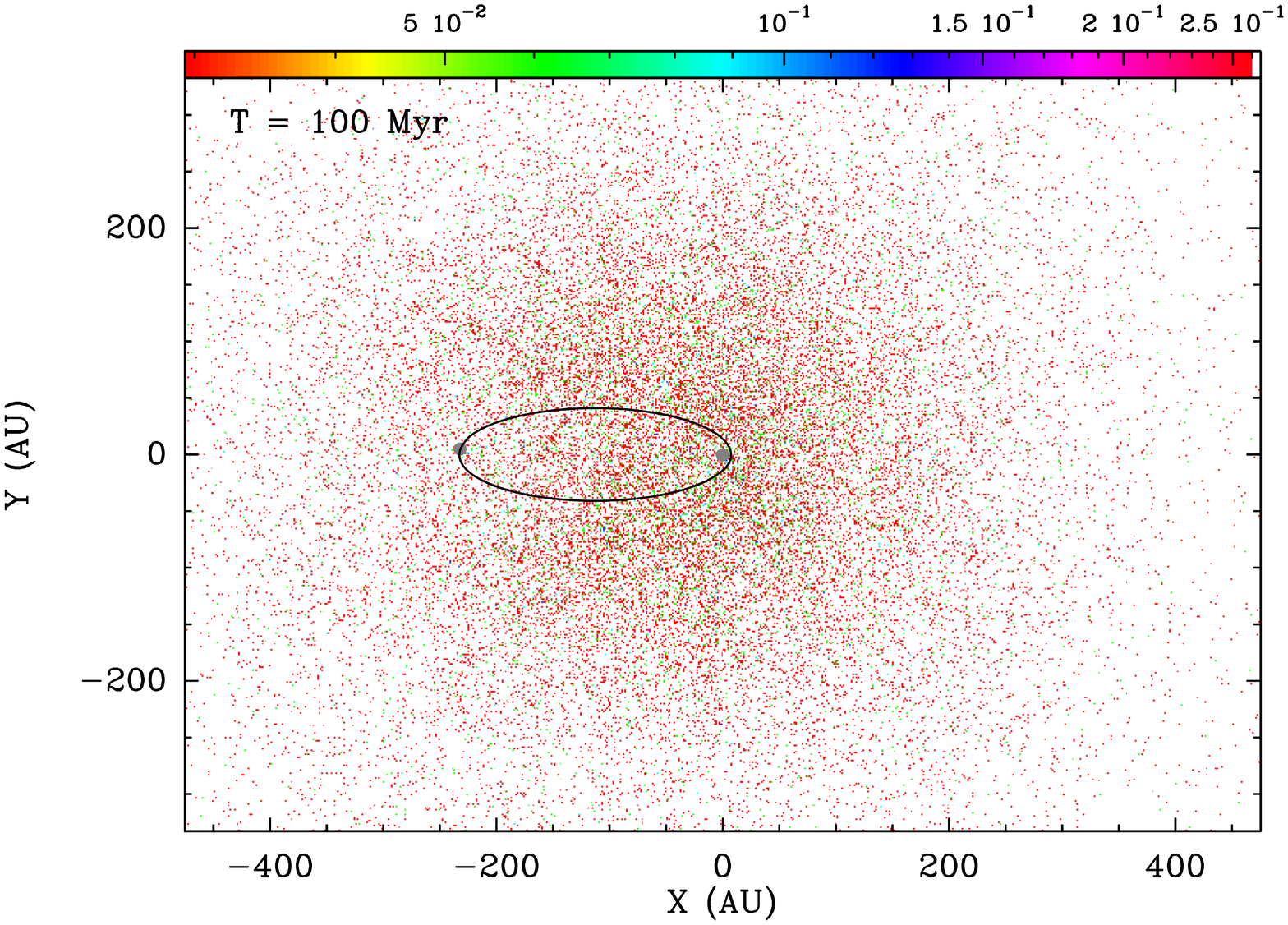}}
\makebox[\textwidth]{
\includegraphics[width=0.33\textwidth]{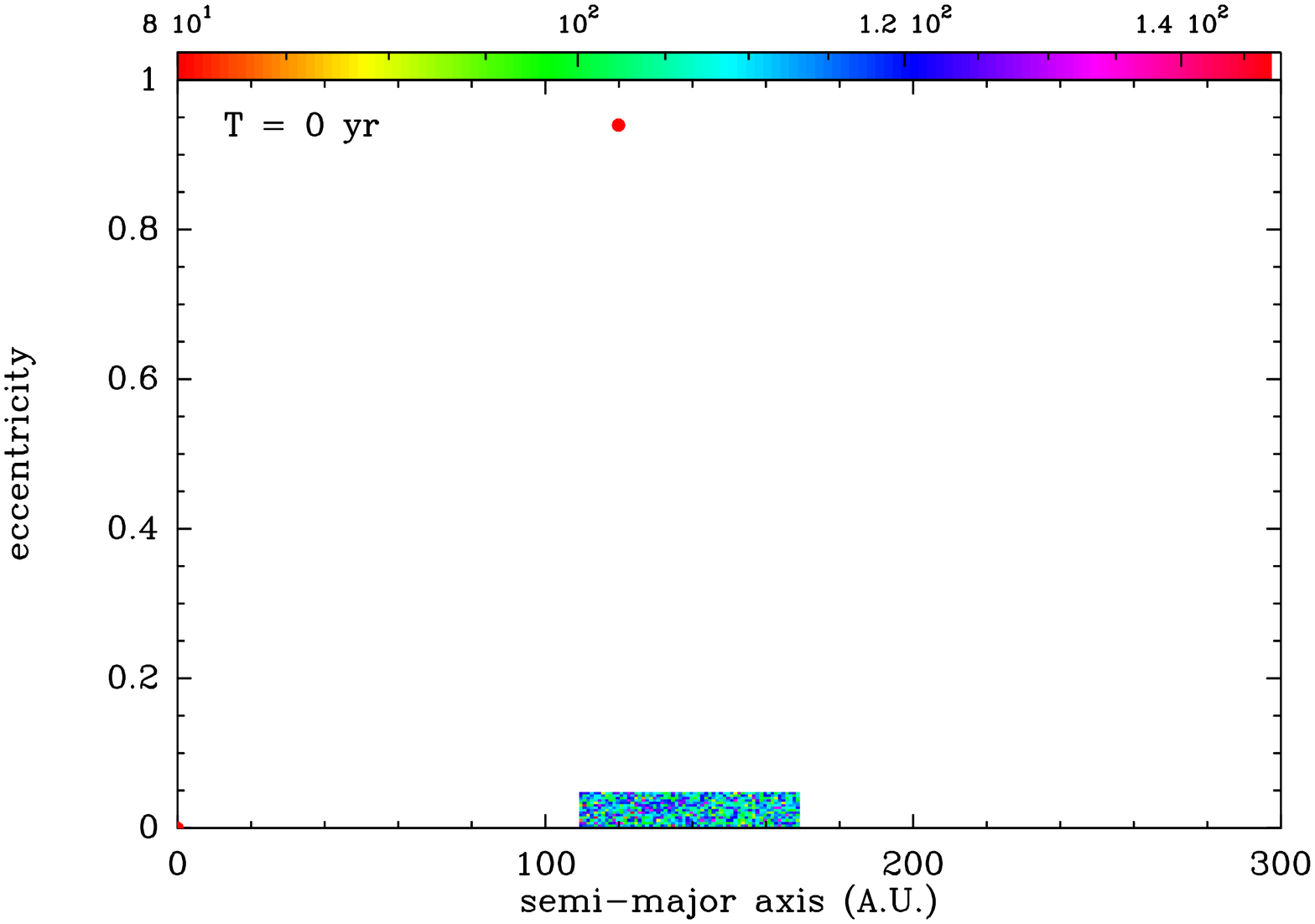} \hfil
\includegraphics[width=0.33\textwidth]{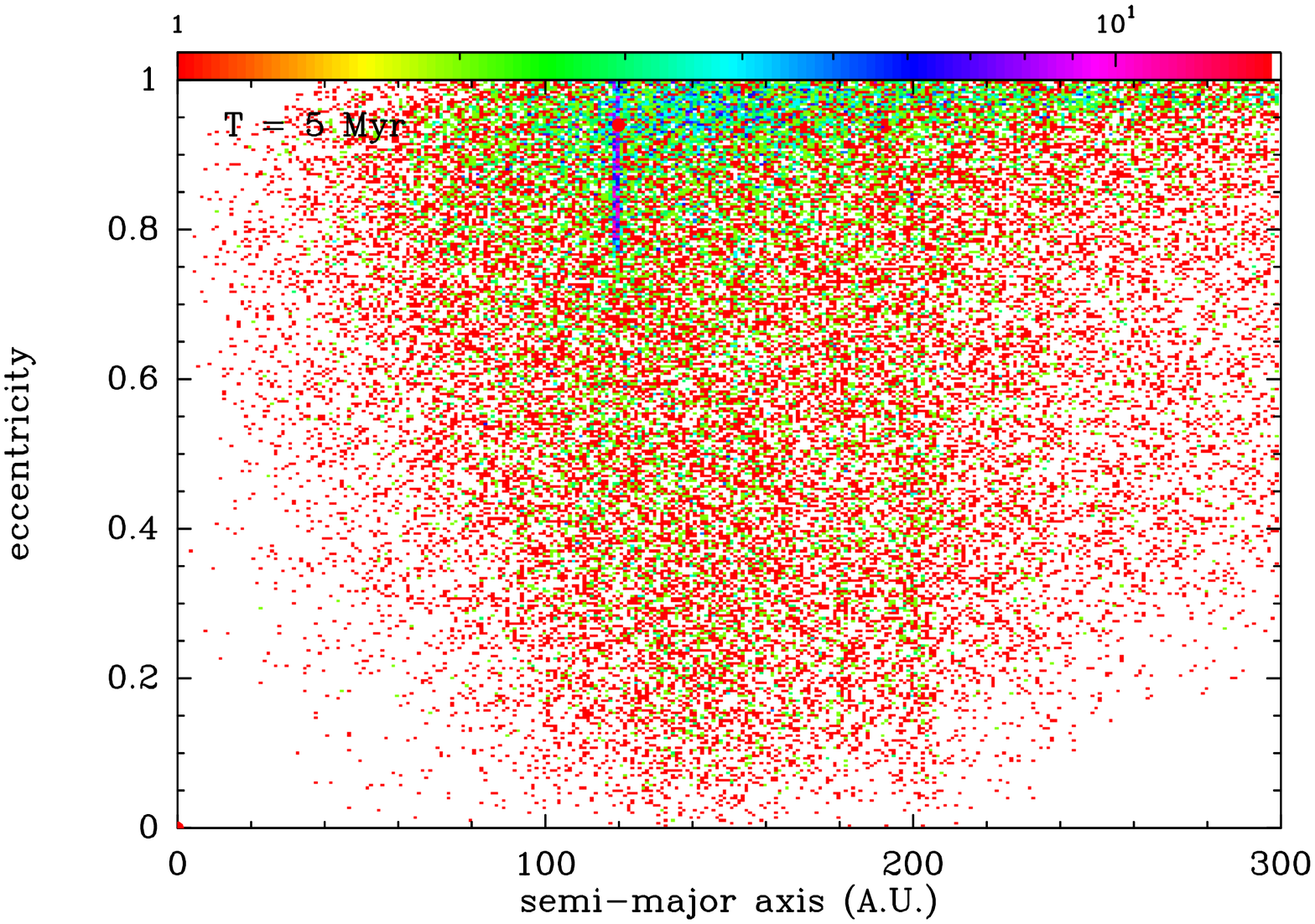} \hfil
\includegraphics[width=0.33\textwidth]{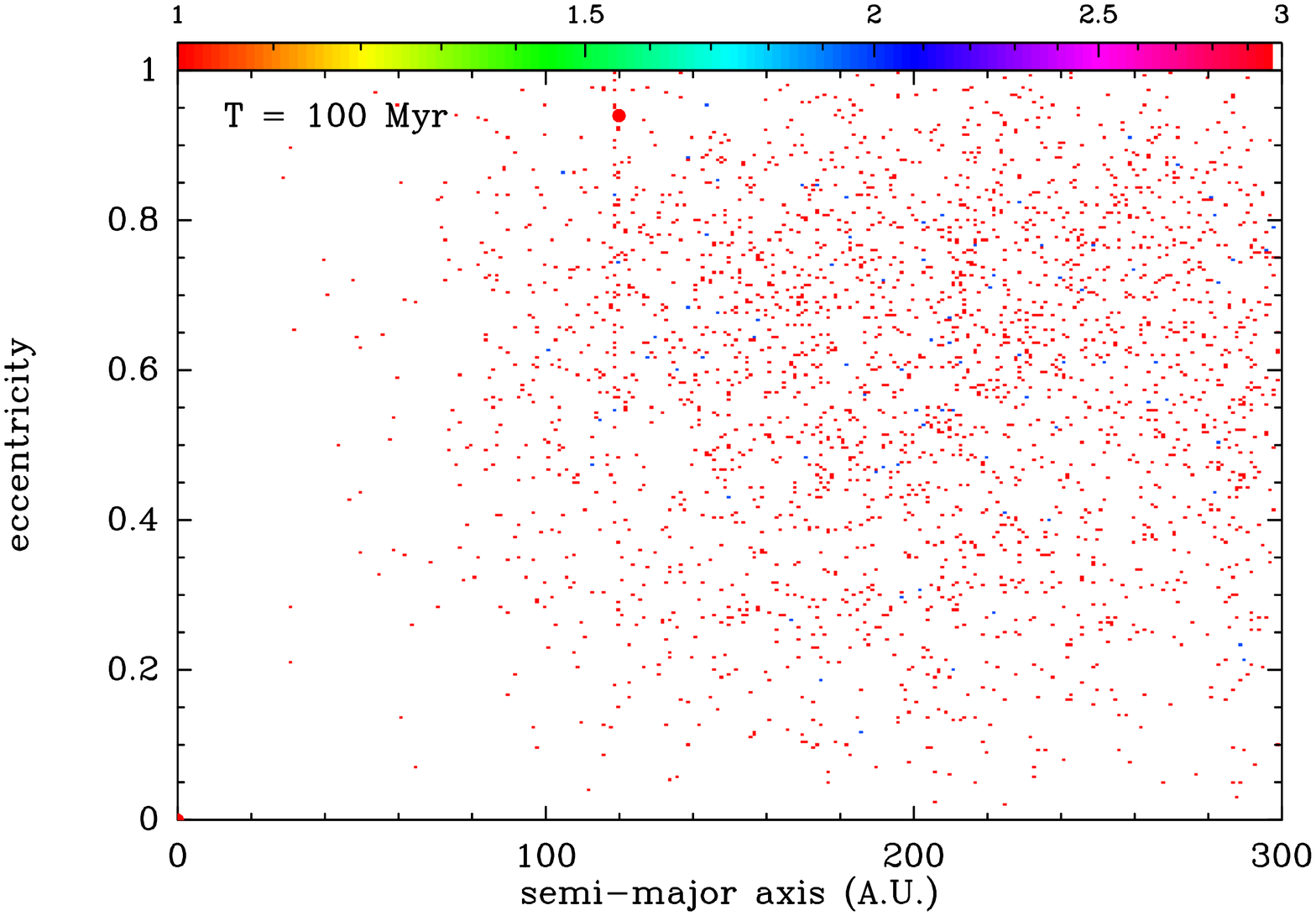}}
\caption[]{Result of the N-body integration with a perturbing planet with
$m=1\,\mjup$. We display here upper views of the planetesimal disk together
with the planet's orbit (top) and semi-major axis -- eccentricity diagrams
of the disk (bottom), at three epochs: beginning of the simulation
($t=0$, left), at $t=5\,$Myr (middle) and $t=100\,$Myr (right). The color
scale is proportional to the projected densities of particles (top plots) and
of orbits in ($a,e$) space (bottom plots). The red circles
represent the location of the star and of the planet. The planet's orbit
is sketched as a black ellipse.}
\label{simu_1mjup}
\end{figure*}
\begin{figure}
\includegraphics[width=\columnwidth]{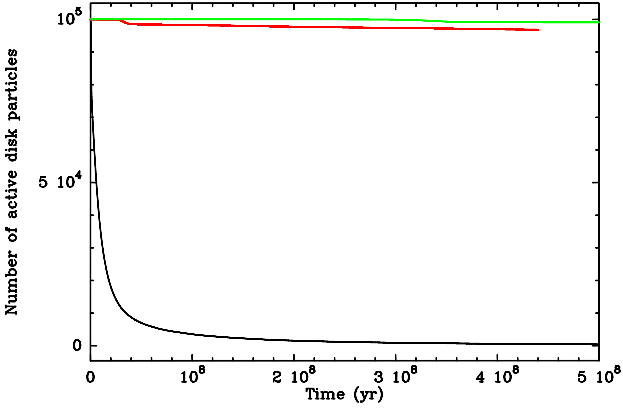}
\caption[]{Evolution of the number of active disk particles as a function
of time in the simulations described in Figs.~\ref{simu_1mjup} (black),
\ref{simu_002mjup} (red) and \ref{simu_0002mjup} (green).
All missing particles have been ejected by close encounters. This phenomenon
mainly concerns the $m=1\,\mjup$ case.}
\label{ntp}
\end{figure}
We first present a run with a massive planet, i.e., $m=1\,\mjup$, but
still fitting the observational constraints \citep{jan12}.
The result is shown in Fig.~\ref{simu_1mjup}. We represent here
upper views of the particle
disk and semi-major axis versus eccentricity plots, at the beginning
of the simulation and at two subsequent epochs: $t=5\,$Myr and
$t=100\,$Myr. As early as $t=5\,$Myr, the disk appears extremely
perturbed and actually no longer assumes a disk shape. Many particles
have already had a close encounter with the planet and have been
scattered. Interestingly, a few disk particles have been trapped in
co-orbiting orbits (or 1:1 resonance) with the planet. At
$t=100\,$Myr, i.e., well below the age of \fom, these are no longer
present. The disk now contains fewer particles. Many of them have
been lost in close encounters with the planet. To illustrate this, we
plot in Fig.~\ref{ntp} the number of remaining disk particles
(i.e., those particles which have not been ejected yet) as
a function of time. Starting from $10^5$, we see that it is reduced to
4000 at $t=100\,$Myr and to 400 at $t=500\,$Myr. We can then safely
claim that this situation does not match the observation, unless the planet
was very recently scattered ($\la 10\,$Myr; see Fig.~\ref{ntp})
onto its present orbit. Over any longer time-scale,
the disk is virtually destroyed by close encounters, which are just
too efficient here with such a massive planet. In fact, even a few Myrs
is already too long. The disk particles reach high eccentricities much
earlier than that. An average eccentricity of 0.1 for the disk particles,
which we should consider as matching the observations, is reached
only $\sim 3\times10^4$\,yr after the beginning of the simulations.
As a result any subsequent configuration must be considered as incompatible
with the observation.
\subsubsection{Super-Earth planet}
\begin{figure*}
\makebox[\textwidth]{
\includegraphics[width=0.33\textwidth]{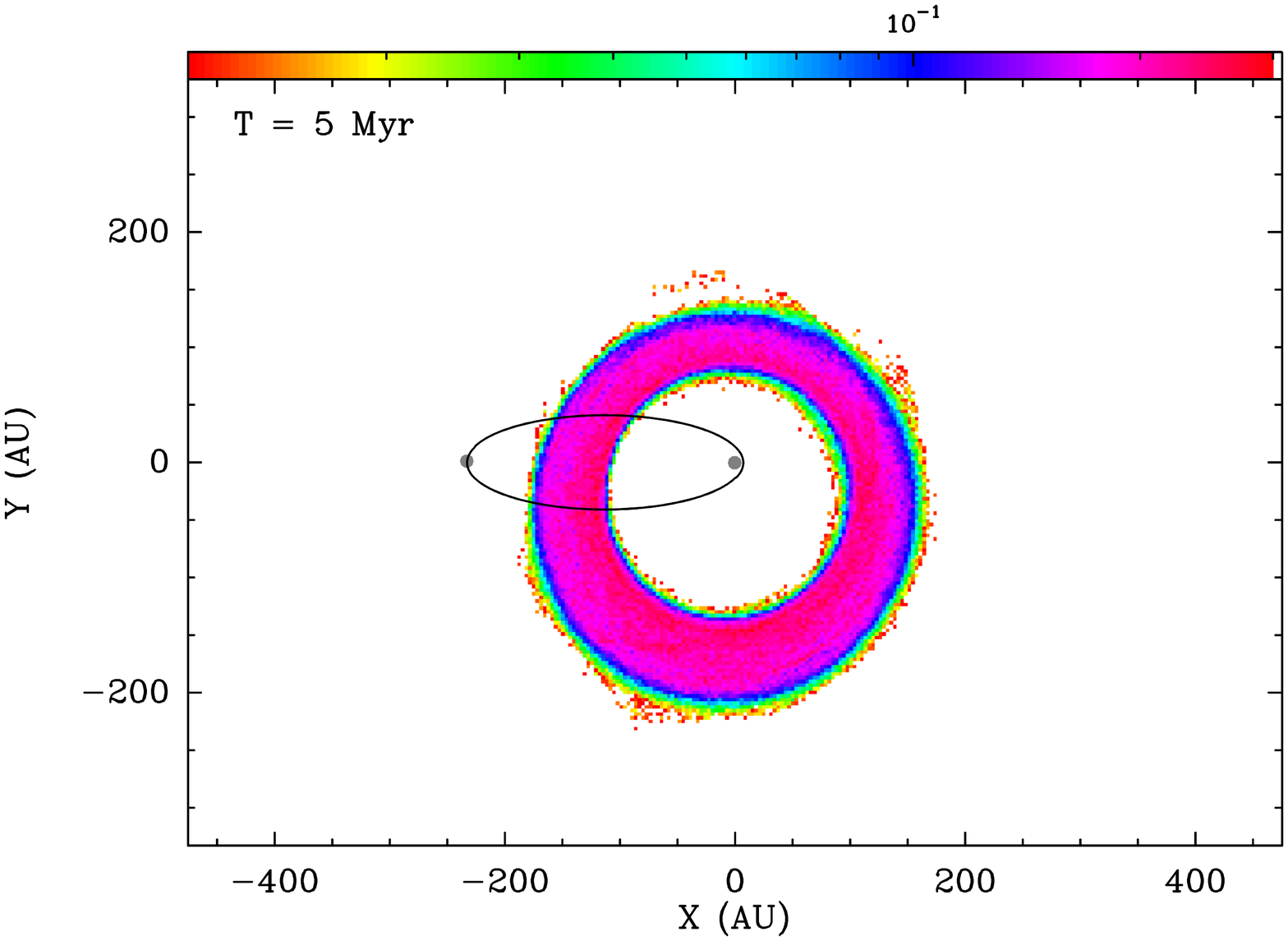} \hfil
\includegraphics[width=0.33\textwidth]{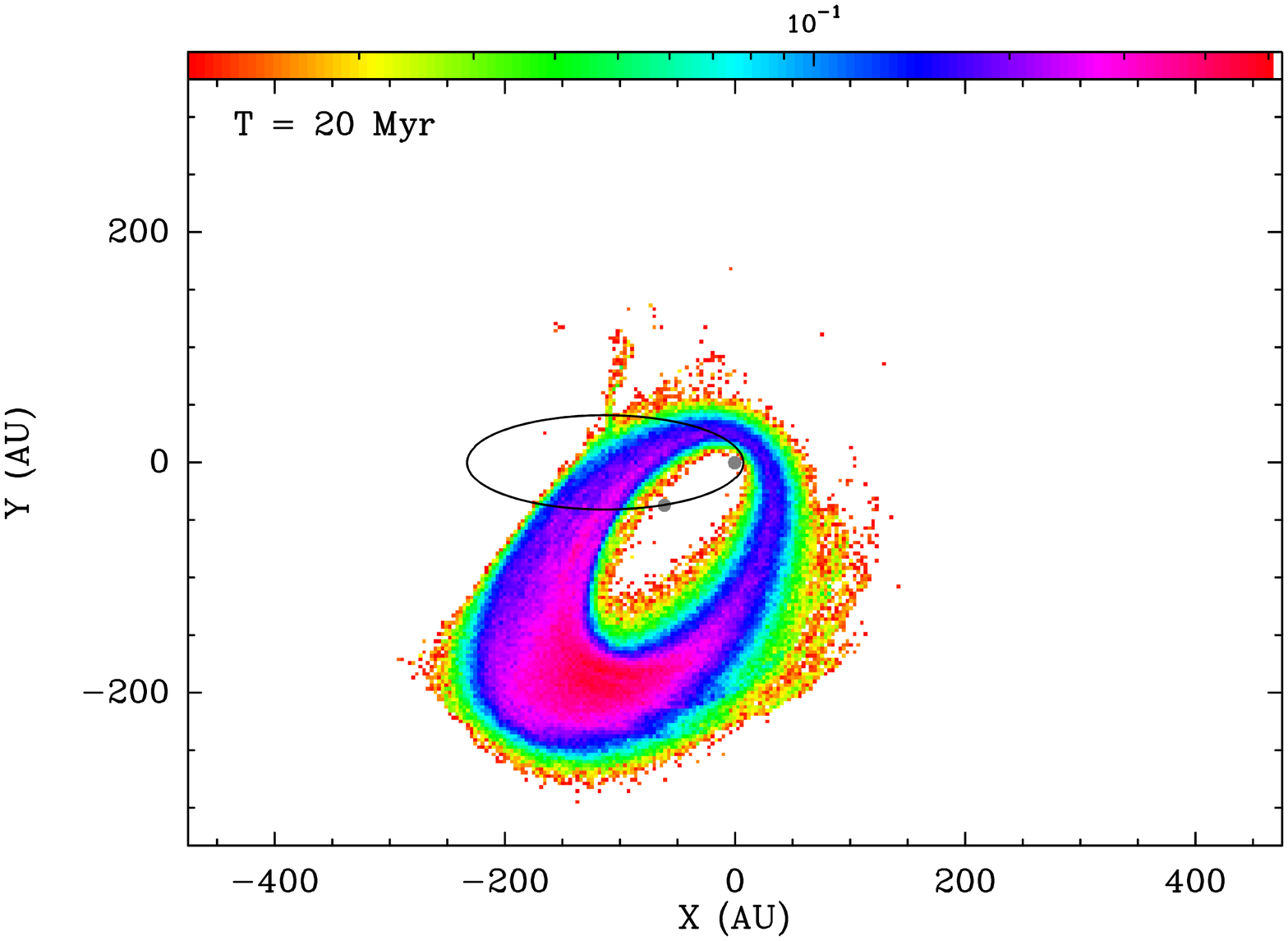} \hfil
\includegraphics[width=0.33\textwidth]{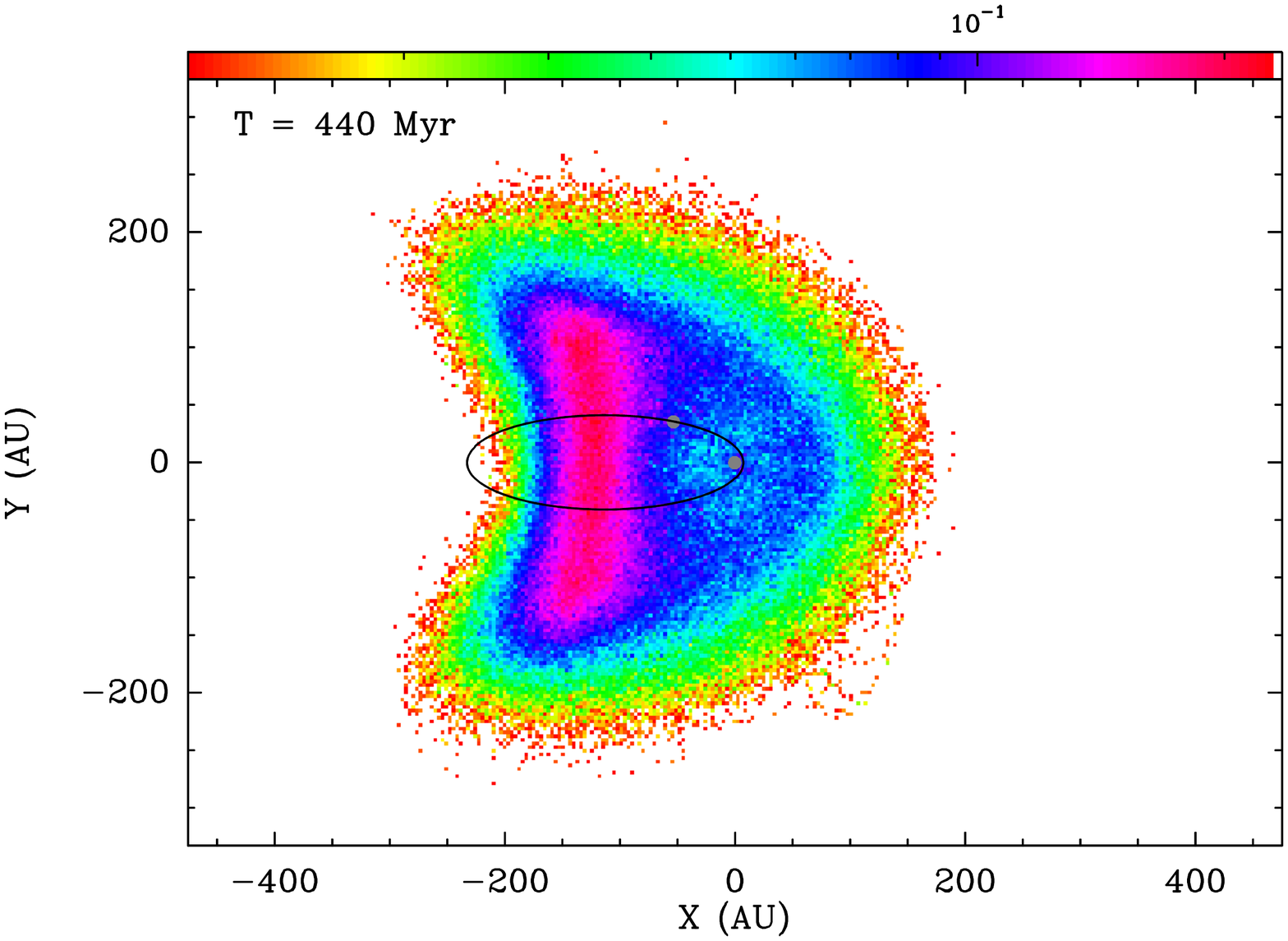}}
\makebox[\textwidth]{
\includegraphics[width=0.33\textwidth]{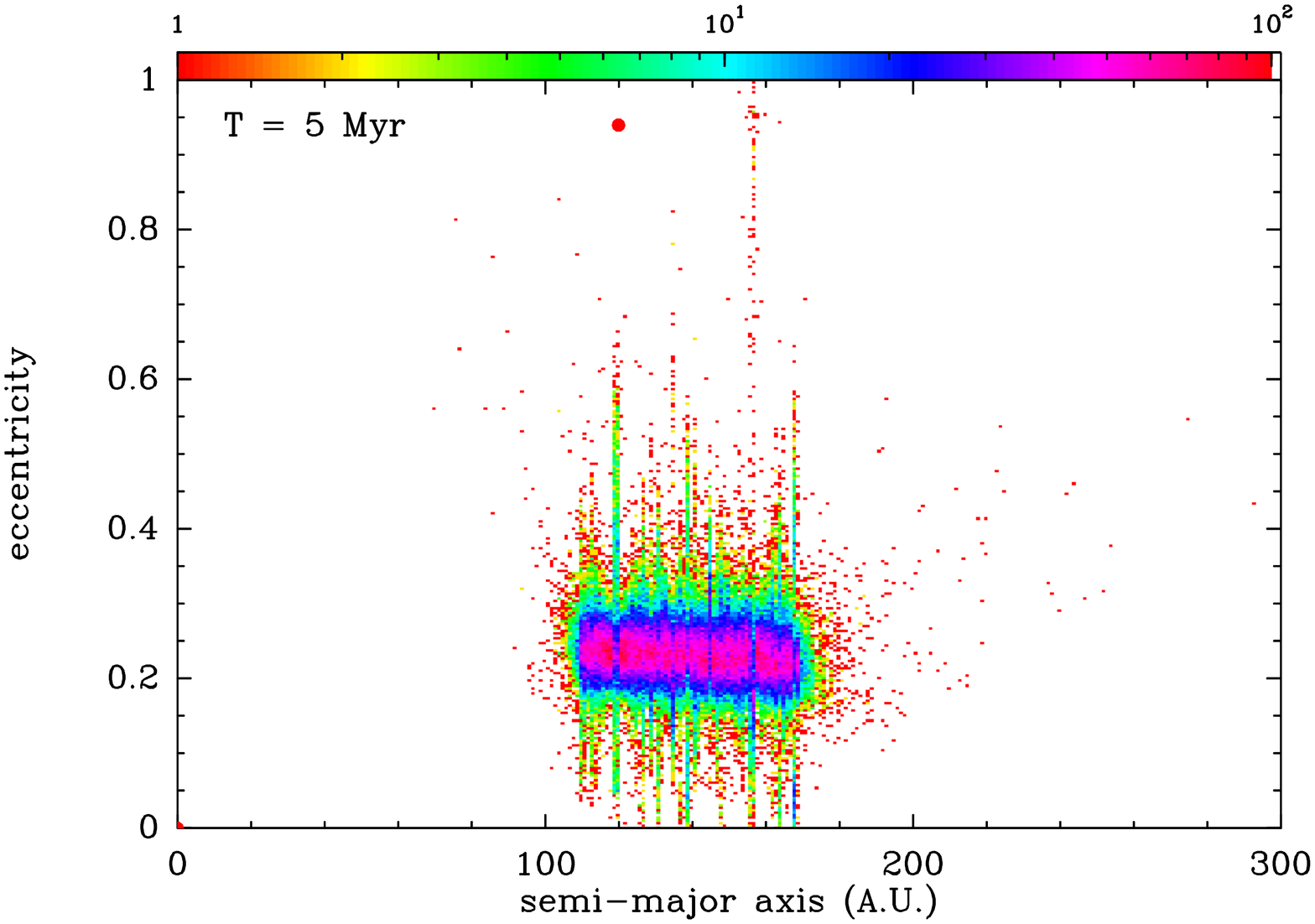} \hfil
\includegraphics[width=0.33\textwidth]{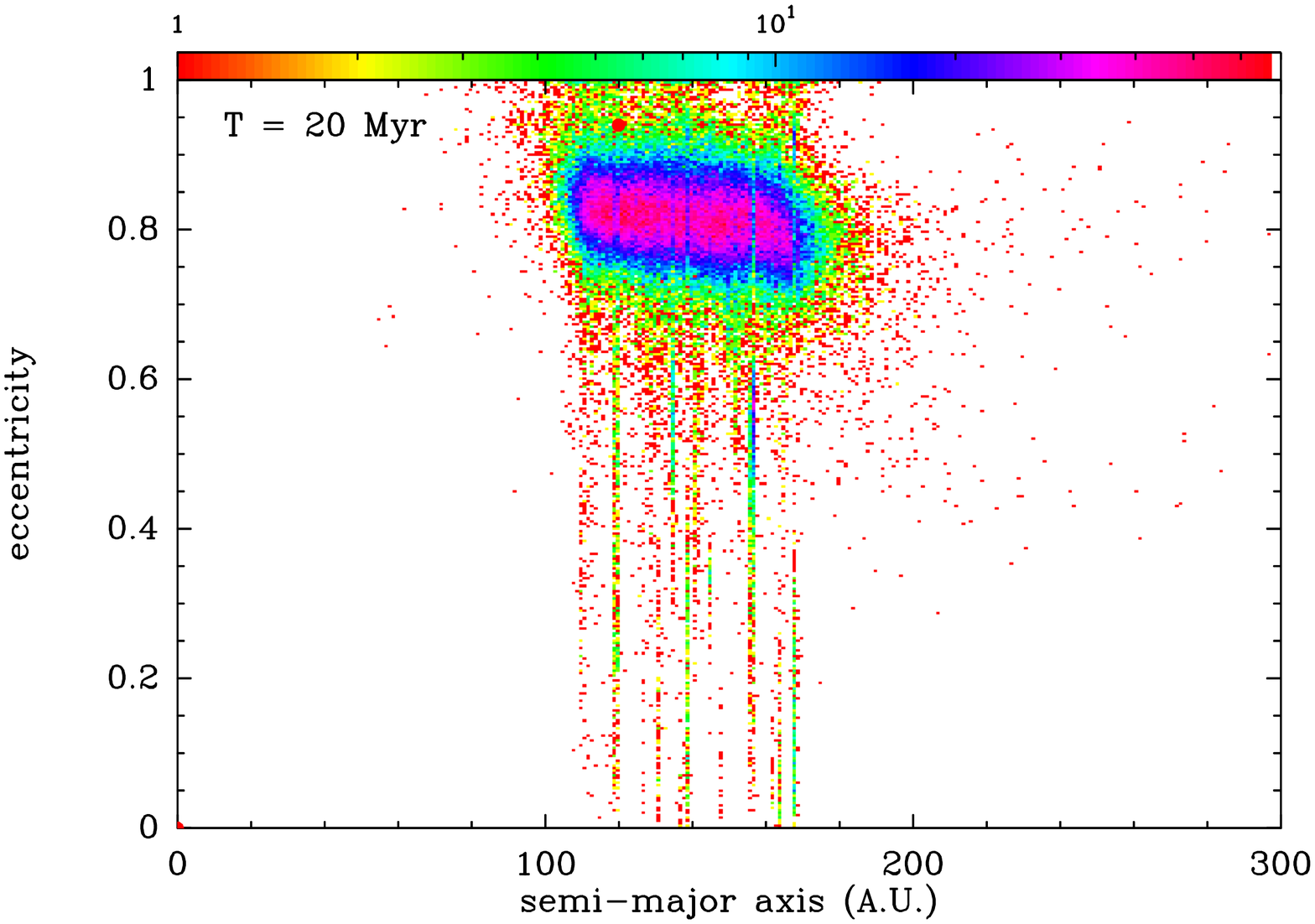} \hfil
\includegraphics[width=0.33\textwidth]{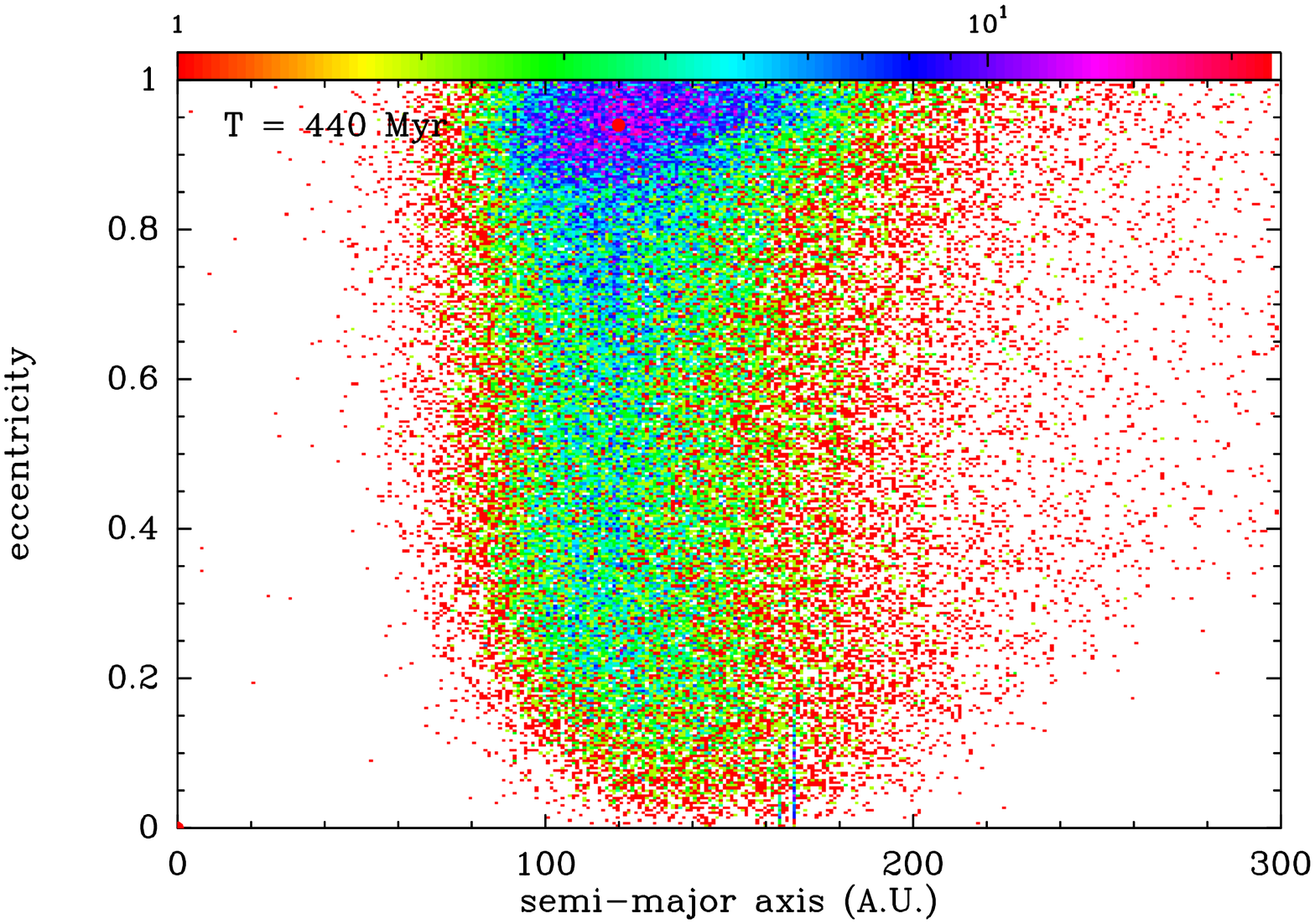}}
\caption[]{Result of the N-body integration with a perturbing planet with
$m=0.02\,\mjup$. The conventions are the same as in Fig.~\ref{simu_1mjup}.
Three epochs are represented:  $t=5\,$Myr (left), $t=20\,$Myr (middle) and
$t=440\,$Myr (right).}
\label{simu_002mjup}
\end{figure*}
We come now to a similar simulation, but with a lower mass for
\fomb. Figure~\ref{simu_002mjup} presents a simulation with a mass
$m=0.02\,\mjup=6.28\,M_\oplus$ (Super-Earth regime).
The disk is represented at three epochs:
$t=5\,$Myr, $t=20\,$Myr and $t=440\,$Myr, i.e., the estimated age of \fom.
We do not show the initial disk, as it is identical to that in
Fig.~\ref{simu_1mjup}. At $t=5\,$Myr, we note a drastic difference
with the previous simulation. The disk now still assumes a disk shape with a
moderate ($e\sim 0.2$) eccentricity. This disk configuration actually
roughly matches the observed disk, but the elliptic disk is not apsidally
aligned with the planet's orbit. It instead appears rotated by $\sim 70\degr$.
This contradicts both our orbital determination, which suggests
apsidal alignment, and the predictions of the standard
pericenter glow theory. This is actually due to the high eccentricity
of \fomb; see explanation in Sect.~4.

At $t=20\,$Myr, the disk still assumes this elliptic shape with a
similar angular tilt with respect to the planet's apsidal line. But
now the disk particles have reached much higher eccentricities ($\sim
0.6$ -- $\sim 1$), causing the disk to no longer resemble the observed
one. In fact, the bulk eccentricity of the disk increases continuously
with time. At $t=5\,$Myr it is $\sim 0.2$, while at $t=20\,$Myr it is $\ga0.6$.
An average disk eccentricity of 0.1, considered as a
good match to the observations, is reached earlier than $t=5\,$Myr, in fact
at $t=2\,$Myr (plot not shown here). But even in that case, the disk appears
tilted the same way as at $t=5\,$Myr.

At $t=440\,$Myr, the particles' eccentricities have spread over all
possible values. The disk no longer assumes a ring shape. This indeed appears
to be the case much earlier in the simulation. After $t=20\,$Myr,
the particles' eccentricity keep increasing up to high values, and the disk
structure is already lost at $t=\sim 80\,$Myr. In fact the
situation at $t=440\,$Myr with $m=0.02\,\mjup$ is comparable to that
at $t=5\,$Myr with $m=1\,\mjup$, except that less particles have
been lost in close encounters.
\subsubsection{Sub-Earth regime}
\begin{figure*}
\makebox[\textwidth]{
\includegraphics[width=0.33\textwidth]{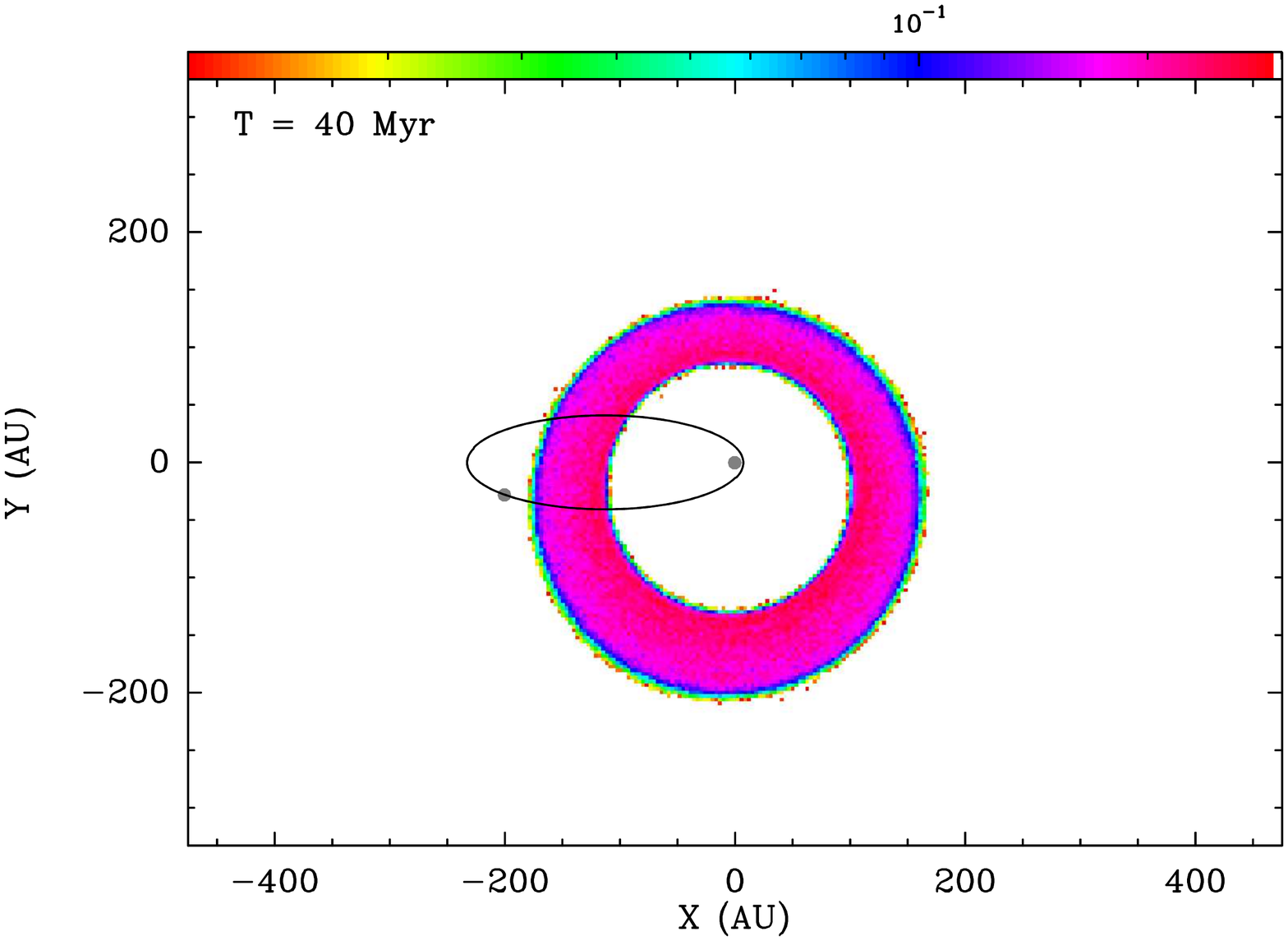} \hfil
\includegraphics[width=0.33\textwidth]{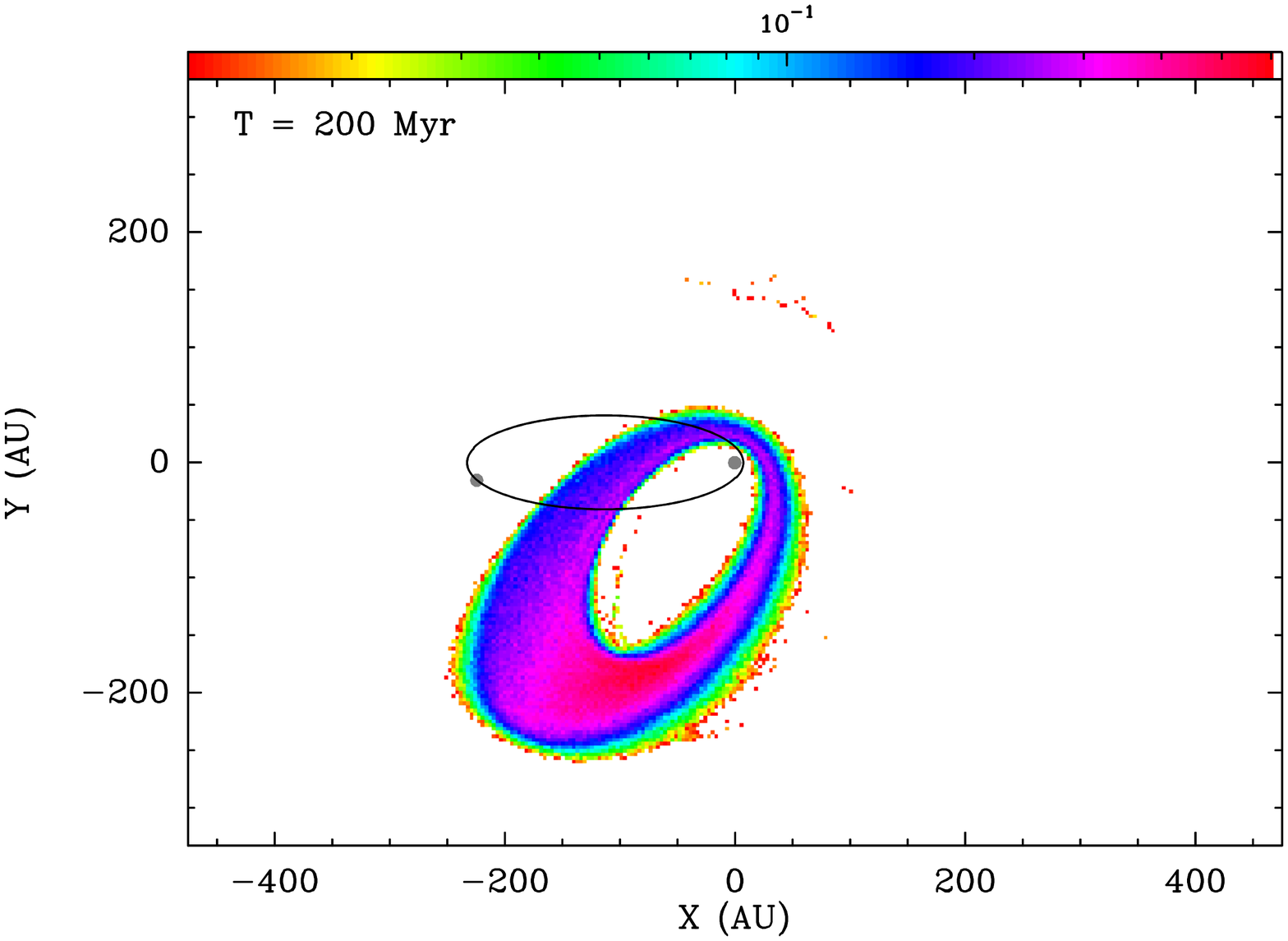} \hfil
\includegraphics[width=0.33\textwidth]{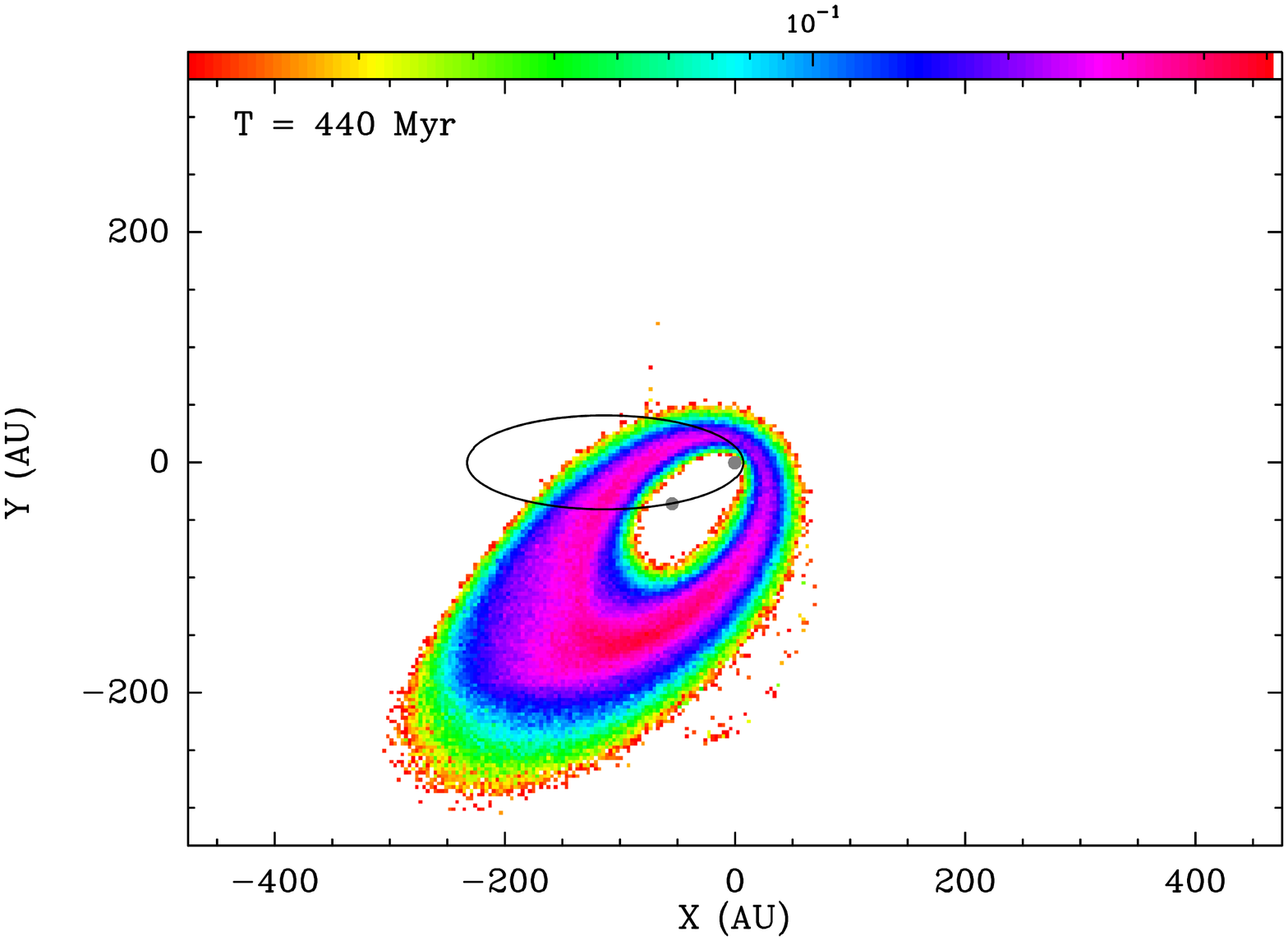}}
\makebox[\textwidth]{
\includegraphics[width=0.33\textwidth]{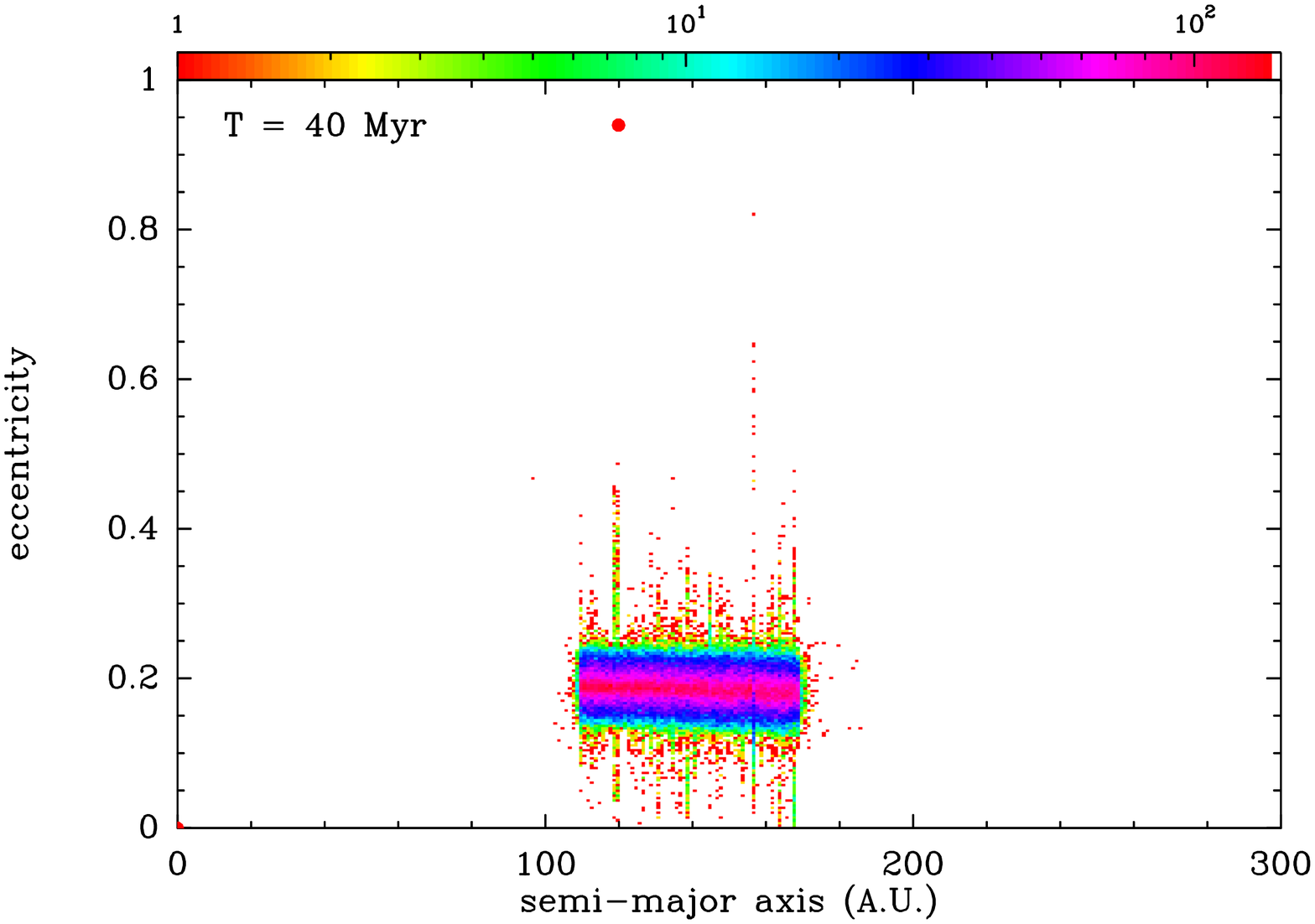} \hfil
\includegraphics[width=0.33\textwidth]{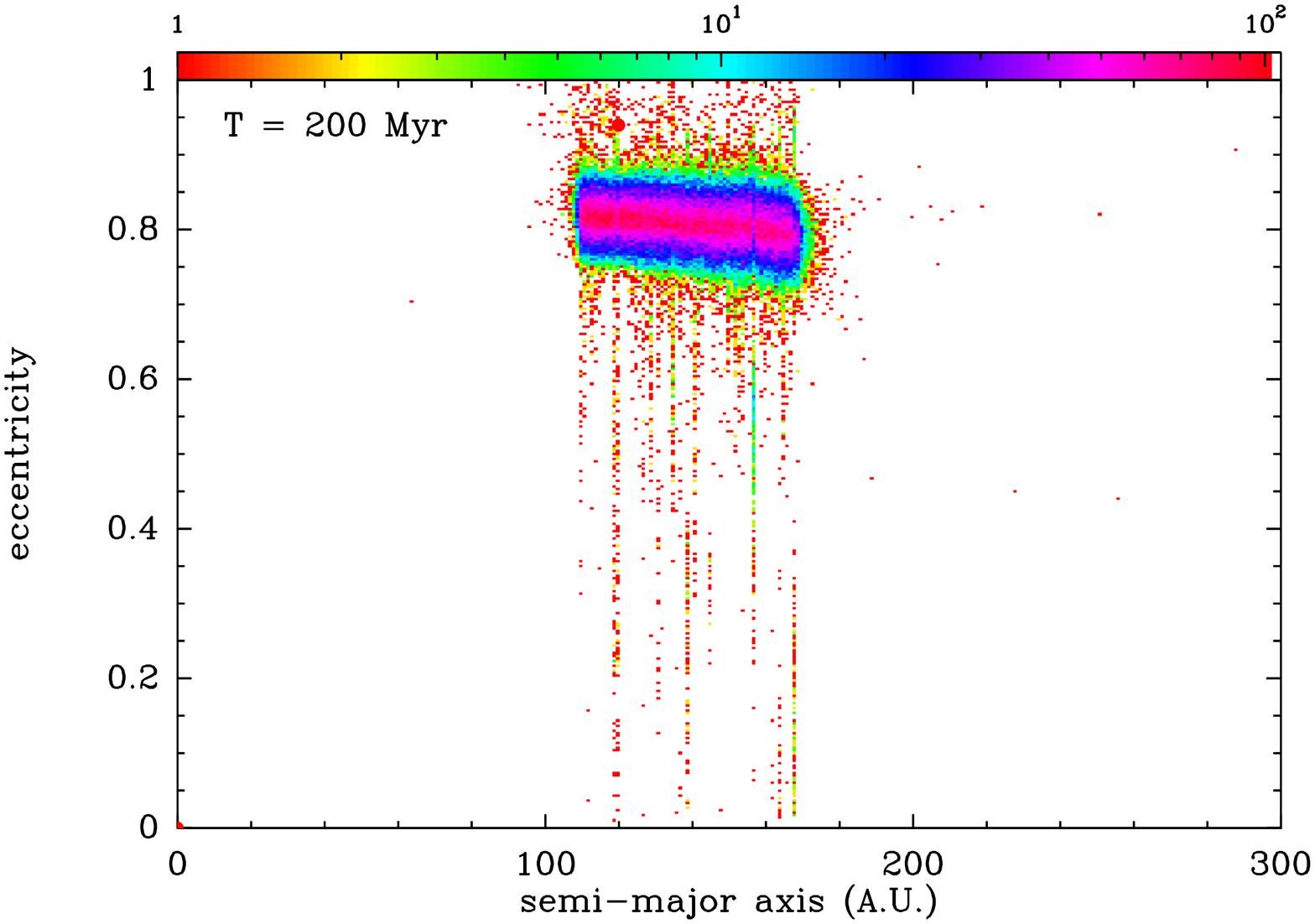} \hfil
\includegraphics[width=0.33\textwidth]{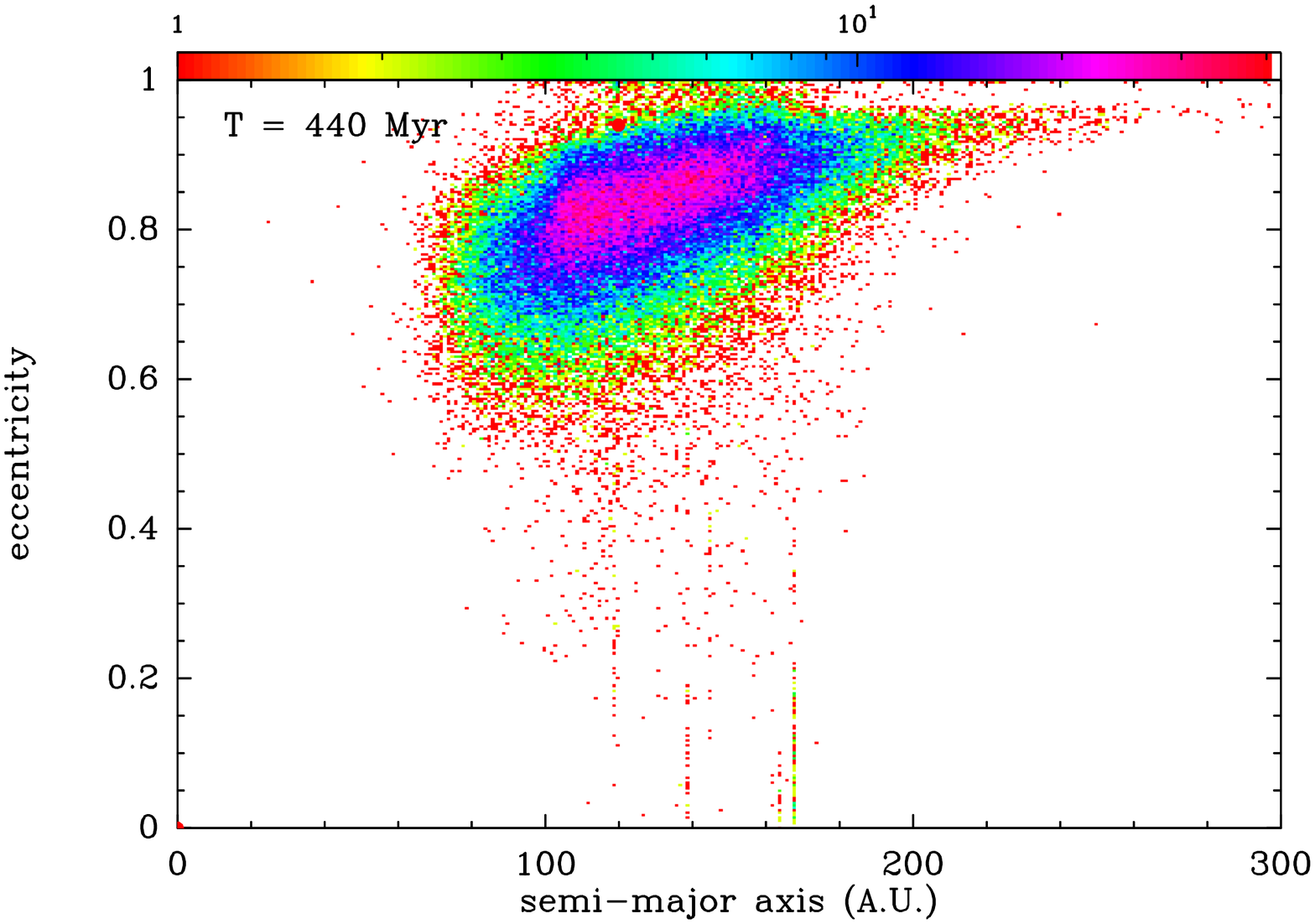}}
\caption[]{Result of the N-body integration with a perturbing planet with
$m=0.002\,\mjup$. The conventions are the same as in Fig.~\ref{simu_1mjup}.
Three epochs are represented:  $t=40\,$Myr (left), $t=200\,$Myr (middle) and
$t=440\,$Myr (right).}
\label{simu_0002mjup}
\end{figure*}
Figure~\ref{simu_0002mjup} presents now a simulation with a mass
$m=0.002\,\mjup=0.628\,M_\oplus$ (Earth or sub-Earth regime).  The
epochs represented are now $t=40\,$Myr, $t=200\,$Myr and
$t=440\,$Myr. The main difference is that at all 3 epochs, the disk
still assumes an elliptic disk shape. But as with $m=0.02\,\mjup$,
the global disk eccentricity increases to reach high values.  This is
of course due to the increase of the eccentricity of the disk particles
which keep being apsidally tilted by $\sim 70\degr$ with respect to
the planet's orbit. In fact, the situation at $t=40\,$Myr with
$m=0.002\,\mjup$ is somewhat comparable to that at $t=5\,$Myr with
$m=0.02\,\mjup$, and the situation at $t=200\,$Myr with
$m=0.002\,\mjup$ also compares with that at $t=20\,$Myr with
$m=0.02\,\mjup$. An average disk eccentricity of 0.1 is reached at
$\sim 20\,$Myr.

It must be specified that this last simulation may be less realistic than
the others, in the sense that the planet's mass is now lower than
the mass of the dust disk. According to \citet{wya02} and \citet{chi09},
a mass of planetesimals ranging between $3\,M_\oplus$ and $20\,M_\oplus$ is
required to sustain the dust disk over the age of the star. This issue
is investigated further in Sect. 4.3.

\subsubsection{Discussion}
The three simulations described above with different masses for
\fomb\ present similarities and differences. The comparison between
the various outputs reveals comparable sequences: the disk is first
perturbed to assume an elliptic shape. This is due to an increase of
the eccentricities of the particles, while their longitudes of
periastron remain more or less constrained to $\sim 70\degr$ with
respect to the apsidal line of the planet. Then the global
eccentricity increases to reach very high values. Afterwards, the
particles spread in eccentricities and the structure of the disk is
lost. The main difference resides in the time-scale of this
process. At $t=5\,$Myr with $m=1\,\mjup$ the particles have already
very high eccentricities, and the structure of the disk is already
getting lost. At $t=440\,$Myr with $m=0.002\,\mjup$ we are barely
reaching this stage after the disk particles have seen their
eccentricities increase. Comparing the three runs, the time-scale of
the process turns out to be roughly inversely proportional to the
planet's mass. This is characteristic for a secular process, as the
secular disturbing function due to the planet is proportional to its
mass while the topology of the Hamiltonian depends only weakly on the
planet's mass (see next section).

Another difference between the three simulations resides in the loss
of particles. Obviously the higher the mass, the more efficiently
particles are lost. Particle loss is due to scattering by close
encounters. As expected, more massive planets are more efficient at
scattering particles. With $m=1\,\mjup$, particle scattering actually
dominates the dynamics after $\sim 5\,$Myr, so that there is virtually
no particle left at the age of the star. This is conversely not the
case for low mass planets. Figure~\ref{ntp} shows that the loss of
particles, although it is present, is not significant over a
time-scale of \fom's age. Thus we may stress that for low mass planets,
the dynamics is essentially secular, and that close encounters are negligible.
Note that this does not necessarily mean that there are no close encounters.
There are inevitably encounters, but they are less numerous, thanks to a
shorter Hill sphere. However, as the Hill radius scales as $m^{1/3}$, the
effect should not be so drastic. The other reason is that for a low mass
planet, it would take many subsequent encounters to actually eject a particle.

We also tried to vary the orbital configuration, in particular to add a
few degrees inclination ($5\degr$)
to the planet with respect to the disk mid-plane.
This appeared not to produce significant changes in the global results
describe here, so that our conclusion still hold and may be regarded as robust.

It turns out that with the orbit we deduced from our fitting
procedure, assuming a low mass for \fomb\ is enough to prevent the
destruction of the disk by scattering close encounters over a
time-scale corresponding to the age of the star, even if the planet
crosses the disk. The secular perturbations by the planet succeed in
rendering the disk eccentric, but they inevitably drive the particles
towards very high eccentricities that do not match the
observation. Depending on the mass assumed for \fomb\ an average
disk eccentricity of 0.1 is reached between a few $10^6$ to a few $10^7$\,yr
after the beginning of the simulations, which is still far below the
age of the system. Moreover, even when its global eccentricity matches
the observation, the disk appears not apsidally
aligned with the planet's orbit, which does not match the
conclusion of our orbital fit (Sect.~2).
This is also in contradiction with the pericenter glow
dynamics, where the particles get their maximum eccentricities when
they are apsidally aligned with the planet, causing the global disk
figure to be aligned similarly. The linear pericenter glow analysis
obviously does no longer apply here. This is a consequence of the very
high eccentricity assumed for the planet, as we detail below.
\section{Semi-analytical study}
\begin{figure*}
\makebox[\textwidth]{
\includegraphics[width=0.33\textwidth]{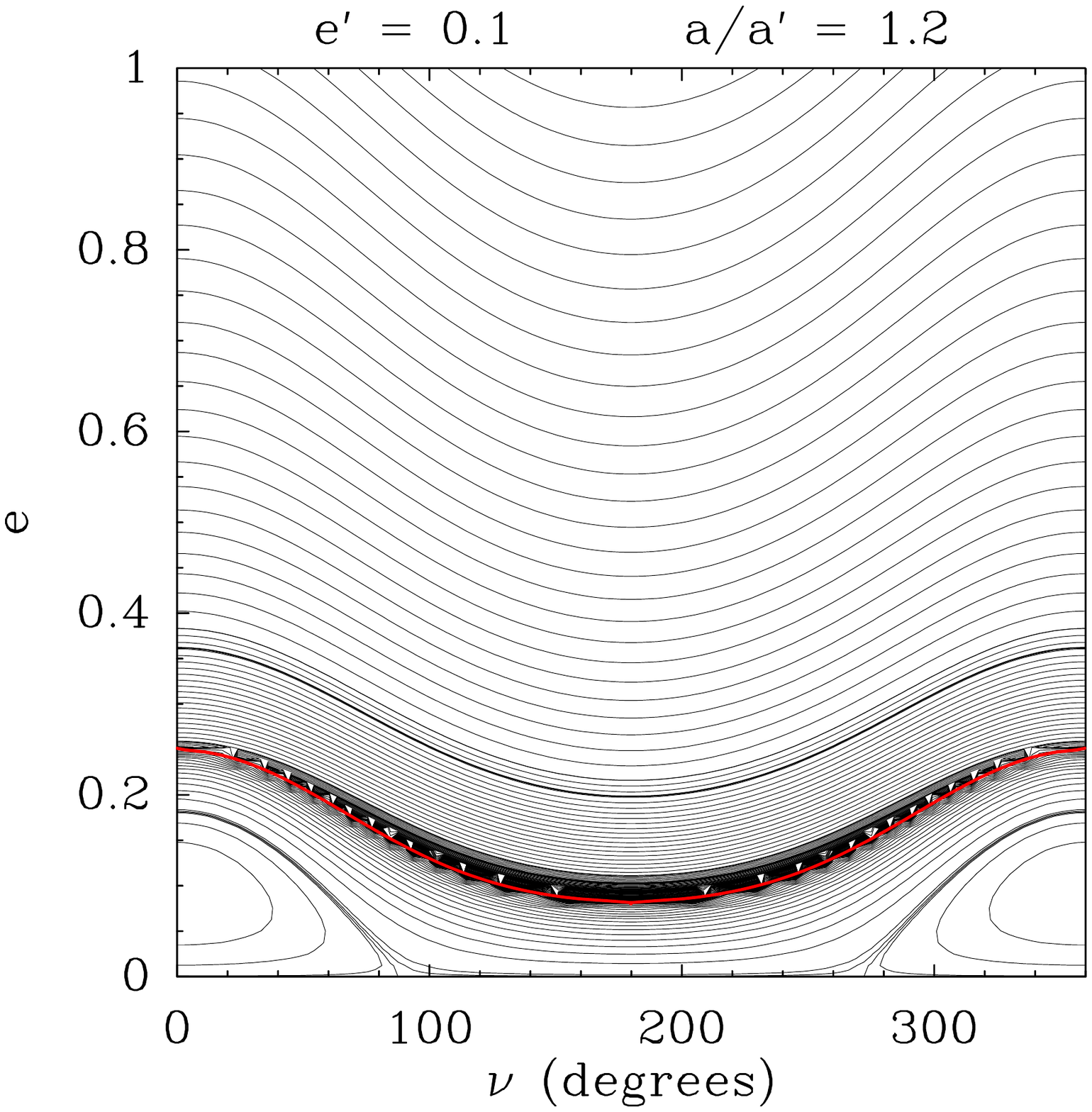} \hfil
\includegraphics[width=0.33\textwidth]{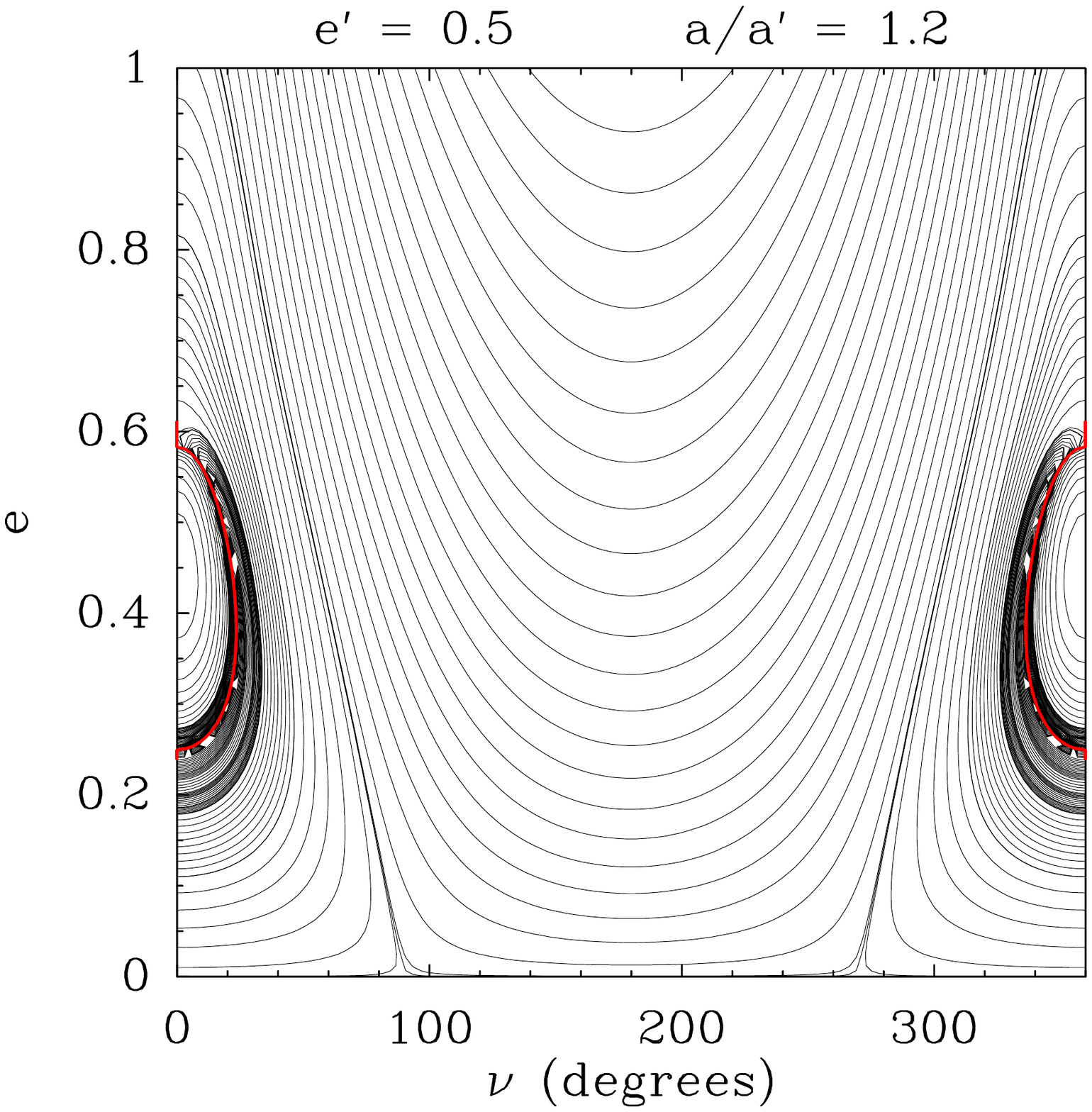} \hfil
\includegraphics[width=0.33\textwidth]{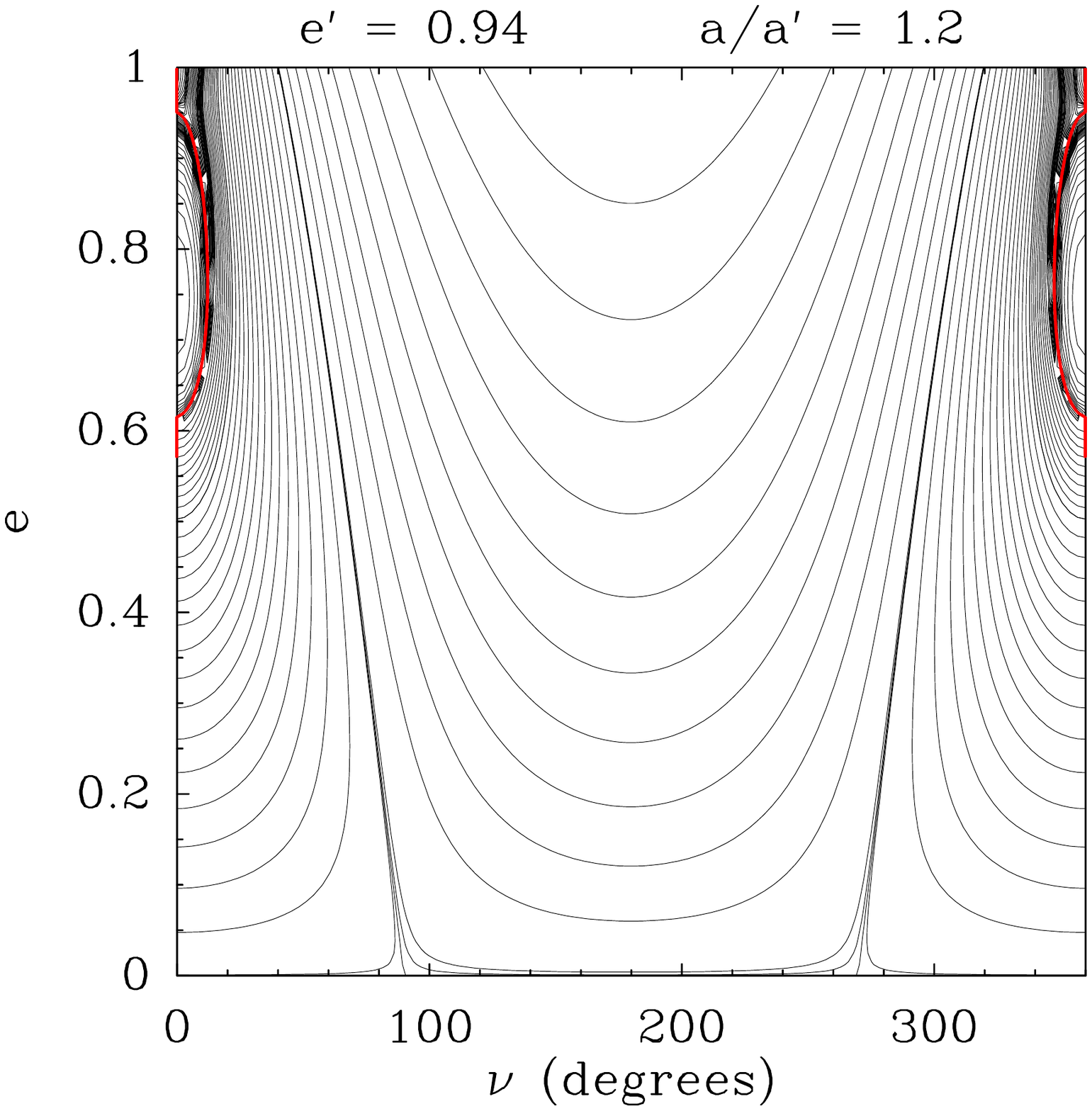}}
\caption[]{Phase portraits of secular averaged Hamiltonian
  $\overline{H}$ for different values of the perturber's eccentricity
  $e'$ and a fixed semi-major axis ratio $a/a'=1.2$, as a function of
  the longitude of periastron of the particle relative to that of the
  perturber $\nu=\varpi-\varpi'$. The red curves separate regions
  where the orbits actually cross from region where they do not. In
  the case $e'=0.1$ (left), the orbits do not cross below the red
  curve, while for $e'=0.5$ and $e'=0.94$, they do not cross inside
  the curves around $\nu=0$.}
\label{hsec}
\end{figure*}
\subsection{Theoretical background}
We consider a massless disk particle moving in the gravitational field of the
star \fom\ with mass $M$ and the planet \fomb\ with mass $m$. The motion of
the particle is thus described in the framework of the restricted three
body system. The Hamiltonian of the particle's motion then reads in
stellocentric reference frame
\begin{equation}
H=-\frac{GM}{2a}-Gm\left(\frac{1}{|\vec{r}-\vec{r'}|}-
\frac{\vec{r}\cdot\vec{r'}}{r'^3}\right)\qquad,
\end{equation}
where $G$ is the gravitational constant, $a$ is the particle's
semi-major axis in stellocentric referential frame, and $\vec{r}$ and
$\vec{r'}$ are the position vectors of the particle and the planet in
the same referential frame. We shall restrict ourselves to the planar
problem, where all three bodies move in the same plane. With this
assumption, the Hamiltonian $H$ reduces to two degrees of freedom.

The secular dynamics of the particle is then studied taking the time
average of $H$ over both orbits
\begin{equation}
\overline{H}=\frac{1}{4\pi^2}\int_0^{2\pi}\!\!\int_0^{2\pi}H(\lambda,\lambda')\,
\rd\lambda'\,\rd\lambda\qquad,
\label{hbar}
\end{equation}
where $\lambda$ and $\lambda'$ are the mean longitudes of the particle
and of the planet respectively.
This averaged Hamiltonian describes accurately the secular motion of
the particle as long as i) there is no close encounter between the
particle and the planet, ii) the two bodies are not locked in a
mean-motion resonance. We will assume that both conditions are
fulfilled, even when both orbits cross each other.
The numerical study showed indeed that as long as we do not take
too high a mass for \fomb, scattering by close encounters remains a
minor phenomenon (Fig.~\ref{ntp}).
Similarly, most planetesimals in our simulation are
very probably not in resonance with \fomb, as mean-motion resonances
usually cover small areas in semi-major axis. In fact, to enhance
resonance structures, additional mechanisms such as planet migration
are required \citep{rec08,wya03}.

The averaged Hamiltonian $\overline{H}$ cannot in general be expressed
in closed form. A full analytical treatment requires first to perform
an expansion of $H$ before averaging. There are two ways to do this. The
first is to assume that both orbits have very different sizes. Then
$H$ can be written in ascending powers of $r/r'$ (or $r'/r$ depending
on which orbit is the wider) using Legendre polynomials. The final
averaging is then written in ascending powers of $a'/a$ (or $a/a'$),
where $a$ and $a'$ are the semi-major axes. The second way to average
is to consider that both orbits may be of comparable sizes, but that the
eccentricities and inclinations are and will remain low. $H$ is then
expanded in ascending powers of eccentricities and inclinations using
Laplace coefficients and then averaged over both orbital motions. This
second technique, once truncated to second order in eccentricities and
inclinations, leads to describing the pericenter glow phenomenon.

We stress that none of these techniques can be applied here. As
\fomb's orbit crosses the disk, the disk particles' orbits cannot be
considered as significantly wider or smaller than \fomb's, and the
very high eccentricity we determine for \fomb\ prevents from using any
technique based on an expansion in ascending powers of eccentricity. A
semi-analytical study is nevertheless possible. As we consider the
planar problem, the Hamiltonian $H$ has two degrees of
freedom. But the averaged Hamiltonian $\overline{H}$ has only one, as
thanks to the averaging process, the semi-major axis is a secular
invariant. Considering then that $a$ is a secular invariant,
$\overline{H}$ is basically a function of only two dependant variables,
namely the eccentricity $e$ and the longitude of periastron
$\varpi$. It is even more relevant to describe it as a function of $e$
and of $\nu=\varpi'-\varpi$, where $\varpi'$ is the longitude of
periastron of the planet. It is then possible, for a given semi-major
axis value $a$, to compute numerically the value of $\overline{H}$ for
various sets of variables ($\nu$,$e$), and to draw level curves of
$\overline{H}$ in ($\nu$,$e$) space. As $\overline{H}$ is itself a
secular invariant, any secular evolution must be done following one of
these level curves. This technique of phase portrait drawing has
already proved efficiency to describe non-linear dynamics, such as in
resonant configurations in the $\beta\:$Pictoris case
\citep{bm96,bp3d}.
\subsection{Application to a test particle perturbed by \fomb}
The result in the case of a disk test particle perturbed by \fomb\
is shown in Fig.~\ref{hsec} for three planet eccentricity
values (from left to right): $e=0.1$, $e=0.5$, $e=0.94$. Of course,
given our orbital determination, the latter value is more relevant for
\fomb. The semi-major axis ratio was fixed to $a/a'=1.2$, as typical
of the situation under study. Assuming indeed $a'=120\,$au for \fomb,
$a/a'=1.2$ leads to $a=144\,$au, i.e., a typical particle in the middle of
the dust belt \citep{kal05}. We also checked nearby $a/a'$ values also
representative for various belt particles. We do not show the
corresponding phase portraits here. The Hamiltonian topology described
in Fig.~\ref{hsec} turns out indeed to be only slightly affected by the
fixed $a/a'$ value, so that the conclusions we derive here with
$a/a'=1.2$ still hold. Similarly, the mass ratio between the planet
and the star was fixed to $m/M=10^{-6}$ to build the phase portraits,
i.e., an Earth-sized planet. Changing $m/M$ appears not to change
anything noticeable to the shape of the Hamiltonian level curves, so
that we do not show corresponding phase portraits which are virtually
identical to those displayed here. This can be understood easily. The
variable part of $H$, which is responsible for the topology, is just
proportional to $m$. Therefore changing $m$ only scales that variable
part accordingly but does not affect the global topology.

In the phase portraits of Fig.~\ref{hsec}, the red curve separates
regions where both orbits not only overlap in distance, but
actually cross each other (assuming they are coplanar)
from regions where
they do not. We first describe the $e'=0.1$ case (left plot). We note an
island of $\nu$-libration around $\nu=0$ surrounded by smooth
$\nu$-circulating curves. We stress that this phase portrait actually
faithfully describes the pericenter glow phenomenon. Any particle
moving along a $\nu$-circulating curve will be subject to a precession
of $\nu$ (i.e., of the longitude of periastron $\varpi$) coupled with
an eccentricity modulation, and the maximum eccentricity will be
reached for $\nu=0$, i.e., when both orbits are apsidally
aligned. This is characteristic for pericenter glow, and this secular
evolution exactly matches the circular path of $z(t)$ in complex plane
described above. The same applies to particles moving in the
$\nu$-libration island around $\nu=0$. This corresponds to cases where
the circular $z(t)$ path does not encompass the zero point. This
situation can be viewed as a secular resonance where $\varpi$ no
longer circulates.

\begin{figure*}
\makebox[\textwidth]{
\includegraphics[width=0.33\textwidth]{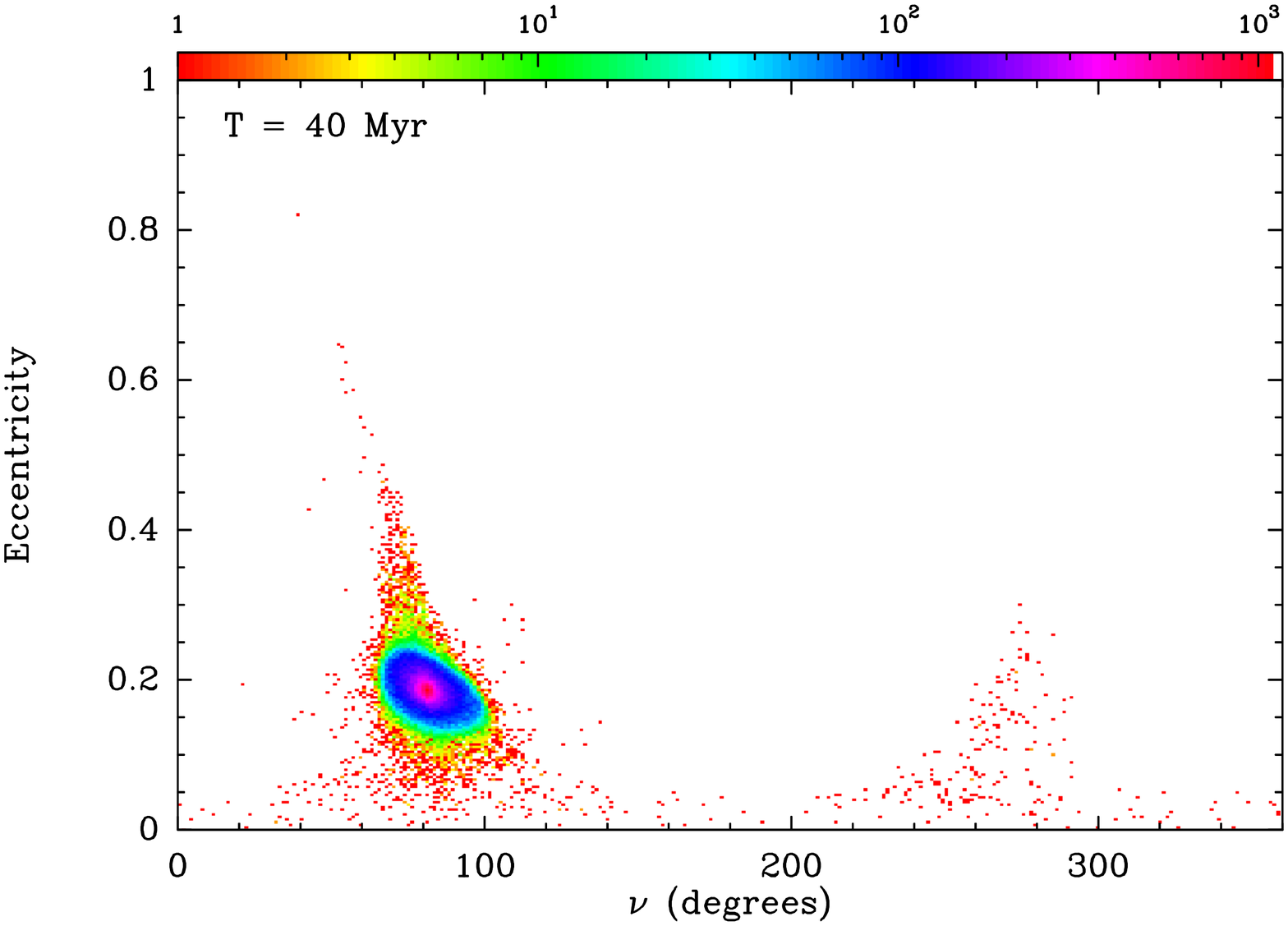} \hfil
\includegraphics[width=0.33\textwidth]{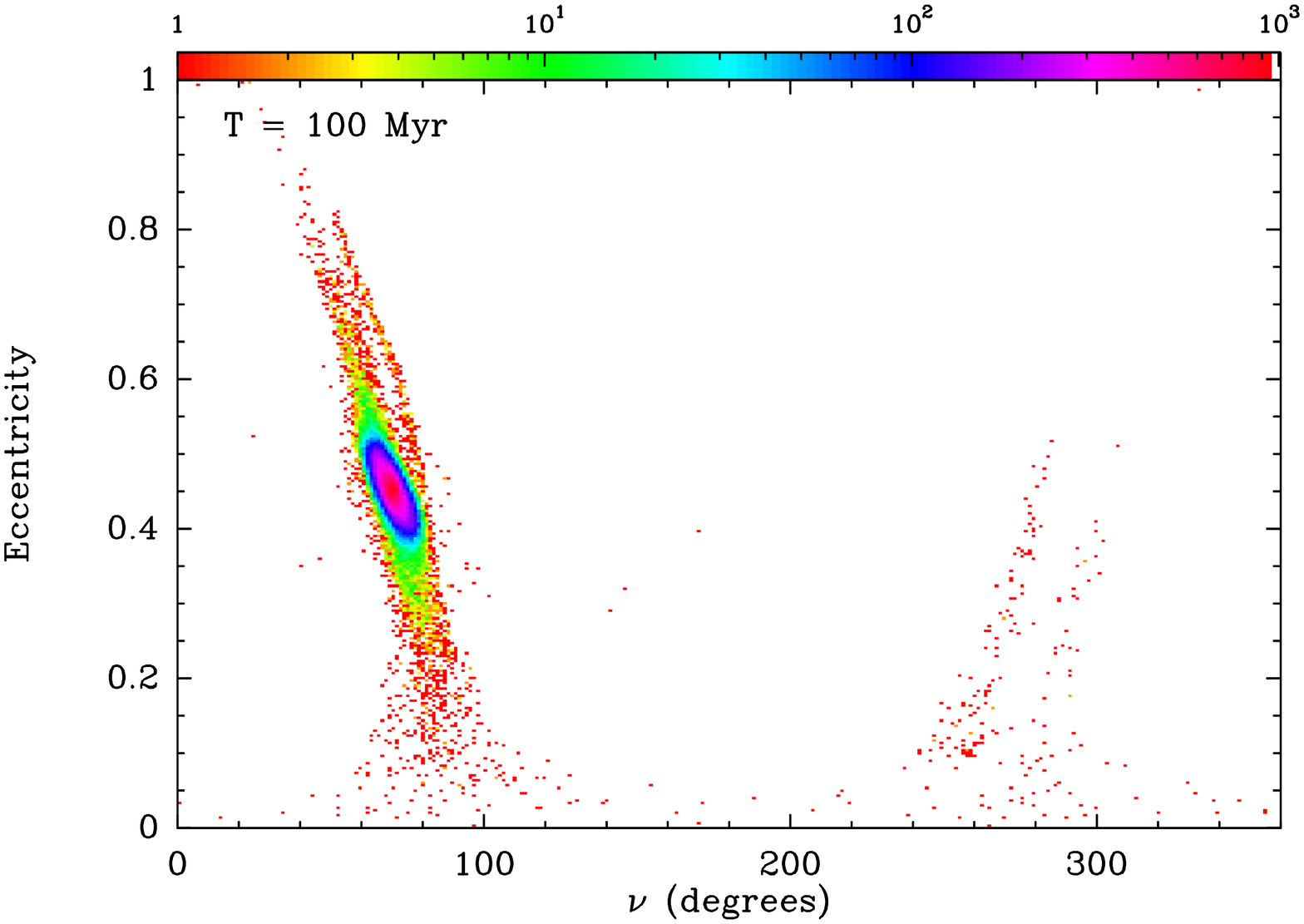} \hfil
\includegraphics[width=0.33\textwidth]{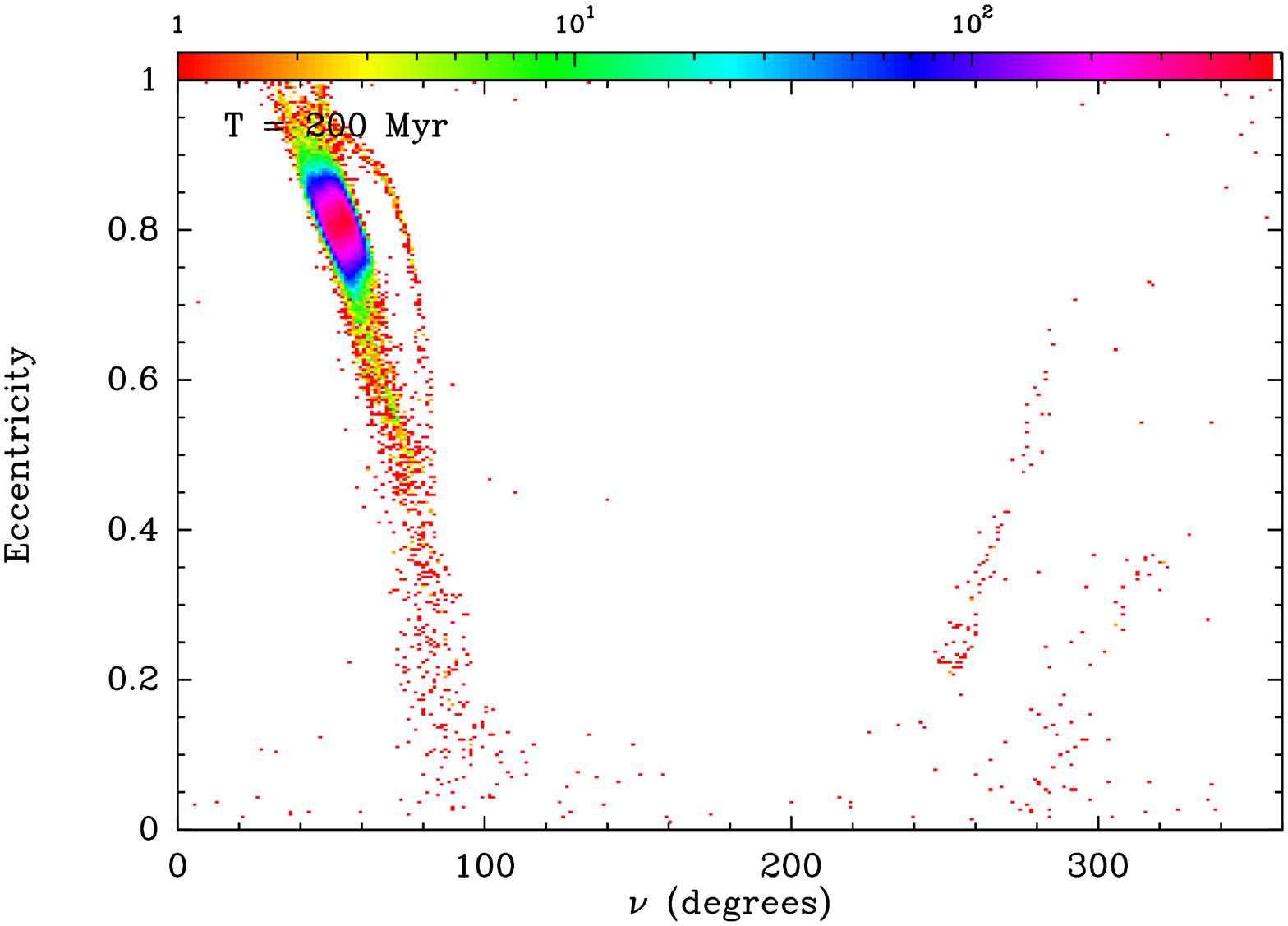}}
\makebox[\textwidth]{
\includegraphics[width=0.33\textwidth,origin=br]{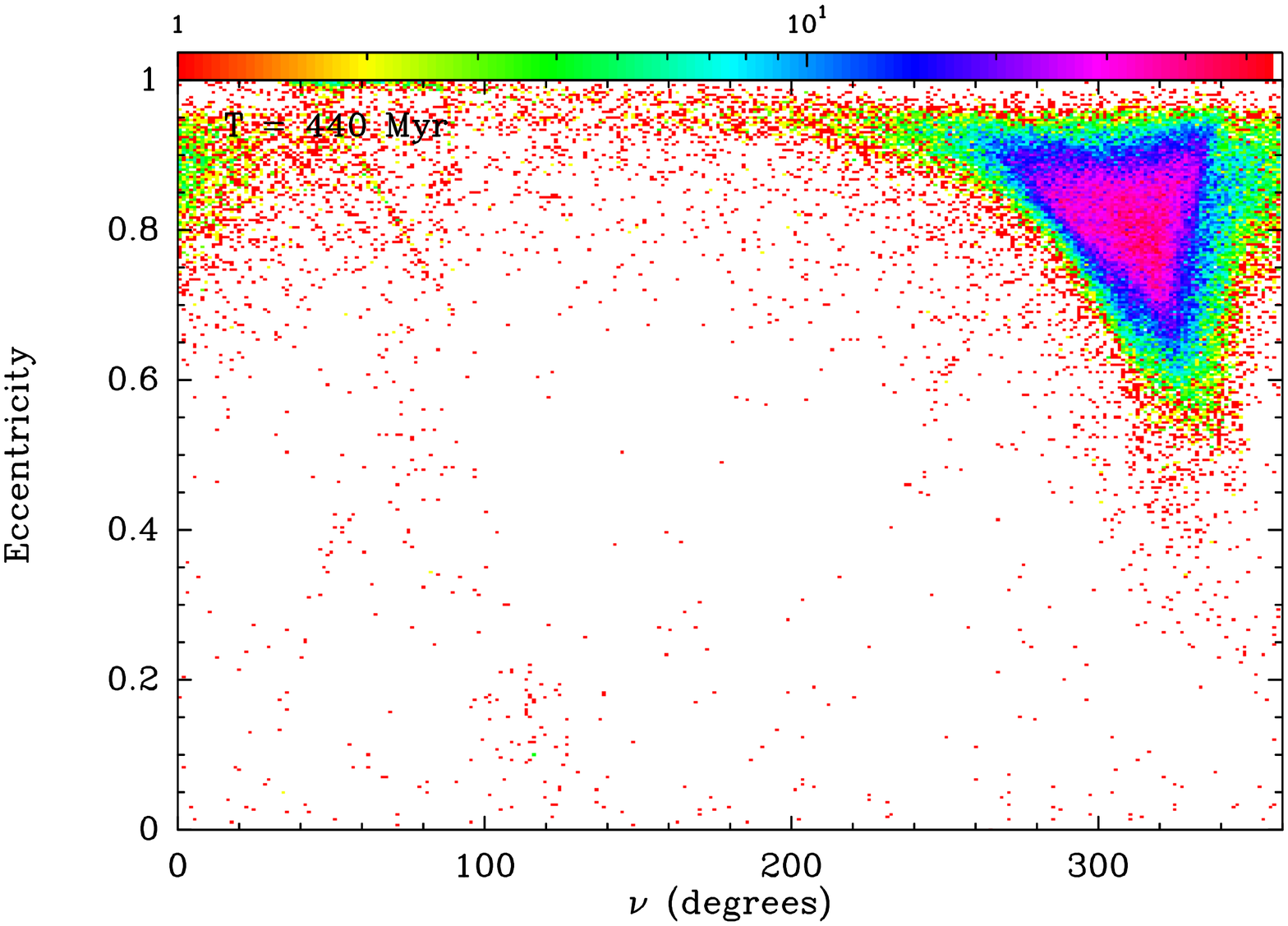} \hfil
\includegraphics[width=0.33\textwidth,origin=br]{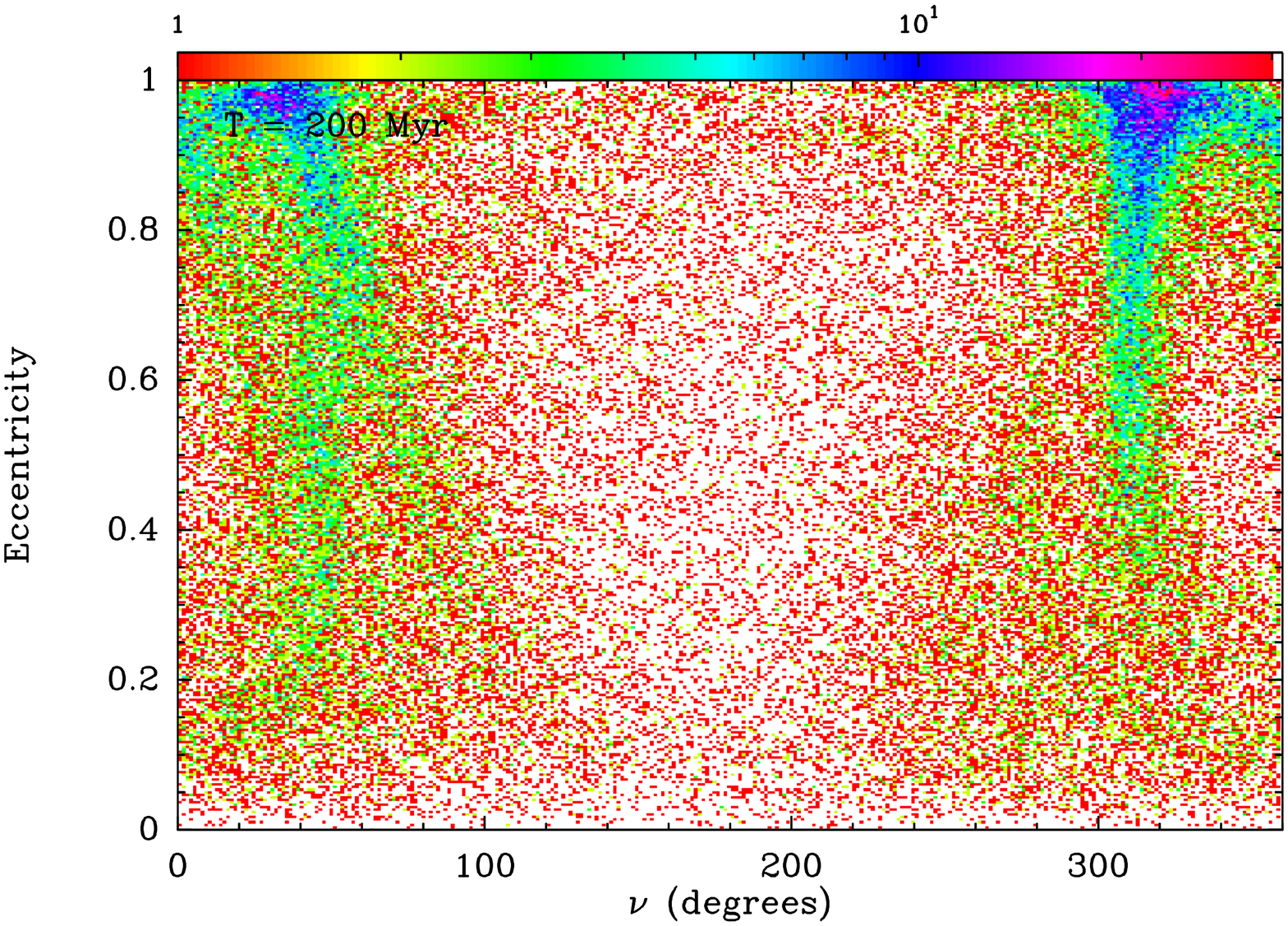} \hfil
\parbox[b]{0.3\textwidth}{
\caption[]{Views of the simulations of Fig.~\ref{simu_0002mjup} and
  \ref{simu_002mjup} in ($\nu,e$) space like in Fig.~\ref{hsec}. The
  three upper plots refer to Fig.~\ref{simu_0002mjup}
  ($m=0.002\,\mjup$) at $t=40\,$Myr, $t=100\,$Myr, $t=200\,$Myr, while
  the lower left plot corresponds to $t=440\,$Myr. The lower middle
  plot refers to Fig.~\ref{simu_002mjup} ($m=0.02\,\mjup$) at
  $t=200\,$Myr.}
\label{nue}}}
\end{figure*}
The situations with $e'=0.5$ and $e'=0.94$ are different. With
$e'=0.5$, the island of $\nu$-libration around $\nu=0$ is still
present, but it reaches now much higher eccentricities. It actually
encircles a small region in ($\nu,e$) space where both orbits do not
cross. But the main difference concerns the circulating curves. They
all reach very high eccentricities, virtually $e=1$. This means that
any particle starting at low eccentricity is about to evolve to this
very high eccentricity regime, unless it is subject to a close
encounter before. Contrary to what could be suggested from the phase
portrait, these particles do not pass beyond $e=1$, i.e., they are not
ejected by the secular process. Our numerical simulations show that
they pass through a very high eccentricity maximum before going down
in the diagram (see below). This does not show up in Fig.~\ref{hsec},
but can be understood in terms of orbital energy. As the semi-major
axis $a$ is a secular invariant, so is the orbital energy $-GM/2a$
(the fixed part of Hamiltonian $H$). It thus remains negative, hence
the particle remains bound to the star. The only way to eject a
particle here is to have a close encounter which has the ability to
affect the orbital energy.Strictly speaking, $\nu$ does
not circulate in this regime, but rather librates around $\nu=180\degr$.
Such $\nu=180\degr$-librating curves are in fact already present
in the $e'=0.1$ case, but only in the very high eccentricity regime (top
of the diagram). With $e'=0.5$, this regime extends down to low eccentricities
and the $\nu$-circulating regime has disappeared.

The situation at $e'=0.94$ is similar to that with $e'=0.5$, except
that it is even more drastic. The island of $\nu$-libration is now
confined to a tiny region close to $\nu=0$ at high eccentricity. As a
consequence, nearly all particles initially at low eccentricity in the
disk must evolve to the very high eccentricity regime. We claim that this
phase portrait exactly describes the dynamics observed in
Figs.~\ref{simu_002mjup} and \ref{simu_0002mjup}. We must specify here
that the level curves of Fig.~\ref{hsec} are explored in a fixed sense
that is imposed by Hamiltonian dynamics. For $e'=0.94$, basically the
left part of the diagram $\nu<180\degr$ corresponds to growing
eccentricities, while the right part $\nu>180\degr$ corresponds to
decreasing eccentricities. Now, consider a disk of particles initially
at low eccentricities and random $\nu$ values.  Following the
$\overline{H}$ level curves, all particles will see their eccentricity
grow when they reach $60\degr\la\nu\la 90\degr$. Irrespective of their
initial $\nu$ value, they will all have similar longitudes of
periastron during their eccentricity growth phase up to $e\simeq
1$. This is the exact origin of the eccentric disk tilted by $\sim
70\degr$ we observe in Figs.~\ref{simu_002mjup} and
\ref{simu_0002mjup}. Remember that in these simulations we had chosen
the perturbing planet in such a way that $\varpi'=0$, so that
$\nu=\varpi$.

To illustrate this, we plot in Fig.~\ref{nue} snapshots of the
simulations described in Figs.~\ref{simu_002mjup} and
\ref{simu_0002mjup}, but in $(\nu,e)$ space to better compare with
Fig.~\ref{hsec}. The first four plots (upper plots and lower left one)
show the evolution with $m=0.002\,\mjup$ (Fig.~\ref{simu_0002mjup}) at
various epochs. The correspondence with the phase portrait in
Fig.~\ref{hsec} is striking. We clearly see the eccentricity growth
phase of the particles with constrained $\nu$.
A discrepancy can nevertheless be noted
in the lower left plot (at $t=440\,$Myr) with respect to the
corresponding plot in Fig.~\ref{simu_0002mjup}, where we note that
the global orientation of
the disk leads to suggest that most particles have
$0\degr<\nu<180\degr$, while in Fig.~\ref{nue}, it turns out that at
the same time, they mostly have $180\degr<\nu<360\degr$. As explained in
Sect.~5, this apparent discrepancy is due to an inclination effect. At
this time, most particles have indeed moved to retrograde orbits, which
does not show up in the projected upper view of Fig.~\ref{simu_0002mjup}.

We also see that once
the particles reach high eccentricities, they start to diffuse in the upper
part of the diagram, before starting to get down to lower
eccentricities in the right part of the diagram. But at this level the
cloud of particles is much less concentrated in ($\nu,e$) space,
resulting in a less sharp eccentric disk. This diffusion is due to the
difference in secular evolution time-scales for the individual
particles. All particles do not rigorously evolve at the same speed in
($\nu,e$) space, so that they inevitably diffuse after a few
cycles. This is illustrated in the fifth plot of Fig.~\ref{nue} (lower
middle), which corresponds now to $m=0.02\,\mjup$
(Fig.~\ref{simu_002mjup}) at $t=200\,$Myr. As pointed out above, the
dynamical evolutions in both cases are almost identical, but with
$m=0.02\,\mjup$ it is just achieved faster, actually in a manner
proportional to $m$, as the variable part of $H$ is $\propto
m$. The situation at $t=200\,$Myr with $m=0.02\,\mjup$ can therefore also
be regarded as virtually corresponding to $t=2\,$Gyr with
$m=0.002\,\mjup$, as long as close encounters can be neglected. At
this stage, we see that the cloud of particles has diffused in all
parts of the diagram. A kind of steady-state regime has been achieved
where individual particles are at random phases of their evolution
tracks. They still gather around $\nu\simeq70\degr$ and
$\nu\simeq290\degr$ when their eccentricities grow or decrease, but the
disk no longer achieves an eccentric ring shape (Fig.~\ref{simu_002mjup}).
This picture does not change drastically if we adopt different orbital
parameters for the perturbing planet. Assuming different eccentricity and
semi-major axis values, the gathering points at $\nu\simeq70\degr$ and
$\nu\simeq290\degr$ appear to move by no more than $\sim 20\degr$.

It could been argued looking at the left plot of
Fig.~\ref{hsec} that a disk of particles starting at zero
eccentricity and perturbed by a planet with $e'=0.1$ may start start
to gather around $\nu\simeq70\degr$ before filling all the available
phase space and generate the pericenter glow phenomenon. This
corresponds indeed to the transient spiral structures noted by
\citet{wya05}. But this transient phase lasts at most a few Myrs
\citep{wya05}, which is very short. The steady-state regime,
characterized by diffusion of particles into the phase space and
subsequent apsidal alignment, sets on more quickly.
\subsection{Disk self-gravity and very low mass regime for \fomb}
As pointed out above, the simulations involving a very low mass
\fomb\ might be unrealistic because of the neglected disk mass. In our
simulations indeed, the disk is made of massless particles which do
not influence \fomb's orbit nor perturb each other. This approximation
remains justified as long as \fomb's mass remains higher than the disk
mass. According to \citet{wya02} and \citet{chi09}, a mass of
planetesimals ranging between $\sim3\,M_\oplus$ and $\sim
20\,M_\oplus$ is required to sustain the dust production in the
debris disk over the age of the star. It is of course hard to derive a
more accurate estimate, but obviously, when we consider a
$6\,M_\oplus$ \fomb, its mass is comparable to that of the disk, and
with $m=0.6\,M_\oplus$ it is clearly below. Consequently the reality of some
of our simulations may appear questionable.

Strictly speaking, addressing this issue would require to perform
simulations with a self-gravitating disk over the age of the star,
which would be extremely computing time consuming. It is nevertheless
possible to derive the effect
of the disk mass using our semi-analytical approach. As long as close
encounters and mean-motion resonances are not considered, which is the
case here, the secular effect of an elliptic disk is basically
identical to that of a planet with the same mass and orbiting on the
same orbit. In fact, the averaging process described in
Eq.~(\ref{hbar}) is virtually equivalent to replacing both bodies with
massive rings spread over their orbits. \citet{ter10} showed
for instance
directly that a planet perturbed by a massive inclined disk is subject to
Kozai effect exactly as if it was perturbed by another planet.

\begin{figure}
\centerline{\includegraphics[width=0.33\textwidth]{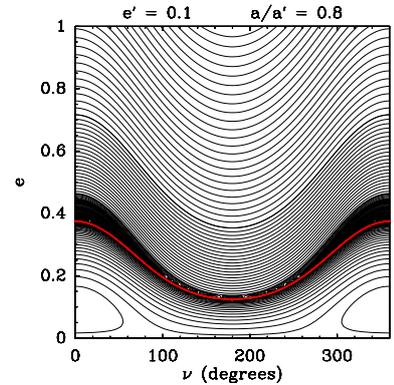}}
\caption[]{A phase portrait equivalent to the left plot of
  Fig.~\ref{hsec}, but with $a/a'=0.8$. This situation mimics the
  dynamics of \fomb\ as perturbed by a massive disk (see text).}
\label{hsec_bper}
\end{figure}
The first thing we need to investigate is the secular effect of a
massive ring on the orbit of \fomb. This situation can be modelled
treating \fomb\ as a test particle initially at $a=120\,$au and
$e=0.94$, perturbed by a planet orbiting at $a'=140\,$--150\,au and
$e'=0.1$. This is in fact very close to the situation depicted in the
left plot of Fig.~\ref{hsec}, except that the semi-major axis ratio
should be now taken as $a/a'\simeq 0.8$ instead of 1.2. The result is
shown in Fig.~\ref{hsec_bper}, which appears indeed very similar to
the left plot of Fig.~\ref{hsec_bper}. The initial configuration of
\fomb\ ($e=0.94$ and $\nu\simeq0$) corresponds to the top curves of
the phase diagram. Following any of these curves, we see that due
to the disk perturbation, the
periastron of \fomb\ is subject to precession, but that in any case,
its eccentricity will never get below $\sim 0.6$. Figure~\ref{hsec}
shows then that the dynamics of disk particles perturbed by a $e=0.6$
\fomb\ is very similar to that with a $e=0.94$ \fomb. We are thus
confident in the fact that even if its orbit is secularly perturbed by
the disk, this does not prevent \fomb\ from perturbing the disk
particles as described above.

\begin{figure*}
\makebox[\textwidth]{
\includegraphics[width=0.33\textwidth]{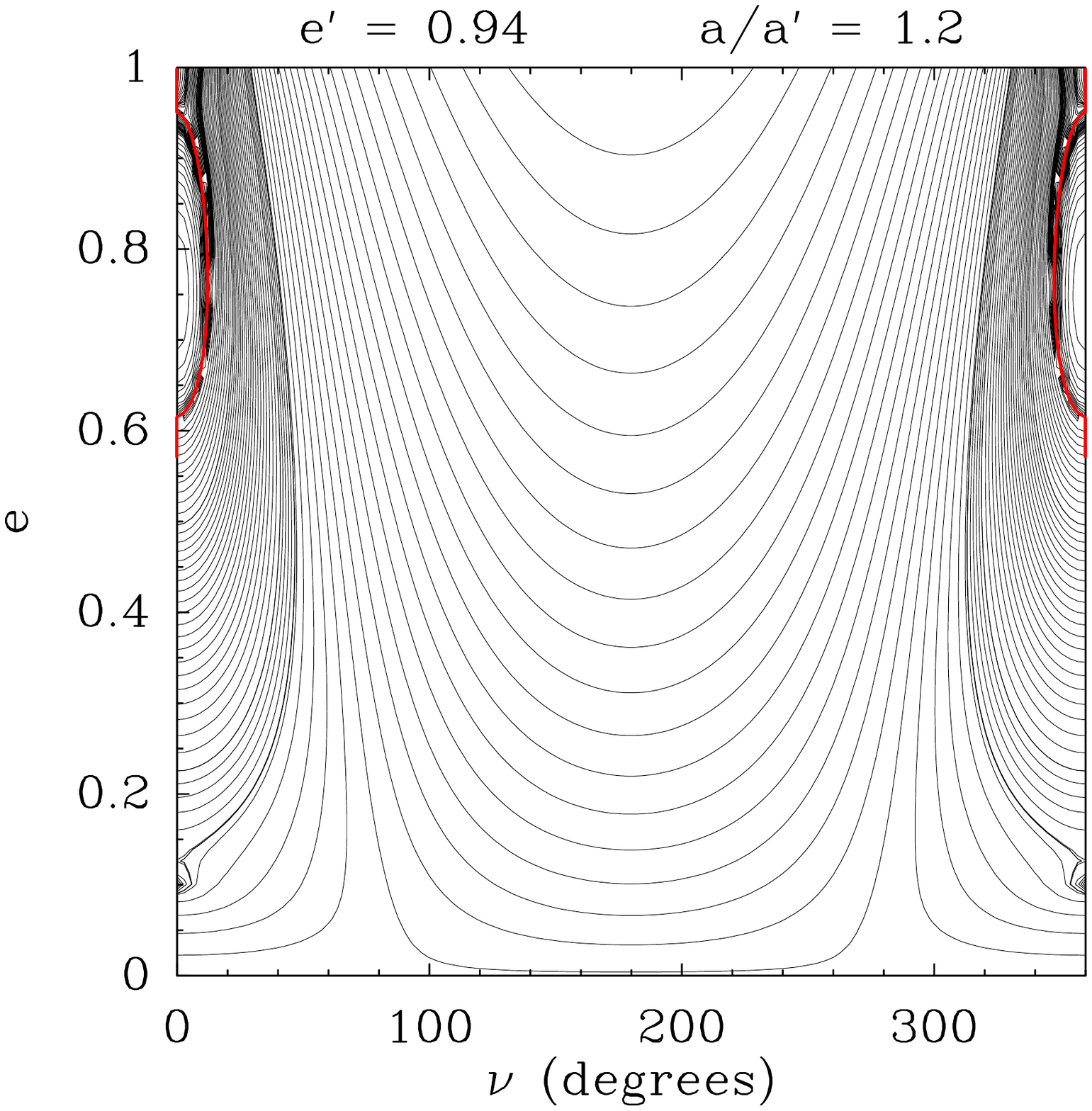} \hfil
\includegraphics[width=0.33\textwidth]{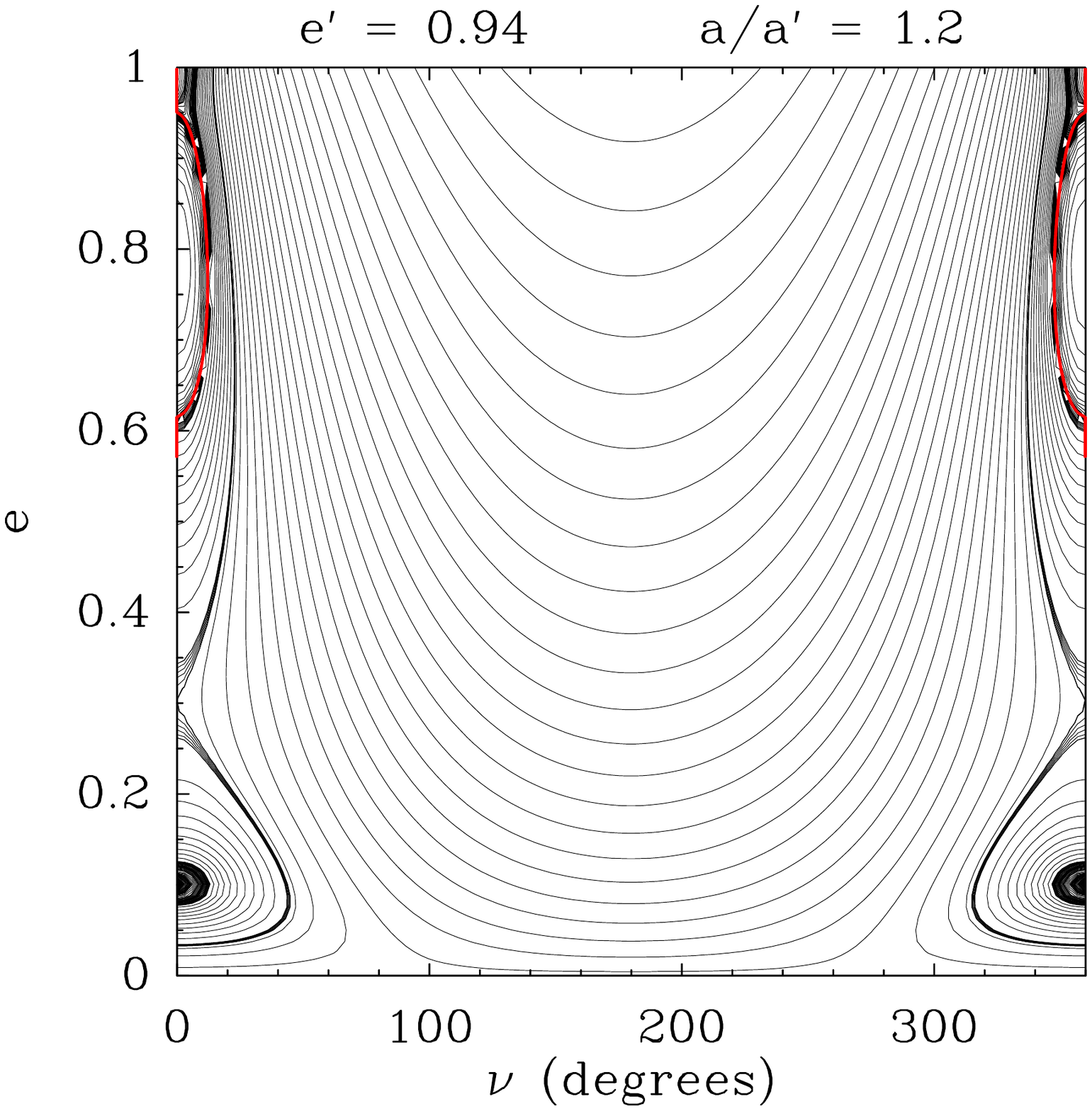} \hfil
\includegraphics[width=0.33\textwidth]{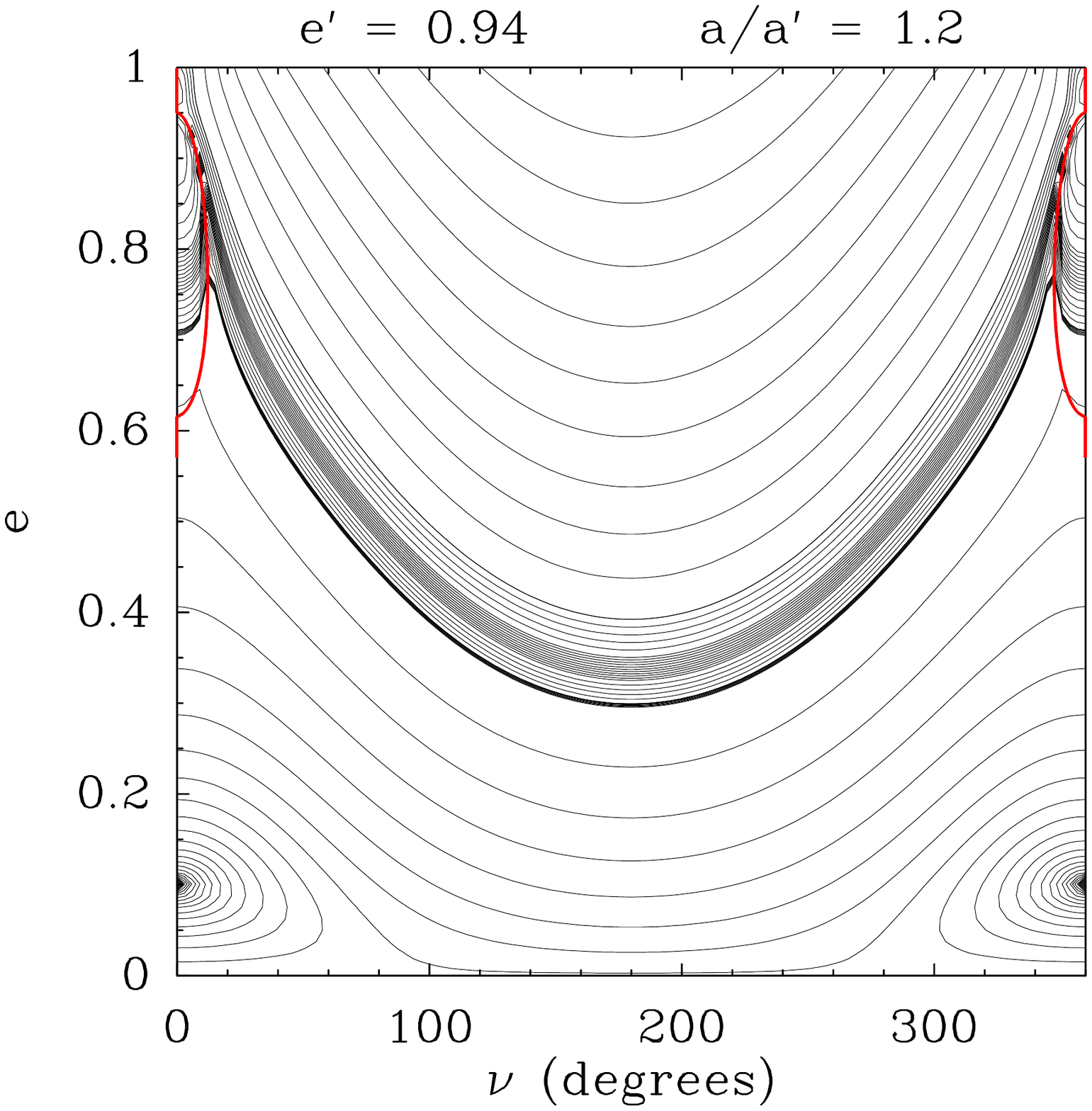}}
\caption[]{Phase portraits similar to those of Fig.~\ref{hsec} but to
  which we have added a second perturbing planet (representing the
  disk) apsidally aligned with the first one (\fomb). The second
  planet is assumed to orbit at the same semi-major axis as the disk
  particle. The important parameter is the mass ratio $\rho$ between
  the two planets. \textbf{Left plot:} $\rho=0.1$, i.e., a disk 10
  times less massive than \fomb; \textbf{Middle plot:} $\rho=1$, disk
  and \fomb\ have equal masses; \textbf{Right plot:} $\rho=10$, i.e.,
  a disk 10 times more massive than \fomb.}
\label{hsec_disk}
\end{figure*}
The second potential effect is the self-gravity of the disk, i.e., the
perturbation the massive disk can rise on disk particles. This can be
investigated adding a second perturbing planet to the situation
depicted in Fig.~\ref{hsec} and averaging the resulting Hamiltonian
over all orbits. This is illustrated in
Fig.~\ref{hsec_disk}. \fomb'eccentricity is taken equal to 0.94. The
planets are taken apsidally aligned to mimic the alignment between
\fomb\ and the disk, and the second planet's semi-major axis is taken
equal to that of the test particle, i.e., 1.2 times that of \fomb.
The important parameter here is the mass ratio $\rho$ between the
perturbing planets, i.e., between the disk and \fomb. In
Fig.~\ref{hsec_disk} we show phase diagrams for $\rho=0.1$ (left plot,
disk less massive than \fomb), $\rho=1$ (middle plot, equal masses),
$\rho=10$, (right plot, disk more massive than \fomb). With
$\rho=0.1$, the situation is very close to that of Fig.~\ref{hsec}
with $e=0.94$, which is not surprising as \fomb\ dominates the
dynamics. With $\rho=1$ (middle plot) the situation is now somewhat
changed. An island of libration appears now at low eccentricity around
$\nu=0$. This island corresponds to a secular resonance pericenter
glow region controlled by the second planet. But not all disk
particles moving at low eccentricity are concerned by this
behaviour. Contrary to a pure pericenter glow configuration (left plot
of Fig.~\ref{hsec}), those which are not trapped in the libration
island actually follow a route that drives them to high eccentricity
almost exactly as if the second planet was not there. Those particles
are stilled controlled by the highly eccentric \fomb. Given the
limited size of the libration island around $\nu$, the latter class of
particles is potentially more crowded that the former.  As a result we may
claim that a disk perturbed by an equal mass \fomb\ would still see a
significant part of its particles evolve towards high eccentricities
and yield a disk figure that does not match the present day
observation.

With $\rho=10$ (right plot), now the bottom part of the phase diagram
closely looks like that of the left plot of Fig.~\ref{hsec}. Only the
particles initially moving at high eccentricity actually feel a
noticeable perturbation by \fomb. Conversely, all particles moving at
low eccentricity follow a route entirely controlled by the disk treated as a
second planet. This does not explain the
eccentricity of the disk, as the second planet was initially given
the suitable eccentricity. All we stress here is that
we expect here the disk to be no longer affected by \fomb, which is not
surprising as it is now 10 times less massive than the disk.

We also checked intermediate values of $\rho$ (not shown here). When
increasing $\rho$ from 1 to 10, the island of libration at low
eccentricity around $\nu=0$ gets higher. The transition between the
regime where a significant part of the low eccentricity particles are
still perturbed towards high eccentricity and that where all particles
remain at low eccentricity occurs around $\rho\simeq 3.5$. As a
result, even if \fomb\ is 3 times less massive than the disk, it can
still perturb it in such a way that many particles are driven towards
high eccentricities. Given the disk mass estimates by \citet{wya02}
and \citet{chi09}, we conclude that a super-Earth sized \fomb\ (like
in the simulation of Fig.~\ref{simu_002mjup}) is very probably capable
of efficiently perturb the disk, while this is certainly no longer the
case for a sub-Earth sized \fomb. Our corresponding simulation
(Fig.~\ref{simu_0002mjup}) can therefore be considered as unrealistic
given the probable mass of the disk. We nevertheless presented it here to
illustrate the mechanism we describe and how its time-scale scales
with the planet's mass.

We may thus distinguish two regimes : For a $\sim$super-Earth sized
\fomb\ and above, the dynamics outlined in the previous section holds,
while for lower masses, the secular effect of \fomb\ is overridden by
the self-gravity of the disk so that its influence of the disk is very
small.
\section{Vertical structures}
\begin{figure*}
\makebox[\textwidth]{
\includegraphics[width=0.33\textwidth]{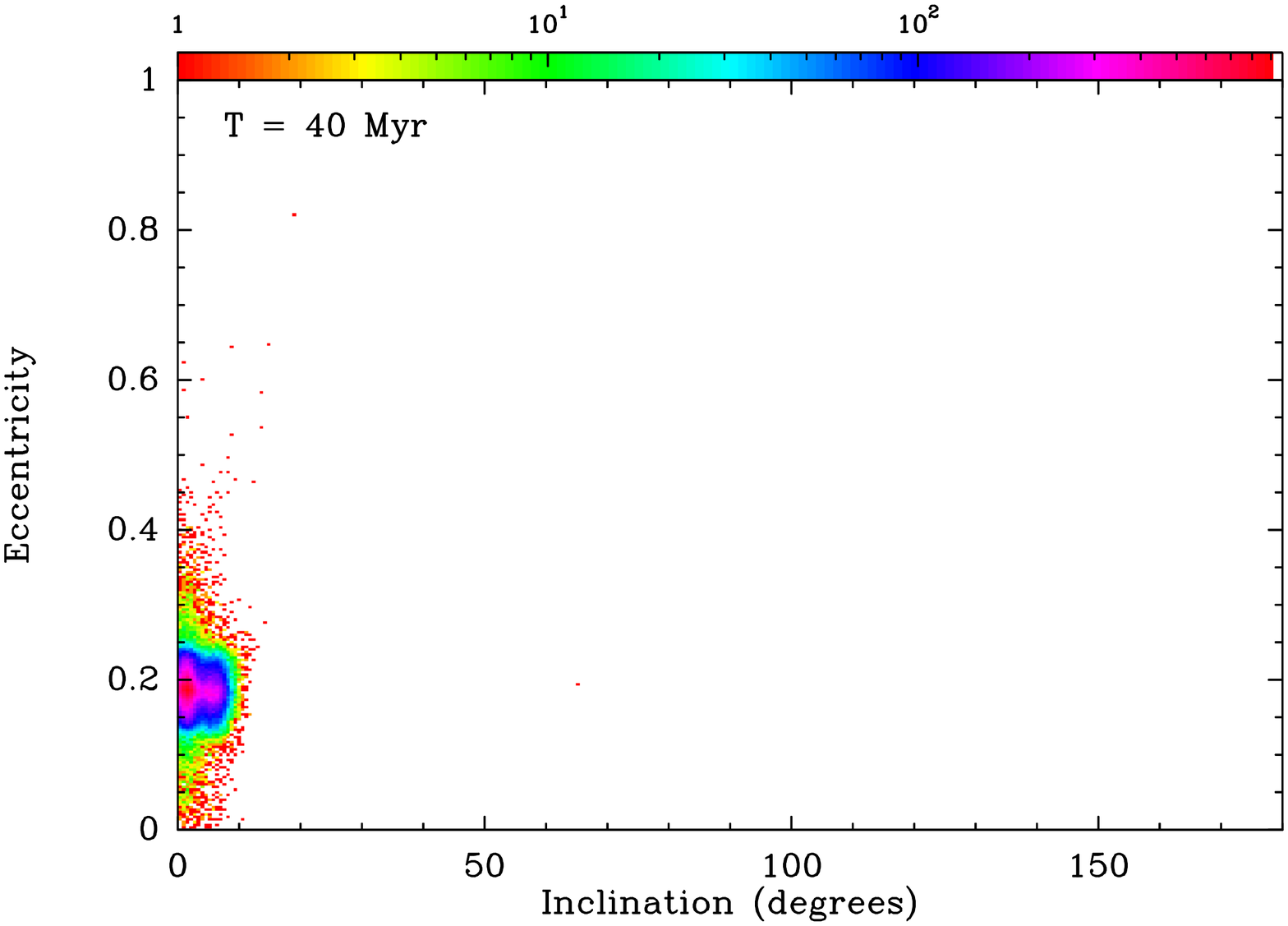} \hfil
\includegraphics[width=0.33\textwidth]{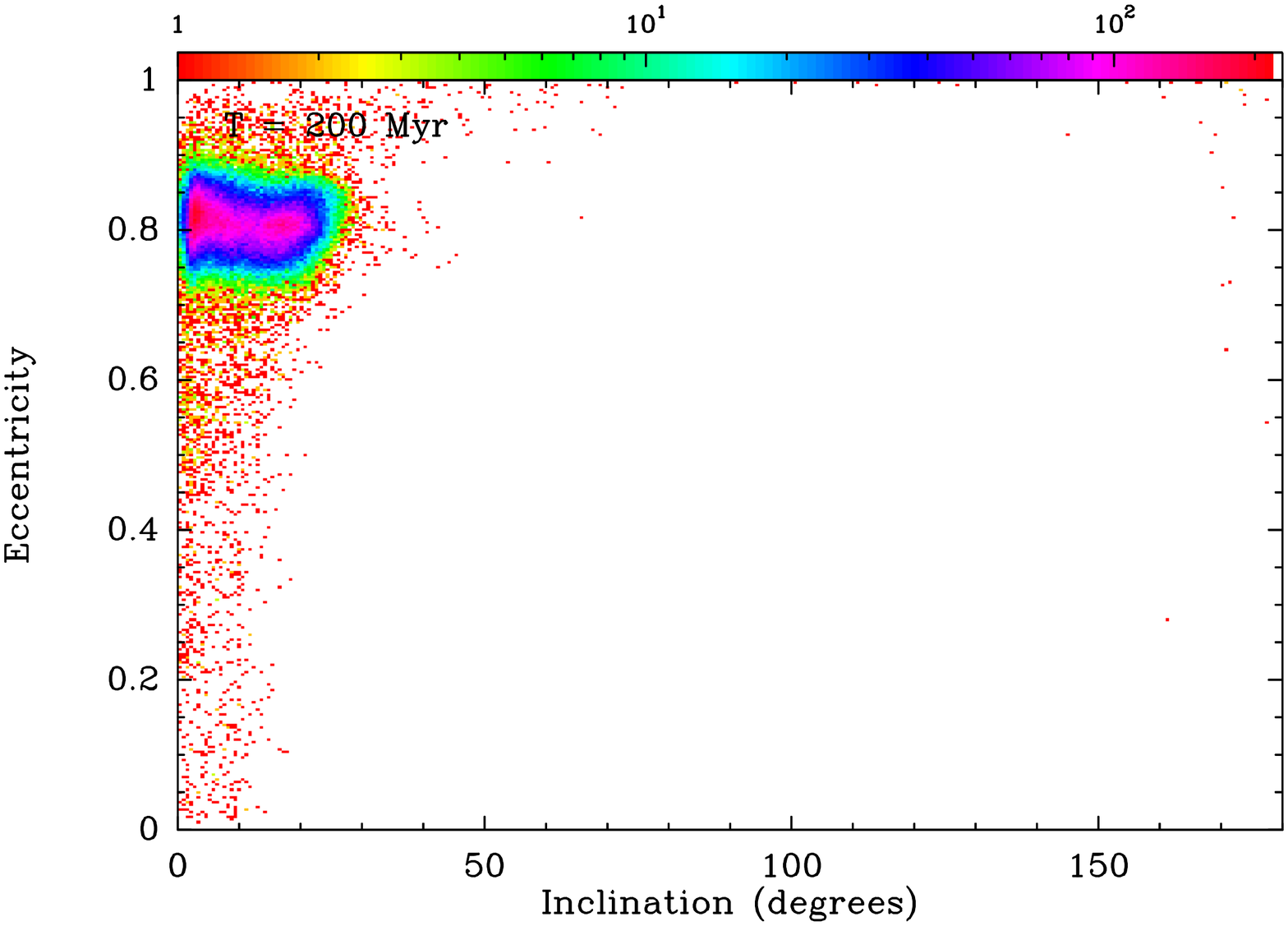} \hfil
\includegraphics[width=0.33\textwidth]{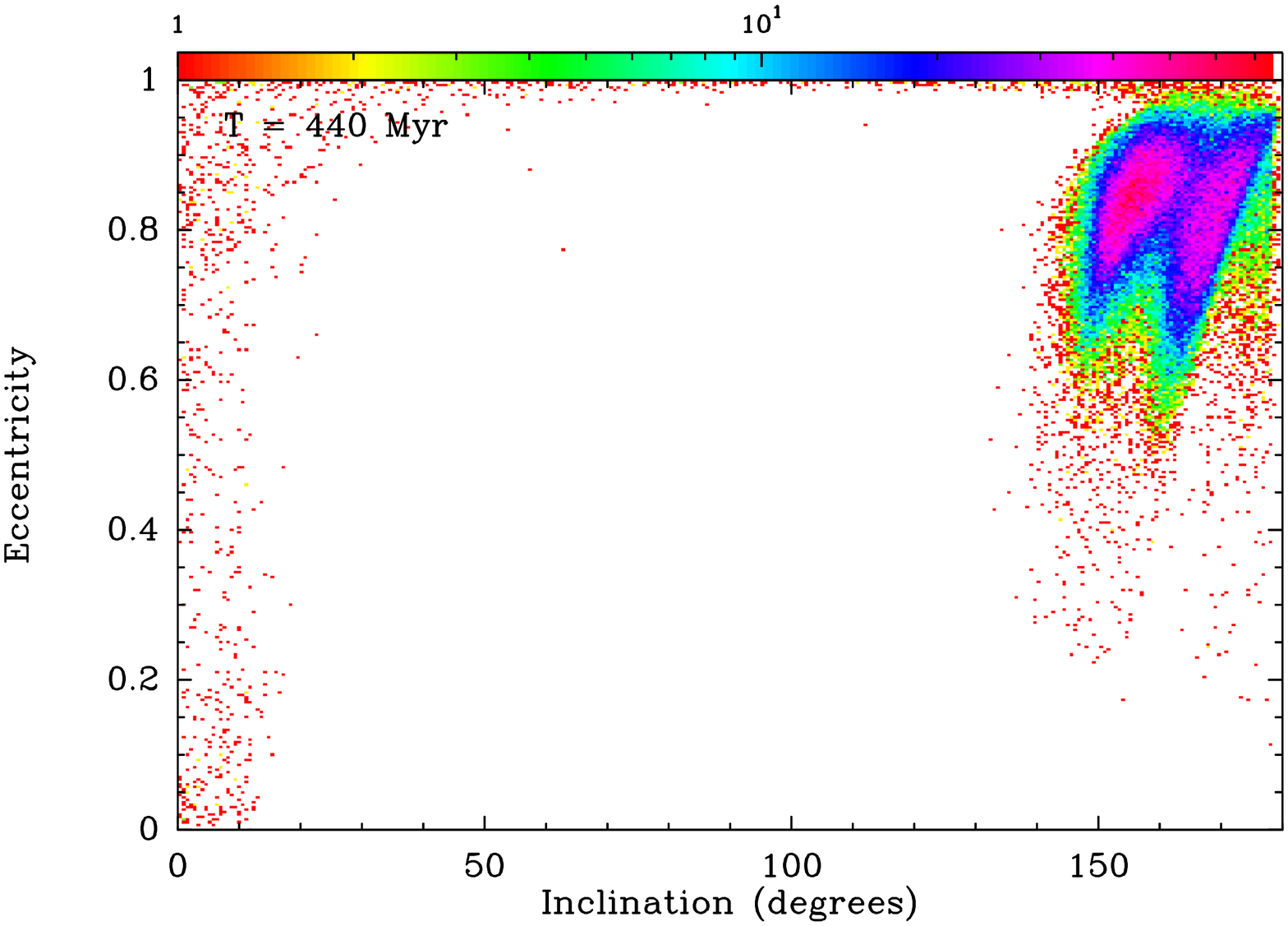}}
\caption[]{Same simulation as presented in Fig.~\ref{simu_0002mjup},
but in inclination--eccentricity space $(i,e)$, at the same
corresponding times~: $t=40\,$Myr (left),
$t=200\,$Myr (middle), $t=440\,$Myr (right)}
\label{inc-ecc}
\end{figure*}
As of yet, we only cared about planar structures, \fomb\ and the disk
were assumed to be coplanar. This choice was indeed guided by the
result of the orbital determination. However, in all simulations
presented above, the disk of particles was not initially strictly
planar. While \fomb\ was assumed to lie in the mid-plane of the disk, a
random inclination between 0 and $3\degr$ was given to the particles
at the beginning, as to mimic a realistic inclination dispersion
within a real disk. Figures~\ref{simu_1mjup}--\ref{simu_0002mjup}
present in fact projected upper views of the disk. We come now to discussing
vertical structures in the disk and their consequences. All the
results presented below concern the simulation with the
$m=0.002\,\mjup$ \fomb, as it is the slowest evolving one, keeping in
mind that this simulation is probably unrealistic if we consider the
self-gravity of the disk. But the secular evolution we present here
holds for any mass regime. For higher masses, the evolution is the
same except that it occurs faster.

Figure~\ref{inc-ecc} shows the same simulation as presented in
Fig.~\ref{simu_0002mjup}, but in inclination--eccentricity space. We
see that at the beginning of the simulation ($t=40\,$Myr), all
particles are as expected still at low inclination while the
eccentricities have started to grow; at $t=200\,$Myr, the
eccentricities are high, but the inclinations are still moderate,
although the peak inclination value of the distribution is now $\sim
30\degr$. Recalling that all inclinations were initially below
$3\degr$, this shows that the inclinations have grown significantly; at
$t=440\,$Myr, the particles have now passed their peak eccentricity
phase (see Fig.~\ref{nue}), but most inclinations have now jumped
close to $180\degr$, meaning they have evolved to retrograde orbits.

Basically, the typical inclination evolution of a disk particle is the
following~: as long as the eccentricity grows, the inclination keeps
increasing while remaining moderate. When the eccentricity nearly
reaches 1, the inclination rapidly jumps close to $180\degr$ and keeps
evolving retrograde afterwards.

This behaviour was of course already present in
Figs.~\ref{simu_1mjup}--\ref{simu_0002mjup}, but somewhat hidden by
the upper view projections. As noted above, at $t=440\,$Myr in
Fig.~\ref{simu_0002mjup}, most particles seem to have
$0\degr<\nu<180\degr$, while in Fig.~\ref{nue}, they obviously have
$180\degr<\nu<360\degr$. This
discrepancy is indeed due to the inclination. At this time, most
particles already have retrograde orbits, so that once projected onto
the $OXY$ plane, the apparent longitude of periastron $\Omega+\omega\cos i$
rather corresponds to $\Omega-\omega$ than to $\Omega+\omega$.
\begin{figure}
\centerline{\includegraphics[width=0.33\textwidth]{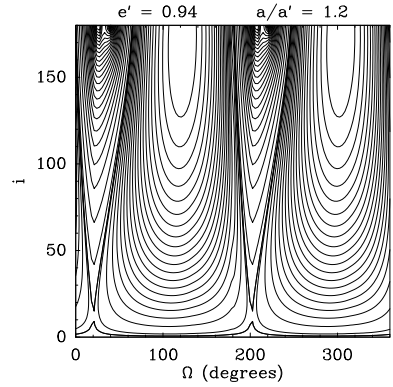}}
\caption[]{A phase portrait in $(\Omega,i)$ space of the averaged spatial
Hamiltonian of a massless particle perturbed by \fomb\ in the same conditions
as in Fig.~\ref{hsec}, computed with constant $\nu=65\degr$ and $e=0.8$.
This approximately describes the secular inclination evolution of the
particle during its eccentricity growth.}
\label{hseci}
\end{figure}
To explain this behaviour, we must get back to our semi-analytical
study. The main difficulty here is that contrary to the planar
problem, the averaged Hamiltonian of the particle has now two degrees
of freedom. The averaged Hamiltonian is usually described by the
classical canonically conjugate Delaunay variables~:
\begin{equation}
\begin{array}{lcl}
\dy \omega & , & G=\sqrt{a(1-e^2)}\\[\jot]
\dy \Omega & , & G\cos i
\end{array}\qquad,
\end{equation}
or similarly, introducing $\varpi=\omega+\Omega$~:
\begin{equation}
\begin{array}{lcl}
\dy \varpi & , & P=\sqrt{a}\,\left(1-\sqrt{1-e^2}\right)\\[\jot]
\dy \Omega & , & G\cos i
\end{array}\qquad,
\end{equation}
where $i$ is the inclination.  As long as the eccentricity does not
reach 1, the fact that the inclination remains moderate actually
validates the planar motion which is described by the canonically
conjugate variables $(\varpi,P)$, or equivalently $(\nu,P)$. Now,
Fig.~\ref{hsec} shows that with $e'=0.94$, a particle starting at low
eccentricity will evolve towards high eccentricity with $\sim$constant
$\varpi$ ($\nu\simeq70\degr$). As $\varpi$ and $P$ are canonically
conjugate, a $\sim$constant $\varpi$ means
$\partial\overline{H}/\partial P\simeq0$, which is equivalent to
$\partial\overline{H}/\partial e\simeq0$. We may thus expect the Hamiltonian
to weakly depend on the eccentricity during this phase.
The dynamics of the particle
during the eccentricity increase will then be approximately well
described drawing level curves of Hamiltonian in $(\Omega, G\cos i)$
space, or equivalently in $(\Omega, i)$ space for a fixed $\nu\simeq
70\degr$ and a fixed arbitrary $e$ value .

\begin{figure*}
\makebox[\textwidth]{
\includegraphics[width=0.33\textwidth,origin=br]{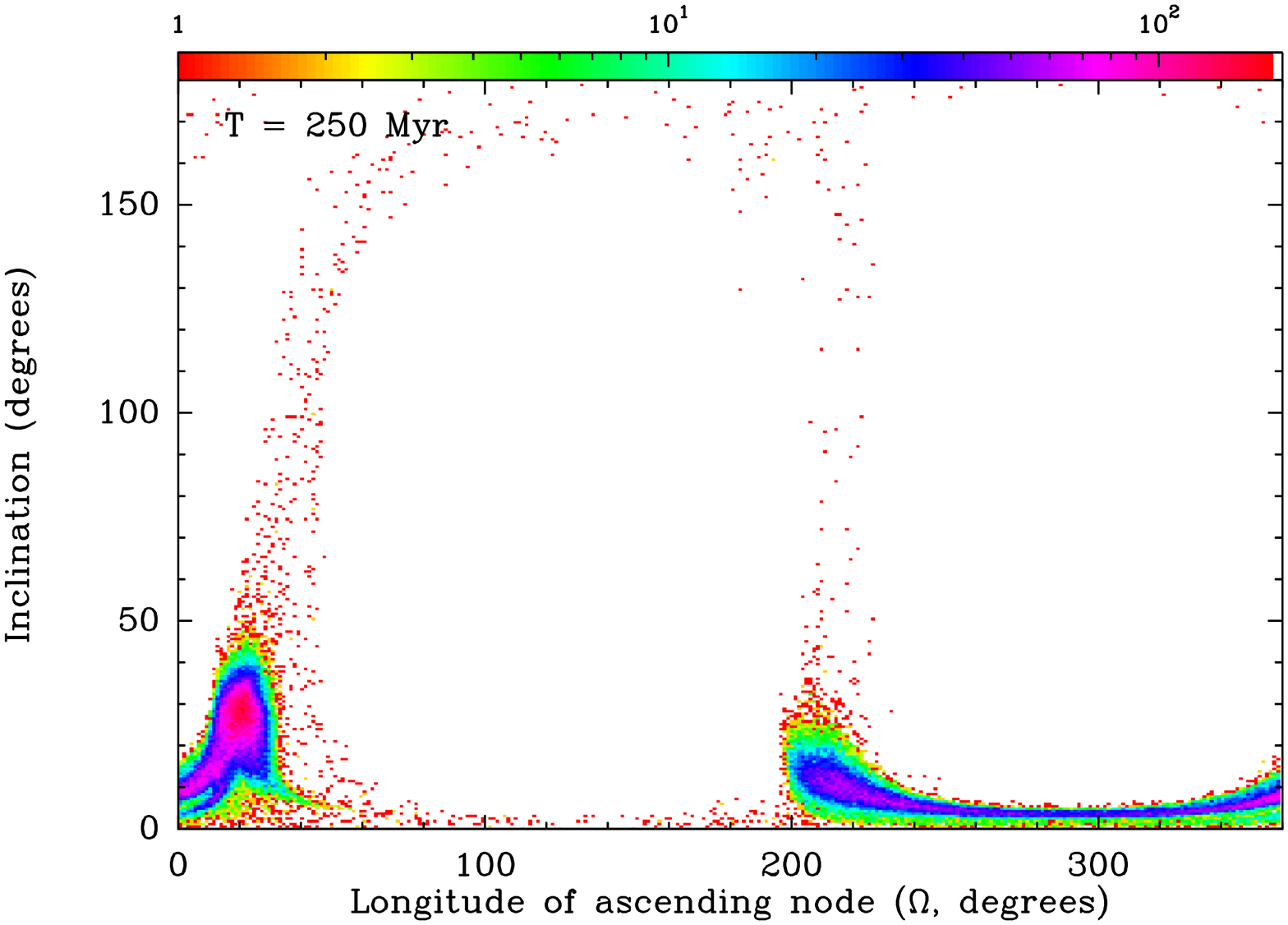} \hfil
\includegraphics[width=0.33\textwidth,origin=br]{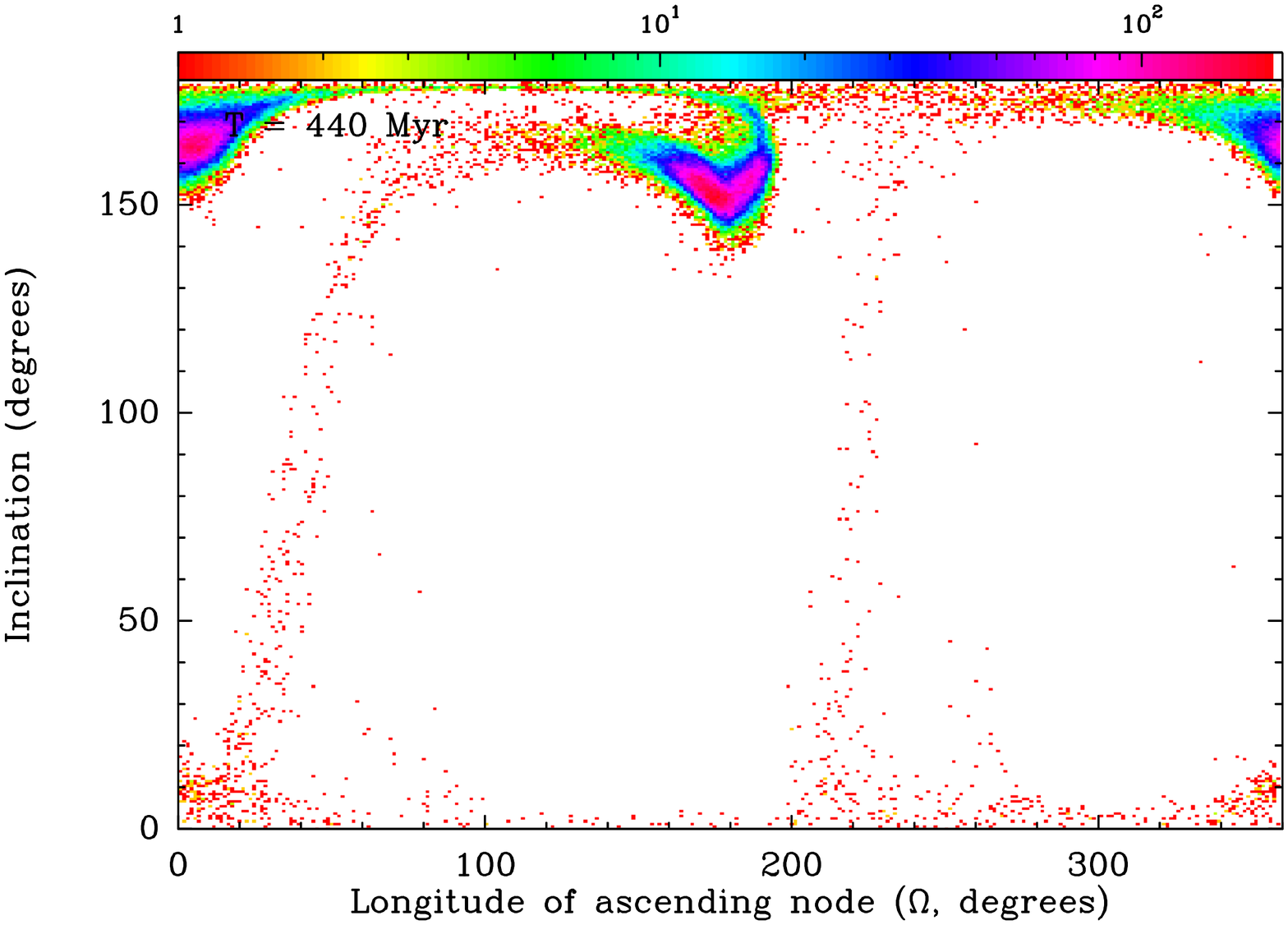} \hfil
\parbox[b]{0.3\textwidth}{
\caption[]{Views of the simulation of Fig.~\ref{simu_0002mjup} and
  in ($\Omega,i$) space like in Fig.~\ref{hseci} at $t=250\,$Myr and
  $t=440\,$Myr}
\label{omega-i}}}
\end{figure*}
Figure~\ref{hseci} shows the result of this computation, performed
with fixed $\nu=65\degr$ and $e=0.8$. We see two major libration
islands that inevitably drive any particle starting at low inclination
towards high inclination. The Hamiltonian curves are here again
explored clockwise. We checked that other choices of $\nu$ and $e$
along the separatrix of the right plot in Fig.~\ref{hsec} lead to
similar diagrams. Following the Hamiltonian level curves, we clearly
see how the particles move towards retrograde orbits.

To check the reality of this analysis, we plot snapshots of our
simulation with $m=0.002\,\mjup$ (Figs.~\ref{simu_0002mjup} and
\ref{inc-ecc}) in $(\Omega,i)$ space like in Fig.~\ref{hseci}. This is
done in Fig.~\ref{omega-i} at $t=250\,$Myr and $t=440\,$Myr. At the
beginning of the simulation (not shown here), all inclinations are
below $3\degr$ while the $\Omega$ values are drawn randomly. All
particles appear thus in the bottom of the diagram in $(\Omega,i)$
space. This remains true for a long time as long as the inclinations
remain low. At $t=250\,$Myr (at this time most particles have already
$e>0.8$, hence our choice of $e$ in Fig.~\ref{hseci}), the
inclinations have started to grow with $\Omega$ value concentrated
around $20\degr$ and $200\degr$. Obviously, the particles follow a
route in $(\Omega,i)$ space that is very close to the level curves of
Fig.~\ref{omega-i}. At $t=440\,$Myr, all particles have moved in the
upper part of the diagram following this route and become
retrograde. Afterwards, the particles get back to low inclinations and
cycle around the two island of libration in $(\Omega,i)$ space.

\begin{figure*}
\makebox[\textwidth]{
\includegraphics[width=0.33\textwidth]{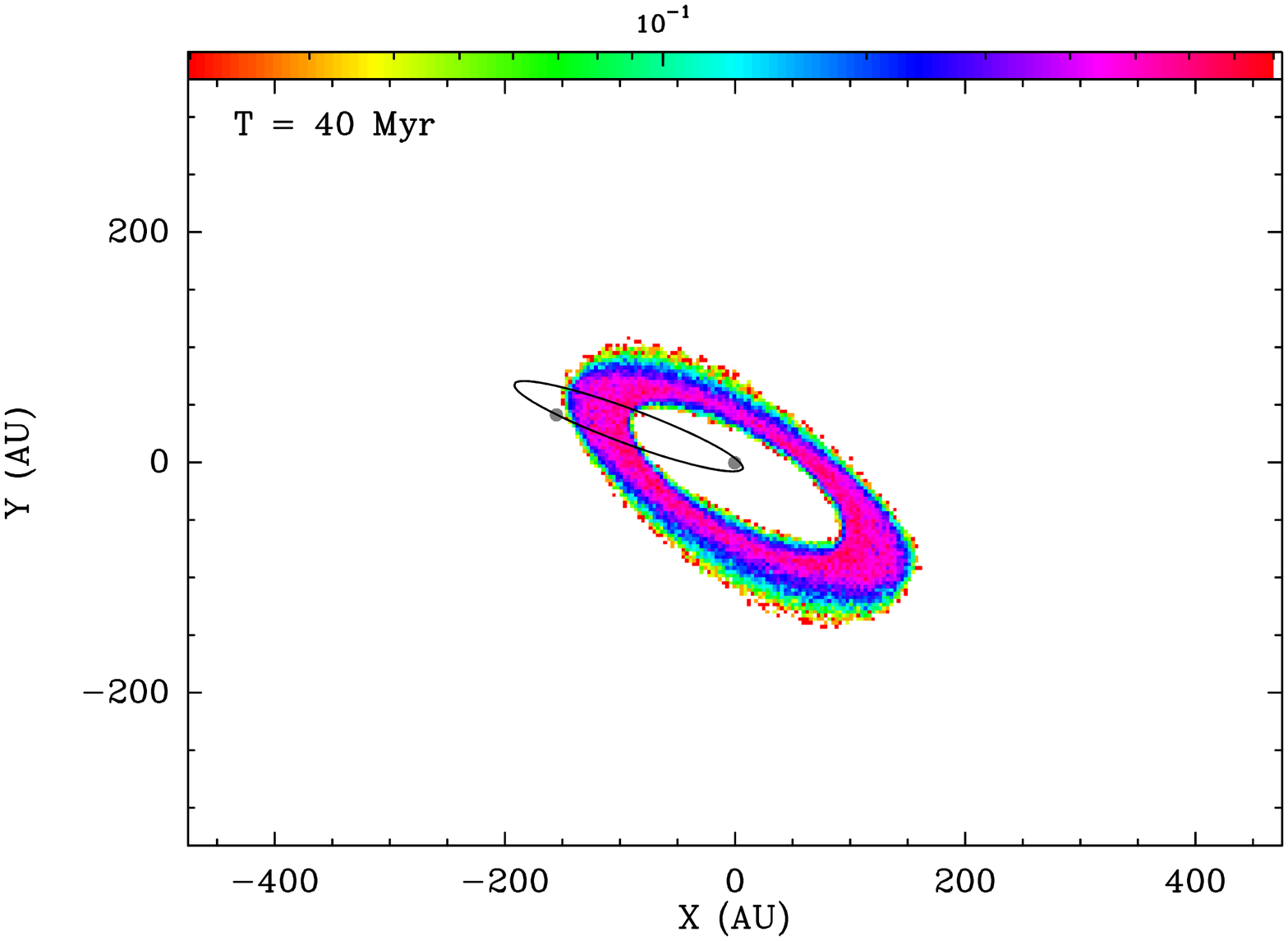} \hfil
\includegraphics[width=0.33\textwidth]{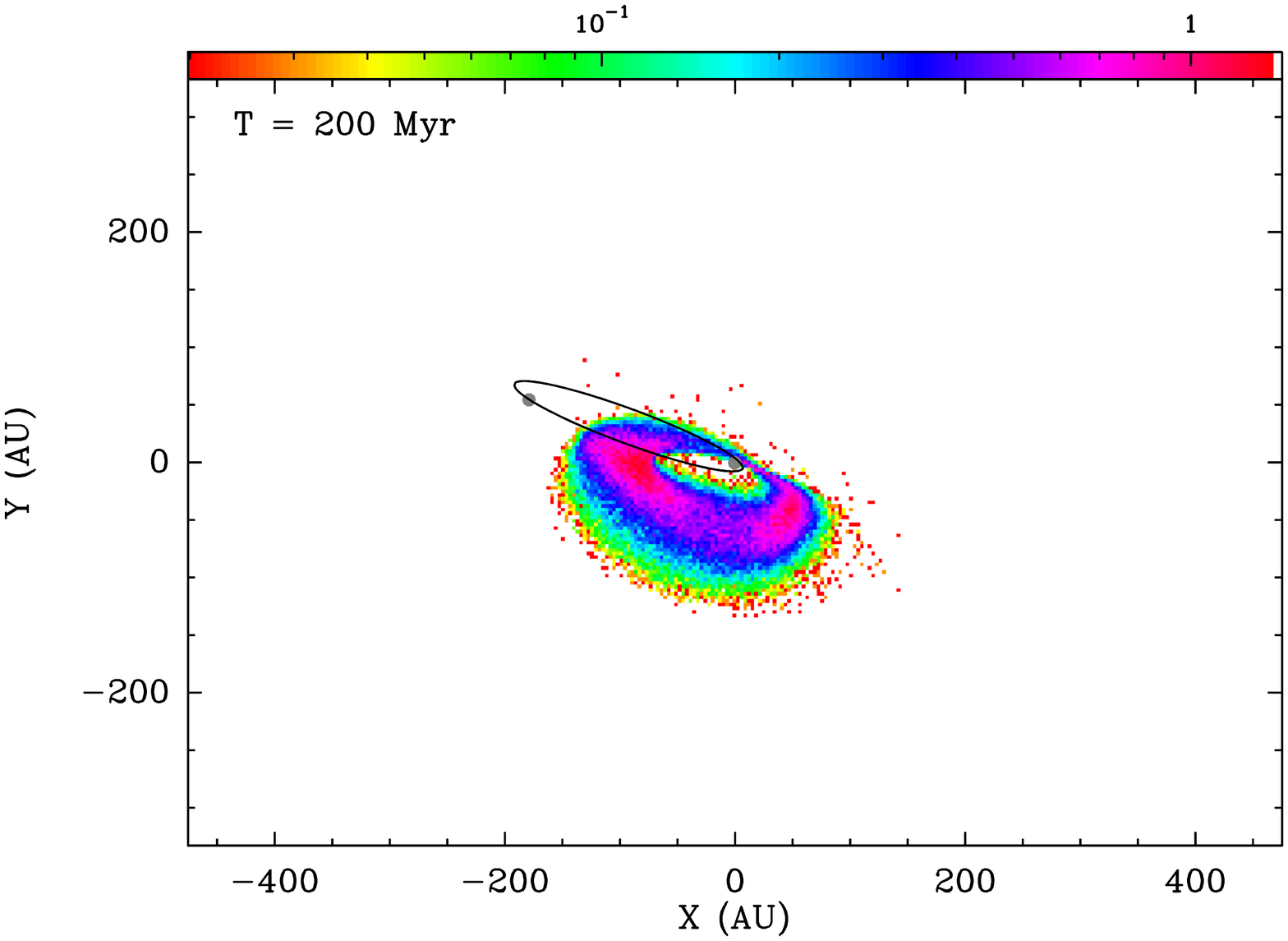} \hfil
\includegraphics[width=0.33\textwidth]{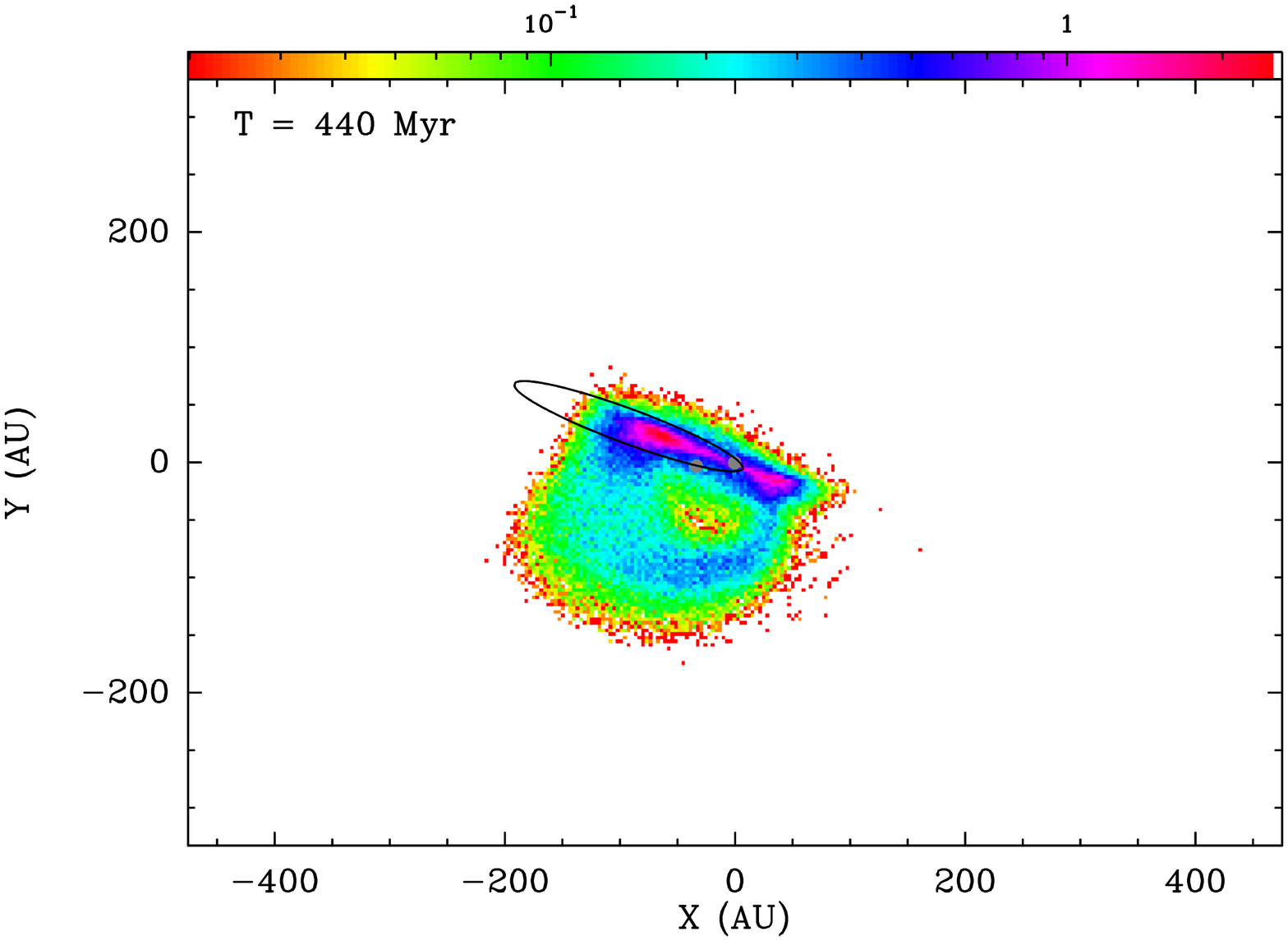}}
\caption[]{Same simulation as presented in Fig.~\ref{simu_0002mjup}, at
the same corresponding times (left: $t=40\,$Myr, middle:
$t=200\,$Myr, right: $t=440\,$Myr)
but the disk is now viewed with a $67\degr$ with respect to pole-on, as
to mimic the viewing conditions of \fom's disk from Earth}
\label{inc-disk}
\end{figure*}
For higher \fomb\ masses like in Fig.~\ref{simu_002mjup}
($m=0.02\,\mjup$), the same dynamics is observed but it occurs
proportionally faster. At the end of the simulation, we have at each
time approximately as many prograde particles as retrograde ones. The
initial disk of particles now assumes a cloud shape rather than a disk
shape. This is illustrated in Fig.~\ref{inc-disk}, which shows the
disk of particles, here again for the $m=0.002\,\mjup$ case and at the
same epochs as in Fig.~\ref{simu_0002mjup}, but viewed with a
$67\degr$ inclination with respect to pole-on. This mimics the viewing
conditions of \fomb's disk from the Earth. At $t=20\,$Myr, the disk
still appears as a clearly eccentric disk with moderate
eccentricities. This is marginally the case at $t=200\,$Myrs and
obviously no longer applies at $t=440\,$Myrs. At this stage (and even
earlier) the simulated disk no longer matches the observed one.

The simulations presented here assumed that the orbital plane of the
perturbing \fomb\ is coplanar with the mid-plane of the disk. As the
coplanarity is not strictly established observationally due to the
uncertainties, we checked other configuration with disks inclined up
to $20\degr$ with respect to the orbital plane of the planet. In all
cases, the behaviour reported in the previous sections remains almost
unchanged. All particles evolve towards high eccentricities and become
retrograde with respect to the planet's orbital plane when reaching
very high eccentricities, so that our conclusions are unchanged~: the
disk inevitably gets a too high eccentricity to match the
observations. Moreover, due to the evolution of the inclinations of
the particles, the disk no longer assumes a disk shape.
\section{Discussion}
\subsection{Disk shaping by \fomb: an unlikely scenario ?}
Our numerical and semi-analytical study shows that
if the perturber is massive enough to efficiently affect the disk,
the pericenter glow
dynamics that applies in the low eccentricity regime cannot be transposed
to the case where the perturber is very eccentric. In that case, we
have a completely different dynamics where the disk particles reach
very high eccentricities and high inclinations.
In a first transient phase, the disk
actually achieves an eccentric disk shape with growing eccentricity,
but afterwards the particles diffuse in phase space and the
steady-state regime does no longer correspond to an eccentric disk
figure. A moderate eccentricity approximately matching the observed
one is in all cases reached shortly after
the beginning of the secular process. The desired time
roughly scales as
\begin{equation}
t_{e=0.1}=\frac{0.04}{m}\qquad,
\label{t01}
\end{equation}
where $m$ is given in Jupiter masses and $t_{e=0.1}$ in Myrs. Reaching
this stage at $t=440\,$Myr would indicate an extremely low planetary mass
($\sim 2.3\,$Lunar masses). As described in the previous section,
such a low mass perturber is unlikely to be able to perturb the dust
ring, given its probable mass. In this regime, the dynamics of the disk
is virtually unaffected by \fomb.

Alternatively, the perturbation of the disk might be recent rather than
primordial. According to that scenario, \fomb's mass should be
closely linked with the date of this event by Eq.~(\ref{t01})
to generate a disk with the suitable bulk eccentricity today.
This situation is nevertheless a transient phase, as the bulk
disk eccentricity is supposed to keep growing. The disk can only
survive in its observed configuration for a short time period comparable to
$t_{e=0.1}$. As a consequence, the higher \fomb's mass, the less
probable this picture is.

In all cases however, the transient elliptic disk is not apsidally
aligned with the perturbing planet, which does not match our orbital
determination for \fomb. However, it is difficult to derive a firm
conclusion on this sole basis, as the determination of the orbital
alignment is only accurate within a few tens of degrees. It must nevertheless
be noted that a $\sim 70\degr$ misalignment would only be marginally
compatible with the data.

Consequently, we come to a contradiction. If we
forget its high eccentricity, the compared orientations of \fomb's
orbit and the dust ring share all characteristics of a pericenter glow
phenomenon. But our analysis revealed that pericenter glow no longer
applies at the eccentricity of \fomb. Even if we consider the lowest
possible eccentricity according to our MCMC distribution
($e'\simeq0.6$, Fig.~\ref{1dmcmc}), Fig.~\ref{hsec} shows that the
topology of the Hamiltonian map is already very different from that
leading to pericenter glow.

We thus have two conclusions. First, \fomb\ is very likely to be a low
mass planet ($\sim$Earth or super-Earth sized). On a $\sim 10\,$Myr
time-scale, a massive planet would destroy the dust ring
(Fig.~\ref{simu_1mjup}). Before that, its secular action would
inevitably drive the disk particles towards high eccentricities
incompatible with the observations. According to Eq.~(\ref{t01}), this
occurs within $\sim 10^5\,$yrs, which is very short. This would
require \fomb\ to have been put on its present day orbit more recently
than that. Given the age of the star, this seems rather unlikely. We
must however note that this is only an order-of-magnitude
estimate. Equation~(\ref{t01}) is actually an empirical fit that hides
some unknown dependencies on the semi-major axis and the eccentricity
of \fomb.  Putting a lower mass limit on \fomb\ is less
straightforward. We have seen that below $\sim$Earth-sized, the planet
has virtually no secular effect on the disk, but this is not
incompatible with the observations. It would just mean that something
else than \fomb\ is responsible for the disk shaping. In fact, a very
low mass \fomb\ would hardly retain enough dust around it to be
detected directly. This was earlier suggested by \citet{ken11} and
more recently by the numerical experiments in \citet{kal13}.
\citet{kal13} propose a lower limit to the mass between Ceres and
Pluto, under the assumption that the more likely models are the most
long lived, and this requires a cloud of dust to be bound to a central
object and have sufficient size to explain the optical luminosity.
Alternatively, \fomb\ could also just be short lived cloud of dust
with no planet mass inside, such as might be created when planetesimal
in the 10--100 km size range collide with each other \citep{kal08,gal13}.

Our second conclusion is that \fomb\ can hardly be responsible for the
shaping of the dust ring into a moderately eccentric ring on its
own. This is actually independent of its mass. If we assume that
\fomb\ is $\sim$sub-Earth sized, then it is just not massive enough to
efficiently influence the ring. According to our semi-analytical
study, this regime holds up to $\sim$Earth-sized planets. If \fomb\ is
more massive, then it has a secular action on the dust ring and
inevitably drives particles towards high eccentricity. This occurs in
any case before the age of the star. Typically, with a super-Earth
sized \fomb, the present-day disk eccentricity is obtained $\sim
10$--$20\,$Myr after the beginning of the simulation. This would imply
\fomb\ to have been put on its orbit that time ago. Here again, given
the age of the star, this seems unlikely, but less unrealistic than
the $10^5$\,yrs required for a massive planet. Therefore we cannot
rule out this possibility. But if \fomb\ was put on its present-day
orbit a few $10^7\,$yrs ago by some scattering event, necessarily this
event was caused by another, more massive planet (see below) which
very probably controls the dynamics of the ring more efficiently than
\fomb\ itself. So, irrespective of its mass, \fomb\ is very probably
not responsible for the sculpting of the observed dust ring.
\subsection{Another planet}
\begin{figure*}
\makebox[\textwidth]{
\includegraphics[width=0.3\textwidth]{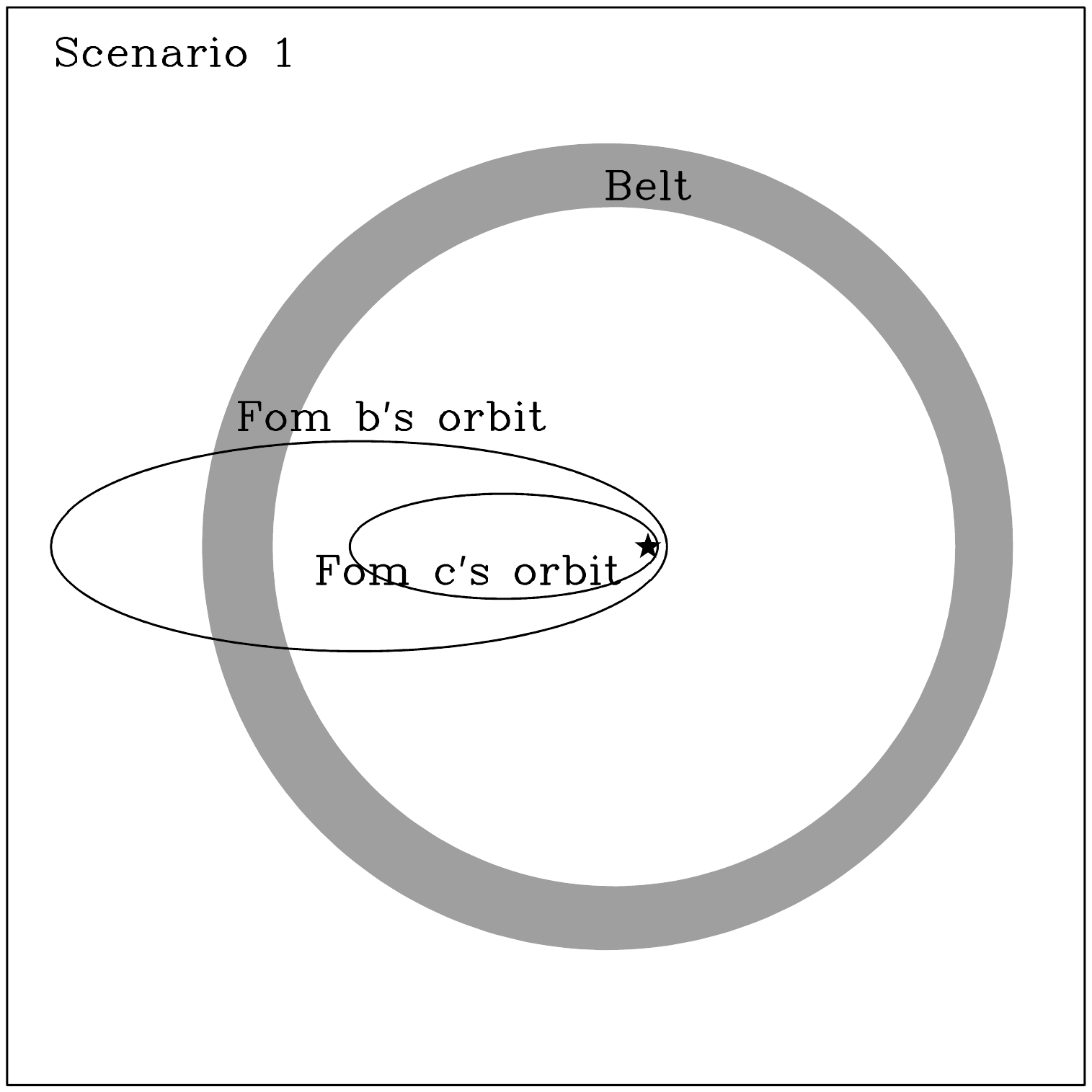} \hfil
\includegraphics[width=0.3\textwidth]{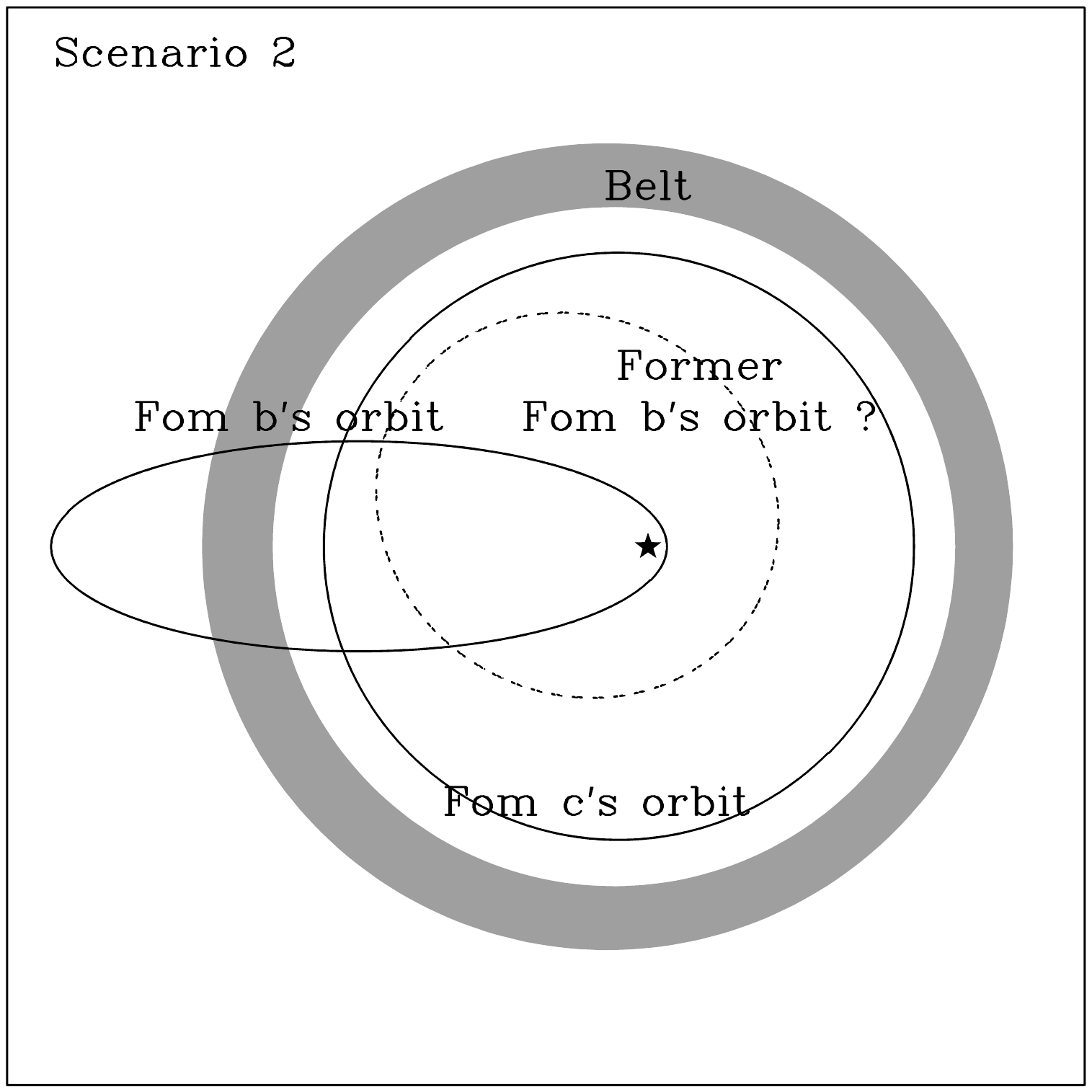} \hfil
\parbox[b]{0.35\textwidth}{
\caption[]{Sketch of the two scenarios for \fomb\ and \fomc\ interaction.
\textbf{Left :} Scenario 1, with non-crossing orbits for \fomb\ and \fomc,
locked in apsidal resonance; \textbf{Right :} Scenario 2, with \fomb\
recently scattered from an inner orbit onto its present day one by a moderately
eccentric \fomc. The latter orbital configuration of \fomb\ is supposed
to be metastable.}
\label{sce}}}
\end{figure*}
If \fomb\ cannot be responsible for the disk sculpting,
a subsequent conclusion is that there must be another, more massive planet
shepherding the dust ring. \citet{kal13} came to the same conclusion
and their numerical experiments assume that a Jupiter mass
planet exists with $a\sim120$\,au, $e\sim0.1$ and serves to dynamically
maintain the inner edge of the belt.
\citet{chi09} actually already invoked the
hypothesis of another planet accounting at least partly for the forced
eccentricity of the belt, concluding that given the residual proper
acceleration of Fomalhaut measured by the \textsc{Hipparcos}
satellite, a $\sim 30\,\mjup$ brown dwarf could be orbiting \fom\ at
$\sim 5\,$au. This possibility was nevertheless ruled out by
\citet{ken13}, who compiled their own observations with other direct
searches for additional companions to \fom\ \citep{abs11,ken09}. They
conclude that no companion more massive than $\sim 20\,\mjup$ is to be
expected from 4\,au to 10\,au and than $\sim 30\,\mjup$ closer. Less
massive companions (Jovian-sized ?) are nevertheless not excluded. The
main problem in this context is to combine the shepherding of the disk
and the survival of \fomb. Basically, to confine the inner
edge of the dust belt at 133\,au as it is observed, a moderately
eccentric Jovian-sized planet must orbit the star between $\sim 90\,$au
and $\sim 120\,$au, depending on its mass \citep[see detailed
calculations by][]{chi09}. But given its high eccentricity, \fomb's
orbit will inevitably cross the orbit of that additional planet
(let us name it \fomc\ hereafter), which
raises the issue of its dynamical stability. Figure~\ref{1dmcmc} shows
indeed that \fomb's periastron is most probably as low as $\sim
8\,$au.

There are two ways to possibly solve this paradox (see
Fig.~\ref{sce}).  The first scenario to suppose that \fomb\ is locked
in a secular (apsidal) resonance with \fomc\ that prevents the orbits
to cross each other. This occurs for instance inside the loops around
$\nu=0$ delimited by the red curves in the $e'=0.5$ and $e'=0.94$
cases in Fig.~\ref{hsec}. Inside these loops, the particle is subject
to a secular resonance where it remains apsidally aligned with the
perturbing planet, while it never crosses its path. Here the particle
would be \fomb\ itself, while \fomc\ would be the perturber. This kind
of locking in secular resonance has already been observed in some
extrasolar systems like $\upsilon\:$Andromedae \citep{chi02}. Although
the eccentricity regime is higher here, this cannot be excluded. It
would have the advantage that it would explain the apsidal alignment
of \fomb\ with the dust ring, as both would be apsidally aligned with
\fomc\ (the belt being apsidally aligned with \fomc\ thanks to
pericenter glow). Figure~\ref{hsec} nevertheless shows that locking in
secular resonance at very high eccentricity, as it is the case for
\fomb, requires a high eccentricity perturber (\fomc). As \fomc\ is
assumed to control the dynamics of the belt instead of \fomb,
one needs to explain now
how the disk remains at low eccentricity, in other words, why a
regular pericenter glow dynamics seems to apply to the disk with respect
to \fomc\ despite its high eccentricity. This is in contradiction with
our previous analysis, but could possibly be due to a wider
  separation between \fomc\ and the disk. Although we cannot firmly
  rule it out, we nevertheless consider this scenario as less
  probable. Obviously a dedicated parametric study is required to
determine in which conditions it could eventually be possible.

The second scenario assumes that \fomb\ is presently on a metastable
orbit (Fig.~\ref{sce}).  In this context, \fomb\ would have resided
initially closer to the star, and it would have been put more or less
recently on its present orbit by a scattering event, possibly
originating from \fomc. We are then back to the hypothesis of a
transient configuration with a more or less recent scattering event.
This scenario would quite naturally explain the very high eccentricity
of \fomb\ and its puzzling belt-crossing orbital configuration. We
could also possibly explain the presence of solid material around this
planet, which actually renders it observable. This material could
actually be captured from the dust belt each time \fomb\ crosses
it. The plausibility of this scenario is basically a matter of
time-scales compared to the masses of both planets.  Figure~\ref{ntp}
shows that after $\sim 100\,$Myr, a particle crossing the orbit of a
Jovian-sized planet has only a few percent chances not to have been
ejected earlier by a close encounter. It can be argued that this
time-scale is not that short compared to the age of the star. This
depends however on the mass of the perturber, here \fomc. Assuming a
more massive \fomc\ would inevitably drastically shorten the ejection
time-scale and render the present day observation of \fomb\ on its
metastable orbit very unlikely. Conversely, a less massive
(Saturn-sized ?) \fomc\ would make it more plausible, but it should
remain massive enough to be able to efficiently sculpt the dust
belt. Another difficulty with this scenario is that it does not
provide a natural explanation for the apsidal alignment between
\fomb\ and the dust belt. As a result of pericenter glow dynamics, the
dust belt would be apsidally aligned with \fomc. It would then be
necessary to explain how \fomb\ would have been put on an apsidally
aligned orbit by the scattering action of \fomc. We must
  however keep in mind that the observed apsidal alignment of
  \fomb\ with the disk is not accurately constrained, so that a
  fortuitous near-alignment within a few tens of degrees is still
  possible. In that scenario, the mass of \fomb\ is less constrained,
as its perturbing action on the disk is recent. We stress however that
a massive \fomb\ ($\sim$Jovian) is rather unlikely, for two
reasons. First, scattering a Jovian-sized planet onto a high
eccentricity metastable orbit would require a very massive \fomc,
which could not fit the observational limits. Second, given the
efficiency of close encounters with a massive \fomb, the scattering
event should have occurred very recently.  Given the age of the star,
we would then be very lucky to witness this event today. For these
reasons, we think that a low-mass \fomb\ is still more likely even in
this second scenario. If \fomb\ is less massive than the
  Earth, then its influence on the disk is damped by the self-gravity
  of the disk so that no constraint can be derived anymore this
  way. The only limitation is then the survival of \fomb\ versus close
  encounters with \fomc.

Both scenarios turn out to present advantages and disadvantages. The
first one is a steady-state configuration where the dynamical
stability of \fomb\ as perturbed by \fomc\ is not ensured, and
where the sculpting of the disk in its present-day shape by a very
eccentric \fomc\ is questionable, but that
would more naturally explain the apsidal alignment between the ring
and \fomb; the second one points towards transient configuration with
a more or less recent scattering event that placed \fomb\ on its
current orbit. The likelihood of the former depends on the
hypothetical dynamical stability of \fomb\ as perturbed by \fomc\
and on the hypothetical existence of configurations allowing
the disk to remain at low eccentricity despite \fomc's high eccentricity,
while that of latter is related to the evolution and survival
timescales of the transient configuration, as compared to \fom's age.
We nevertheless consider that scenario as more likely
than the first one.

An alternative scenario would be that the dust
confinement in \fom's disk
is due to its interaction with gas without any \fomc,
such as suggested recently by \citet{lyr13}. As of yet this cannot be
confirmed nor ruled out. As pointed out by \citet{lyr13}, the key point
is the hypothetical presence of gas in \fom's disk, moreover at such a
long orbital distance.
We know that younger debris disks like $\beta\:$Pictoris actually
contain gas \citep{bran04,xxi}, but for an older system like \fom, it is
less obvious. Only upper limits are available \citep{lis99}.

Both scenarios imply the presence of planets with very short
periastron values. \fomb\ itself has a probable periastron in the
7--8\,au range. If scenario 1 holds, then \fomc\ has an even shorter
periastron than that. \citet{leb13} attributed the near- and
mid-infrared interferometric excesses of Fomalhaut
\citep[see also][]{men13,abs09} to an asteroid belt at about 2\,au
producing a mid-infrared excess, which subsequently produces even hotter
dust detected in the near-infrared. To produce the observed amount of dust,
\citet{leb13} argue that the inner belt had to be somehow excited. The
presence of planets with such short periastron values could actually
provide the suspected source of excitation, or, more generally, may be
related to the process that placed \fomb\ on its peculiar. In scenario
1, \fomc\ would have a periastron in the 2-3\,au range, which would be
enough to excite a belt at 2\,au. The dynamical stability of this belt
would even be questionable and render this scenario unlikely. In
scenario 2, the scattering event that more or less recently put
\fomb\ on its present-day orbit causes it to suddenly approach the
inner belt thanks to its low periastron. This could explain the
excitation of the inner belt and the enhanced dust production.

In all cases, our main conclusions are that \fomb\ is very probably
a low mass planet, possibly orbiting on a metastable orbit, and that another,
more massive planet (\fomc) is required to control the disk dynamics and to
be possibly responsible for the transient orbital configuration of \fomb.
The interplay between both planets is still an open issue. Further work
that continues to investigate and quantify the masses and orbits of the
planets are clearly required. This will be the purpose of forthcoming work.
\begin{acknowledgements}
All computations presented in this paper were performed at the \textsl{Service
Commun de Calcul Intensif de l'Observatoire de Grenoble} (SCCI) on the
super-computer funded by \textsl{Agence Nationale pour la Recherche} under
contracts ANR-07-BLAN-0221, ANR-2010-JCJC-0504-01 and ANR-2010-JCJC-0501-01.
We gratefully acknowledge financial support from the French \textsl{Programme
National de Plan\'etologie} (PNP) of CNRS/INSU, and from the ANR through
contract ANR-2010-BLAN-0505-01 (\textsc{Exozodi}). P. Kalas and J.R. Graham
acknowledge support from NSF AST-0909188 and NASA Origins NNX11AD21G.
\end{acknowledgements}
\end{document}